\documentclass[preprint,3p,sort&compress]{elsarticle}
\usepackage{psfrag}
\usepackage{graphicx}
\usepackage{amssymb}
\usepackage{amsbsy}
\usepackage{amsfonts}
\usepackage{epsfig}
\usepackage{wrapfig}
\usepackage{url}
\usepackage{hyperref}
\usepackage{multirow}
\usepackage{color}     
\usepackage{rotating}  
\usepackage{fancyhdr}
\usepackage{tensor}
\usepackage{inputenc}
\usepackage{tabularx}
\usepackage{hhline}
\usepackage{stix}
\usepackage{miller}
\usepackage{xparse}
\usepackage{caption}
\usepackage[labelformat=simple]{subcaption}

\newcommand{\epsdot}{\dot{\epsilon}}

\newcommand{\Evm}{$\BE_{vm}$}
\newcommand{\Eel}{$\BE_{vm}^{el}$}
\newcommand{\drot}{$\Delta\psi$}
\NewDocumentCommand{\codeword}{v}{%
\texttt{\textcolor{blue}{#1}}%
}

\newcommand{\sref}[1]{\emph{section}~\ref{#1}}%
\newcommand{\fref}[1]{\emph{fig}.~\ref{#1}}%
\newcommand{\Fref}[1]{Fig.~\ref{#1}}%
\newcommand{\tref}[1]{\emph{table}~\ref{#1}}%
\newcommand{\Tref}[1]{Table~\ref{#1}}%
\newcommand{\eref}[1]{\emph{eq}.~(\ref{#1})}%
\def\BE{\boldsymbol{\mathnormal{E}}}
\def\BF{\boldsymbol{\mathnormal{F}}}
\def\BR{\boldsymbol{\mathnormal{R}}}
\def\BU{\boldsymbol{\mathnormal{U}}}
\def\Bq{\boldsymbol{\mathnormal{q}}}
\def\Bx{\boldsymbol{\mathnormal{x}}}
\def\Bone  {{\boldsymbol{ 1}}}
\def\half{\frac{\textmd{1}}{\textmd{2}}}%
\def\calN{\mathcal{N}}
\newcommand{\limsum }[2]{\displaystyle\sum\limits_{#1}^{#2}}
\def\OPexp{\textrm{exp}}%
\newcommand{\fracdis}[2]{\displaystyle\frac{#1}{#2}}

\begin{document}
\title{Automated analysis of continuum fields from atomistic simulations using statistical machine learning}
\author[mimm]{A.~Prakash\corref{cor}}
\ead{arun.prakash@imfd.tu-freiberg.de}
\author[mimm, fzj,rwth]{S.~Sandfeld}
\cortext[cor]{Corresponding author}
\address[mimm]{Micromechanical Materials Modelling Group (MiMM), Institute of Mechancis and Fluid Dynamics, \\Technical University Bergakademie Freiberg, Lampadiusstra{\ss}e 4, 09599 Freiberg, Germany}
\address[fzj]{Institute for Advanced Simulation -- IAS-9: Materials Data Science and Informatics, \\Forschungszentrum Juelich GmbH, 52425 Juelich, Germany}
\address[rwth]{Chair of Materials Data Science and Materials Informatics,\\ Faculty 5 -- Georesources and Materials Engineering, \\RWTH Aachen University, 52056 Aachen, Germany}

\begin{abstract}

Atomistic simulations of the molecular dynamics/statics kind are regularly used to study small scale plasticity. Contemporary simulations are performed with tens to hundreds of millions of atoms, with snapshots of these configurations written out at regular intervals for further analysis. Continuum scale constitutive models for material behavior can benefit from information on the atomic scale, in particular in terms of the deformation mechanisms, the accommodation of the total strain and partitioning of stress and strain fields in individual grains. In this work we develop a methodology using statistical data mining and machine learning algorithms to automate the analysis of continuum field variables in atomistic simulations. We focus on three important field variables: total strain, elastic strain and microrotation. Our results show that the elastic strain in individual grains exhibits a unimodal log-normal distribution, whilst the total strain and microrotation fields evidence a multimodal distribution. The peaks in the distribution of total strain are identified with a Gaussian mixture model and methods to circumvent overfitting problems are presented. Subsequently, we evaluate the identified peaks in terms of deformation mechanisms in a grain, which e.g., helps to quantify the strain for which individual deformation mechanisms are responsible. The overall statistics of the distributions over all grains are an important input for higher scale models, which ultimately also helps to be able to quantitatively discuss the implications for information transfer to phenomenological models.

\end{abstract}


\maketitle

\section{Introduction}
Over the past decade, data science and informatics (DSI) has evolved as the fourth paradigm of scientific research \cite{hey2009}, in addition to the traditional paradigms of experiments/empirical reasoning, theory/modeling and computation/simulation, and has shown great potential for accelerated materials development \cite{sumpter2015,ramprasad2017}. A characteristic feature of approaches and predictive methods from DSI is that they focus strongly on the data itself while still allowing to consider physical domain knowledge \cite{prakash2018}. This essentially makes it particularly attractive for increasing the synergy between, e.g., experiments and simulations. Challenges remain, nonetheless, concerning the availability of data -- usually summarized through the ``four Vs'' (volume, variety, veracity and velocity) --  since algorithms and methods in DSI typically require well-curated and sufficiently large datasets \cite{kalidindi2015}. In recent years, high-throughput experiments with improved imaging techniques as well as large-scale computations, e.g., atomistic simulations, with high performance computing resources have helped significantly in this regard \cite{prakash2018}.

Atomistic simulations have now become an invaluable tool in the field of  computational materials science and have been particularly useful in advancing our knowledge on the mechanical behavior of materials \cite{Farkas2013}. Such simulations are regularly used to study defect-defect interactions \cite{chassagne2011,pan2014,vaid2019,lee2020}, elasto-plasticity \cite{zepeda2017,Brandstetter2006,Prakash2017NC}, fracture \cite{cao2007,bitzek2015}, irradiation \cite{aidhy2015,holmstrom2010,guenole2016,guenole2017}, and other phenomena in crystalline materials. Large scale computations have allowed researchers to study fundamental processes in nanoscale objects like thin films \cite{kumar2011a,li2020}, nanoparticles \cite{shreiber2015,sharma2018,houlle2018} and nanowires \cite{weinberger2012,xie2020,lee2020}, and have furthermore, facilitated experimentally informed simulations \cite{moody2014,prakash2015,prakash2017}, resulting in an improved fundamental understanding of small-scale plasticity.

Studies with atomistic simulations on nanocrystalline materials -- the focus of the current work -- have been successful in elucidating the deformation mechanisms and the role of grain boundaries (GBs) in polycrystalline materials with grain sizes below $100$\,{nm} \cite{Hahn2015,Panzarino2016}. Such studies have demonstrated the role of GBs as sources and sinks for dislocations, due to the lack of nucleation sources like Frank-Read or spiral sources in individual grains \cite{schiotz2004a,vanswygenhoven.etal.2006}. Furthermore, the inverse Hall-Petch effect characterized by a decrease in strength with decreasing grain size below roughly 10$\sim$20 nm is attributed to a change in deformation mechanism from that dominated by intra-grain dislocaitons to grain boundary mediated deformation \cite{van-swygenhoven2001a}.

Efforts have been made to link atomistic data with continuum scale data and simulations \cite{dewald2006,tucker2011,weinberger2012, Gunkelmann2017_ComputMaterSci135, KOSITSKI2018_ComputMaterSci149}. This is particularly important for the development of robust and reliable constitutive models that are able to consider emergent properties from the nanoscale, where, e.g., plasticity is governed by a paucity of dislocation sources and an increased fraction of grain boundaries, interfaces and surfaces. Continuum scale modeling of complex material phenomena at the nanoscale hinges not only on the availability of numerical values of  parameters for constitutive models, but also on information regarding the relative contribution of individual mechanisms to the total deformation, and statistical data pertaining to the distribution of fields like strain, stress, texture etc. Obtaining such information is generally tedious and time consuming due to a lack of automation~\cite{bitzek.etal.2008,Prakash2017NC,gupta2020}.

Data mining and the use of machine learning algorithms have become extremely popular in materials science covering many different problem classes ranging from analysis of microscopy data \cite{Roberts2019, Ma2020, Trampert2021} to design of alloys and meta-materials, accelerated materials discovery and machine learning-guided  theory development \cite{Kalidindi2020, Steinberger2019, Zheng2020}. Such machine learning methods can also help overcome the problem of obtaining the above mentioned information in an automated manner and facilitate knowledge transfer between the atomistic and continuum scales \cite{gomberg2017,vimal2022}. Additionally, one of the challenges in machine learning of ``never enough data'' is easily overcome since every atom is essentially a data point, and typical simulations involve millions to billions of atoms. Mining such data will help understand the complex relationships between the emerging local fields -- such as strain, stress, texture -- with the macroscopic response, and support in the formulation of microstructure-property relationships \cite{ceriotti2019,bock2019}.

In this work, we develop a methodology for automated analysis and visualization of continuum fields like strains and rotations in large scale atomistic simulations. In particular, we use data mining and statistical machine learning algorithms to extract key features from distributions of total strain, elastic strain and rotation in individual grains. Generally, machine learning tasks are classified into two broad categories \cite{tan2016}: descriptive and predictive. The former approach is of explanatory nature and aims to identify patterns in the underlying data based on correlations and trends, whilst the latter approach is used to predict and foresee events induced by certain known factors. 

Herein, the focus in the development of this methodology is primarily of descriptive nature. We develop an understanding of unimodal distributions by identifying the mathematical form of the distribution. For multimodal distributions, the values of individual peaks are identified using a Gaussian mixture model, wherein methods to circumvent overfitting problems are presented. The individual peaks are then correlated with the deformation mechanisms observed in a grain. Finally, approaches to make the methodology a predictive one are briefly discussed.

\section{Details on the data used}\label{sec:DataSource}
\begin{figure}[htbp!]
\centering
\includegraphics[width=0.9\textwidth]{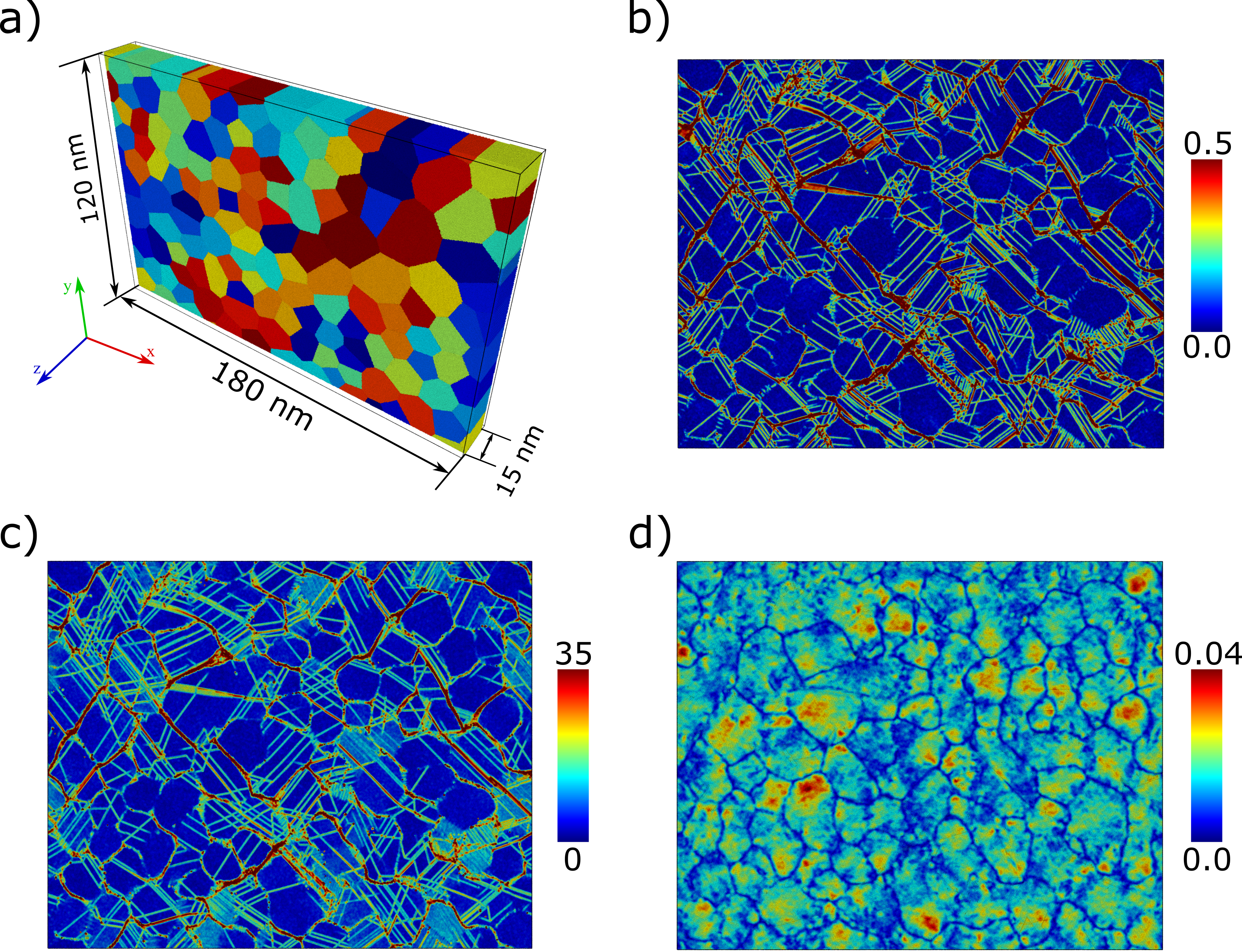}
\caption{Atomistic sample used as data source in the current work: a) Thin film sample in the undeformed configuration along with sample dimensions and global directions. Atoms are colored according to their grain number; b-d) Map of \Evm{}, \drot{} and \Eel{}, respectively, in individual grains in the deformed state after 10 \% applied strain. The rotation angle in c) is in degrees. For further statistical analysis, the distributions of these three quantities are analysed over the domain of an individual grain.} \label{fig:DataSource}
\end{figure}

The data used in the current work is sourced from a tensile simulation of a nanocrystalline thin film sample. The initial structure, with a mean grain size of 15 nm, is generated by means of a constrained Voronoi tessellation (CVT) \citep{Prakash2017NC,Xu2009} so as to reduce non-equilibrium junctions \cite{serrao2021}. The initial dimensions of the thin film are 180 nm x 120 nm x 15 nm (see \fref{fig:DataSource}a). The atomistic sample subsequently generated using the open source toolbox \emph{nano}\textsc{sculpt} \cite{prakash2016b} contains approximately 19 Mio atoms. The interatomic forces are modeled with an embedded atom method (EAM) potential for Al \cite{mishin.etal.1999} using a stable time increment of 1 fs. The structure is relaxed using the \textsc{fire} algorithm in standard molecular statics simulations up to a force norm of 10$^{-7}$ eV/\AA, and subsequently equilibrated at 300~K for 20 ps, followed by constant pressure simulations at 300~K using a Nos\'{e}-Hoover thermo-barostat to ensure zero pressure. 

The relaxed and equilibrated structures are then subjected to uniaxial tension at a constant strain rate of $\epsdot = 10^9 s^{-1}$ at 300~K in the NPT ensemble. Uniaxial strain is modeled by continuously scaling the atomic coordinates and the box-length along the $y$ direction at a fixed rate, whilst simultaneously allowing for contraction along the $x$ direction using a Nos\'{e}-Hoover thermo-barostat. Periodic boundary conditions are imposed along the $x$ and $y$ directions; free boundaries exist in the $z$ direction along the thickness of the film.

The molecular dynamics (MD) simulations are performed with the atomistic simulation code IMD \cite{roth2019}. Snapshots are written out every $0.05\%$ strain increment (time increments of 5 ps). Each snapshot is a separate file in a simple ASCII file format containing information on the position and index of each atom. The index has no physical meaning and has no bearing on the simulations itself. It is merely used to identify each atom uniquely and can hence be used to map an atom to a particular grain in the initial configuration.

The so-obtained snapshot files are subsequently analyzed using the open-source visualization tool \textsc{Ovito} \cite{ovito}. Defect structures are identified using the common neighbor analysis (CNA) modifier \cite{honeycutt.andersen.1987}. Local atomic strains are determined using the atomic strain modifier, which calculates the deformation gradient tensor $\hat{\BF}$ using the displacement of individual atoms, and subsequently, the \emph{Green-Lagrange}an strain tensor $\hat{\BE}$ as follows:
\begin{equation}
\hat{\BE}=\half  \left( \hat{\BF}^T \hat{\BF} - \Bone \right).
\end{equation}
This atomic strain tensor is then averaged over the neighborhood of each atom using a cutoff radius of 1~nm  to obtain an averaged \emph{total strain} tensor per atom:
\begin{equation}
\BE=\frac{ \limsum{i}{N} \hat{\BE}_i \, \hat{V}_i }{ \limsum{i}{N}\hat{V}_i },
\end{equation}
where $\hat{V}$ denotes the volume of each atom as determined by a Voronoi construction. $\BE$ provides a measure of the total strain. The von Mises equivalent total strain $E_{vm}$ is then computed using the averaged strain tensor as follows \cite{Shimizu2007}:
\begin{equation}\label{eq:evm}
E_{vm} = \sqrt{E_{xy}^2+E_{yz}^2+E_{zx}^2+\frac{1}{6}\left[\left(E_{xx}-E_{yy}\right)^2 + \left(E_{xx}-E_{zz}\right)^2 + \left(E_{yy}-E_{zz}\right)^2 \right]}.
\end{equation}
Since $\hat{\BF}$ is already known, we may perform the polar decomposition to obtain the rotation tensor $\hat{\BR}$ and the stretch tensor $\hat{\BU}$ as $\hat{\BF} = \hat{\BR} \, \hat{\BU}$. The so-computed rotation tensor is then averaged over the a shell of next nearest neighbors to obtain the averaged rotation tensor $\BR$, and is expressed as a quaternion $\left( \Bq, q_0\right)$. We now compute the incremental angle of rotation $\Delta\psi$, called \emph{microrotation}, as 
\begin{equation}
\Delta\psi=2 \cdot \cos^{-1}\left(q_0 \right),
\end{equation}
which provides a measure of the rotation experienced by an atom in the deformed sample, with respect to the initial configuration.

\noindent
The local elastic strain $\hat{\BE}^{el}$ is computed using the \emph{elastic strain modifier} in \textsc{Ovito} and averaged with an identical procedure as for the \emph{Green-Lagrange}an strain tensor $\BE$. Subsequently, a von Mises equivalent elastic strain is computed using an equivalent relation as in \eref{eq:evm}.

The analyzed snapshots are now written out as individual files in the IMD ASCII file format. Each file contains the original snapshot data (position and index) appended with the \emph{a-posteriori} calculated data via Ovito. Each column represents the distribution of a field quantity evaluated at discrete points given by the atomic coordinates. For subsequent analysis, we restrict ourselves to three columns: \texttt{ShearStrain} (von Mises equivalent total strain \Evm{}), \texttt{DeltaRot} (angle of rotation \drot{}) and \texttt{ElasStrVM} (von Mises equivalent elastic strain \Eel{}). \Fref{fig:DataSource}b)-d) shows the distribution of the three field quantities on a slice of the thin film after 10\% applied strain.

For further statistical analysis, these field quantities are analyzed over the domain of an individual grain. All atoms are provided with a unique grain number to identify the grain the belong to. This is done by using the index of each atom and mapping it to the grain they belong to in the initial configuration. As a result, the definition of the domain of the grain in the deformed configuration is identical to that in the initial configuration.

\begin{figure}[htbp!]
\begin{center}
\includegraphics[width=0.99\textwidth]{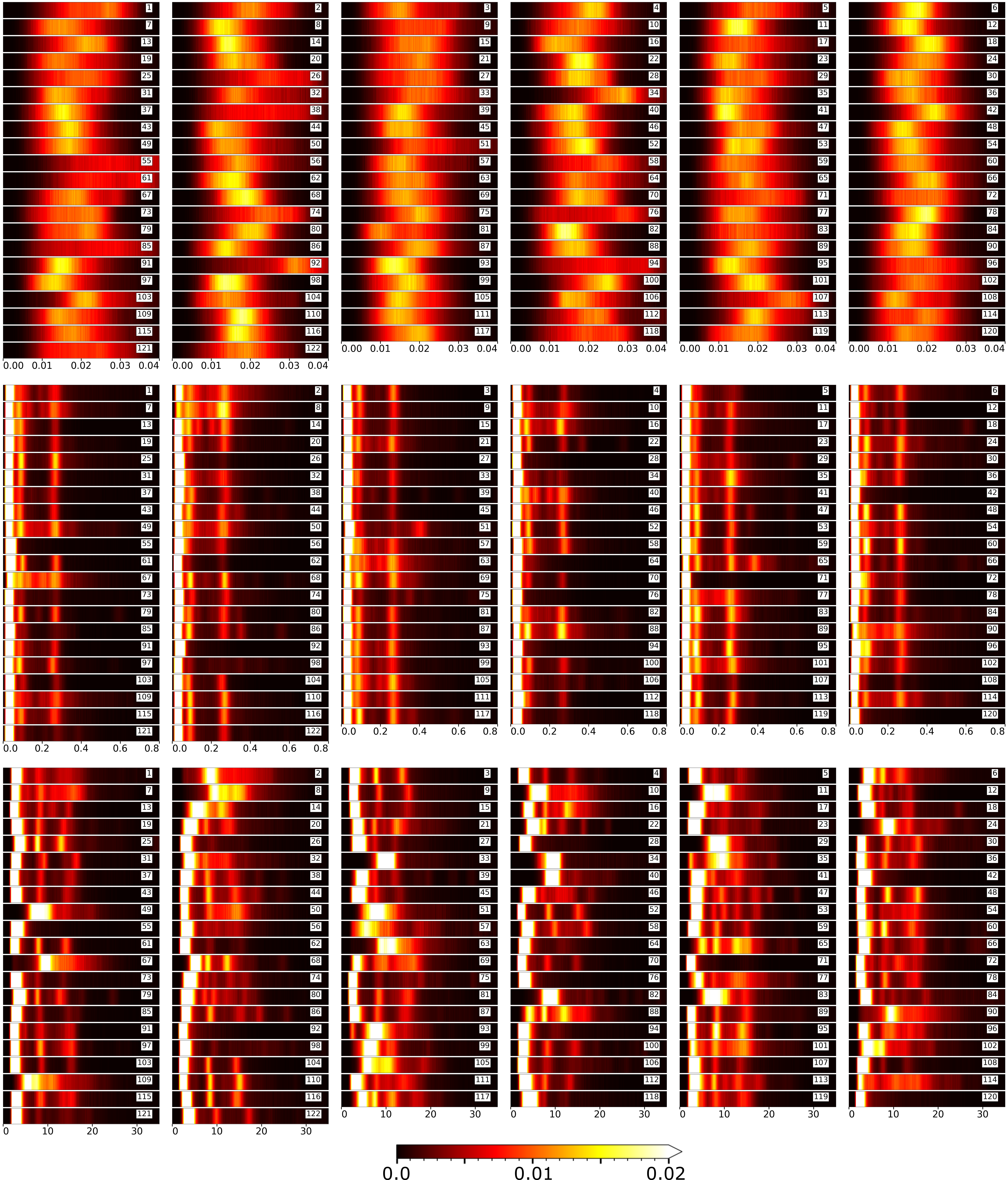}
\end{center}
\caption{Distributions of elastic strain \Eel{} (top row), total strain \Evm{} (middle row) and microrotation \drot{} (bottom row) as heat maps, calculated using 200 bins over the corresponding data range. The colormap shows the volume fraction, i.e. the number of atoms in a bin as a fraction of the total number of atoms in the grain. Each horizontal heat map shows the distribution of the corresponding field in a single grain. The grain number is noted in the white box in each heat map.} \label{fig:HeatMap}
\end{figure}

\Fref{fig:HeatMap} shows the distribution of the three field quantities \Eel{}, \Evm{} and \drot{} in individual grains as one-dimensional (1D) heat maps. Representation of distributions as heat maps is an alternative to the traditional methods of representation as histograms and kernel density estimates (KDE). The latter methods require significant amount of space for plotting the data of all grains, which makes comparison of distributions over multiple grains a cumbersome operation. For instance, in the case of the current sample, which corresponds to a particular snapshot at 10\% strain from a single simulation, we obtain 366 plots for the three field distributions in 122 grains. Plotting as heatmaps ensures minimal space usage for all the plots -- whilst qualitatively keeping all the details -- and facilitates easy comparison of distributions. Nonetheless, for applications that go beyond mere visualization, e.g., determination of the mathematical form of the distribution or a correlation of the distribution with deformation mechanisms in the grain, representation of distributions as histograms or KDEs becomes inevitable.

In general, comparison of plots of multiple grains demands that the set of units used on the ordinate and abscissa axes be identical in all plots. Since the distributions are siginficantly inhomogeneous, manual tuning of the plot limits, which is usually done by inspecting the distributions of only a few grains, can result in truncation of outlying peaks in multimodal distributions. On the other hand, merely using the maximum and minimum value of the data array would result in tails of a distribution with frequency values close to zero dominating the plots of the distributions. This is a non-trivial problem and needs to be solved in an objective manner; otherwise the comparability between plots would suffer.
Here, we propose a method for a reasonable, automated determination of the plot range, the details of which are provided in the supplementary material. Using this automated approach, the distributions are plotted as histograms in \emph{supplementary figures} \ref{sfig:EelHist}, \ref{sfig:EvmHist} and \ref{sfig:DrotHist}. The plots show multiple peaks in the distributions of \Evm{} and \drot{}, evidencing the multi-modal nature of the fields. By comparison, the distribution of \Eel{} is, by and large, unimodal.

The multi-modal nature of the distributions of \Evm{} and \drot{} can be explained as follows. In nanocrystalline materials, dislocations are nucleated at grain boundaries due to the paucity of intra-granular dislocation sources. Such dislocations then traverse through the grain before being absorbed by the opposite grain boundary. As a result, atoms that experience the movement of such dislocations evidence higher strain than those located elsewhere in the grain. The same argument also holds true for the orientation distribution within a single grain. Consequently, the distributions of the total strain and the orientation display a multi-modal character. By contrast, all atoms undergo elastic deformation. As a result, the distribution of elastic strain is likely to be unimodal.

\section{Methodology}\label{sec:Method}
In order to statistically analyze the field distributions in individual grains, we differentiate between uni-modal and multi-modal distributions. Our aim is to determine the location of individual peaks in the multi-modal distributions of total strain (\Evm{}) and microrotation (\drot{}) fields. This information is then used to understand the mechanistic source of the inhomogeneity in the distribution. For the case of unimodal distributions, we aim to identify the mathematical form of a function(al) that can statistically represent the unimodal distribution under consideration, over all grains in the polycrystal. At a later stage, such statistics over multiple samples, strain states and materials can be compared in order to assess the transferability and tractability of such a mathematical representation to higher length-scale simulation frameworks like crystal plasticity or strain gradient models.

\subsection{Mathematical form of uni-modal distributions}
To identify the mathematical function that best describes a uni-modal distribution, we first fit an array of functions to the data of individual grains. The \texttt{stats} library implemented in the SciPy package \cite{2020SciPy} is used to define various functions and to fit them to the given dataset using the maximum likelihood estimation \cite{bishop2006}. For the purpose of this work, we only use a subset of all available functions in the \texttt{stats} library. The chosen candidates are those that are likely to reflect the uni-modal distribution under consideration. A total of 32 functions are used in the current work (see \fref{fig:FuncFit} and \tref{tab:SSEallFunc} for the function names). The reader is referred to the official documentation of the \texttt{stats} library \cite{2020SciPy} for the mathematical formulation of the individual functions.

Once the parameters of individual functions that result in the best fit of the given distribution (in an individual grain) are identified, an assessment of the function that provides a good description over all grains needs to be made. For this purpose, we compute the sum of squared errors (SSE) of the fits with individual functions with respect to the data of individual grains. A decision on the best fit function is then made by evaluating the statistics of SSE of individual functions over all grains.

\subsection{Gaussian mixture model for multi-modal distributions}
We employ a Gaussian mixture model to identify the peaks in a multi-modal distribution. 
Gaussian mixture models (GMMs) belong to the class of unsupervised machine learning algorithms that are generally used to cluster data points that share certain common characteristics \cite{bishop2006}. It is a probabilistic model in which the clustering function is a mixture of several Gaussian functions. As a result, we obtain the probability with which each data point can be assigned to a particular Gaussian function or cluster. GMMs can be deemed as a generalization of the well-known $k$-means clustering which is a so-called hard clustering method where each point is assigned definitively to a cluster without an uncertainty measure to qualify this association \cite{bishop2006,press2007}. By contrast, GMMs incorporate information on not just the probability with which a data point can be associated with a cluster, but also the covariance structure of the data, in addition to the centers of the individual Gaussians.

Formally, a GMM is defined as the weighted sum of $K$ component Gaussian densities as follows:
\begin{equation}\label{eq:GMM}
p(\Bx) = \limsum{i=1}{K}w_i \, \calN\left(\Bx|\mu_i,\sigma_i \right),
\end{equation}
where $p(\Bx)$ is the probability of a data point $\Bx$, $w_i$ are the weights, and  
$\calN\left(\Bx|\mu_i,\sigma_i \right)$ are the component Gaussian densities, which in $D$-dimensions are given by:
\begin{equation}\label{eq:GaussFn}
\calN\left(\Bx|\mu_i,\sigma_i \right) = \fracdis{1}{\left( 2\pi\right)^{D/2} \left| \Sigma_i\right|^{1/2} } \OPexp \left\{ - \half \left( \Bx - \mu_i\right)^T \Sigma_i^{-1} \left( \Bx - \mu_i\right) \right\}, 
\end{equation}
with the mean vector $\mu_i$ and the covariance matrix $\Sigma_i$. The mixture weights $w_i$ are defined as the partition of unity, i.e. 
\begin{equation}\label{eq:wts}
\limsum{i=1}{K} w_i = 1.
\end{equation}
In the case of a 1-D data set, \eref{eq:GaussFn} reduces to 
\begin{equation}
\calN\left(x|\mu_i,\sigma_i \right) = \fracdis{1}{\sigma_i \sqrt{2\pi} } \OPexp \left( - \fracdis{\left( x - \mu_i\right)^2}{2\sigma_i^2} \right).
\end{equation}
\vspace{1ex}

For a fixed number of Gaussian components, the parameters $\mu_i$, $\sigma$ and $w_i$ can be learned using the expectation-maximization (EM) algorithm. EM is a numerical realization of maximum likelihood estimation and has the convenient property that the algorithm tends towards a local optimum with every iteration.

In the current work, the classical non-Bayesian GMM together with the EM algorithm as implemented in scikit-learn \cite{scikit-learn} is used. To identify the individual peaks, we fit GMMs with different number of components to the multi-modal distribution under consideration, and choose the best fit using   certain criteria. The value of individual peaks is then simply the mean of the corresponding Gaussian component. More details on the application of GMM to the atomistic dataset under consideration are provided in \sref{ssec:GMMfit}.

\section{Results}\label{sec:Results}

\subsection{Mathematical form of elastic strain distribution}
Unimodal distributions, like that of \Eel{} here, can be evaluated and better understood via the functional form defining the distribution in individual grains. To identify the mathematical function that best fits the distribution of \Eel{}, we fit various functions to the distribution of \Eel{} in each grain using a maximum likelihood estimation of the distribution parameters. Note that the domain of a grain is defined as in the initial configuration. A total of 32 mathematical functions are used for the fitting purpose.

\begin{figure}[htbp!]
\centering
\includegraphics[width=0.99\textwidth]{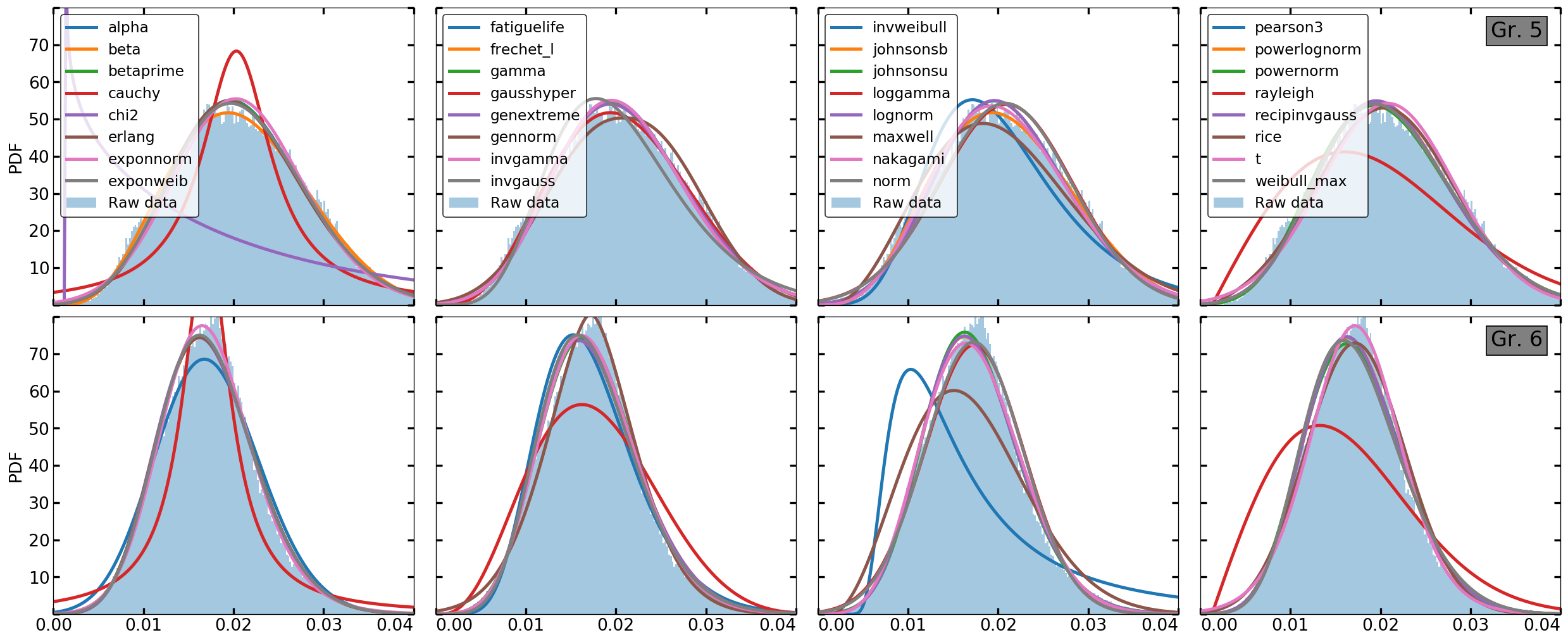}
\caption{Results of fitting mathematical functions to the distribution of \Eel{} in two grains 5 and 6. For the exact mathematical expressions of the individual functions, the reader is referred to \cite{2020SciPy}.} \label{fig:FuncFit}
\end{figure}

\begin{figure}[htbp!]
\centering
\includegraphics[width=0.99\textwidth]{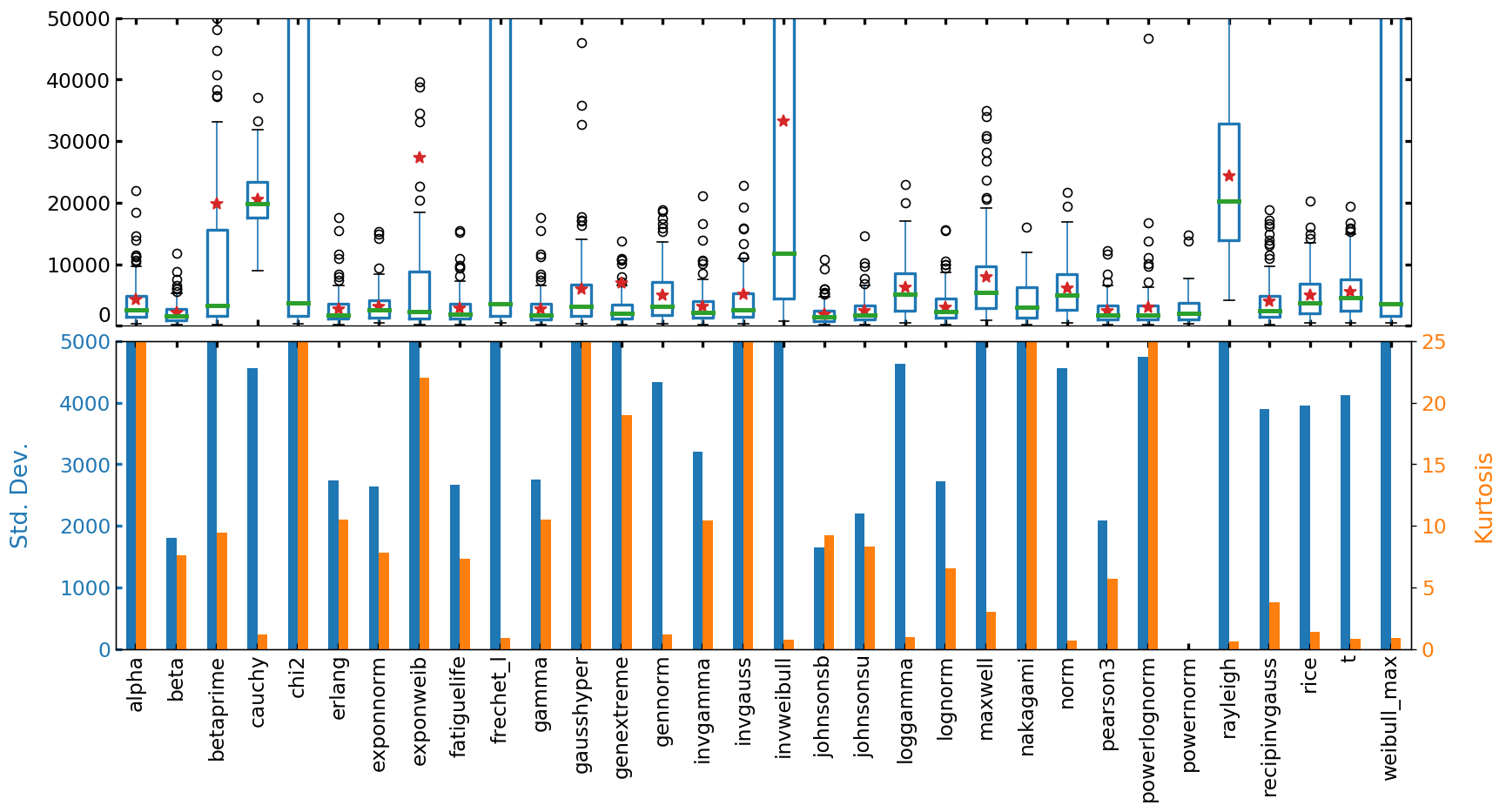}
\caption{Statistics of SSE of individual functions fit to the distributions of \Eel{} in each grain. \emph{Top row}: Distribution of SSE (sum of squared errors) obtained from the fit of each function over all grains as a box plot. The points marked in red denote the mean of the corresponding distribution. \emph{Bottom row}: Standard deviation and kurtosis of the SSE distribution of individual functions.} \label{fig:FuncFit_SSE}
\end{figure}

\Fref{fig:FuncFit} shows exemplarily the result of the fitting procedure for grains 5 and 6. The results for all grains is shown in supplementary figure \fref{sfig:EelWFit}. To assess the best fit among all grains, we use the sum of squared errors (SSE) between the actual distribution and the mathematical function as the metric. The distribution of SSE values over all grains, for the 32 mathematical functions used in the current work, is shown as a box plot in \fref{fig:FuncFit_SSE}. 

The mathematical functions that best fit the distribution of \Eel{} must be those that have low mean (denoted as $\overline{M}$) and median (denoted as $\widetilde{M}$) values of the SSE obtained from the fit to all grains. We combine the two values as $\overline{M} \widetilde{M}/(\overline{M}+\widetilde{M})$, so as to equally weight both the mean and the median values, in order to arrive at a decision on the best fit. \Tref{tab:SSEstatsTop5} lists the top five functions that result from the decision making process, together with their standard deviation and kurtosis. The best fit is evidently obtained with the \codeword{johnsonsb} function, which is the bounded $\textrm{S}_\textrm{b}$ distribution in the Johnson \cite{johnson1949} family of distributions. A complete list of statistical values of SSE for all functions is provided in \tref{tab:SSEallFunc} in the appendix.

\begin{table}[htbp!]
\caption{Statistics of SSE for the top 5 functions that result in the best fit of the distribution of \Eel{} in individual grains}\label{tab:SSEstatsTop5}
\centering
\begin{tabular}{lccccc}
\hline \\[-2ex]
\textbf{Function Name} & \textbf{Mean} ($\overline{M}$) & \textbf{Median} $\widetilde{M}$& $\fracdis{\overline{M} \widetilde{M}}{\overline{M}+\widetilde{M}}$ & \textbf{Std. Dev.} & \textbf{Kurtosis} \\[2ex] \hline \hline \\[-1ex]
\codeword{johnsonsb} & 1869.76 & 1406.28 & 802.62  & 1653.27 & 9.27  \\
\codeword{beta}      & 2103.13 & 1594.28 & 906.84  & 1806.70 & 7.61  \\
\codeword{johnsonsu} & 2465.08 & 1716.04 & 1011.73 & 2203.57 & 8.34  \\
\codeword{pearson3}  & 2471.07 & 1786.44 & 1036.86 & 2089.86 & 5.73  \\
\codeword{gamma}     & 2732.12 & 1780.97 & 1078.16 & 2755.70 & 10.49 \\ \hline
\end{tabular}
\end{table}

\subsection{Peak values in a multimodal distribution via GMM}\label{ssec:GMMfit}
For the analysis of multimodal distributions, like that in \drot{} and \Evm{}, we need to obtain the values of individual peaks. This is achieved by fitting a GMM with multiple clusters -- here, varying from 1 to 6 -- to the distributions in individual grains. 

\begin{figure}[htbp!]
a) 
\begin{center}
\includegraphics[width=0.95\textwidth]{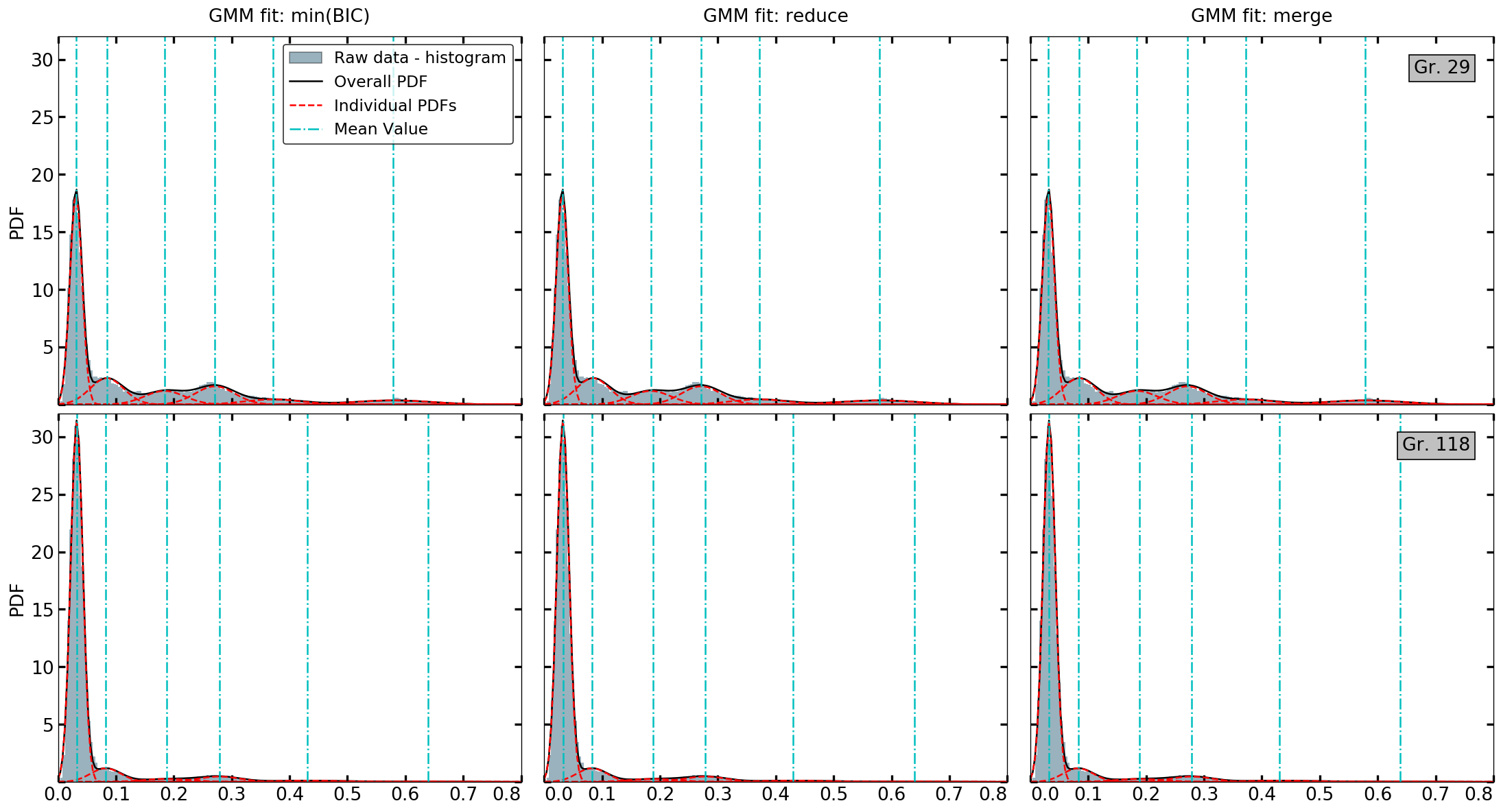}\\
\end{center}
b) 
\begin{center}
\includegraphics[width=0.95\textwidth]{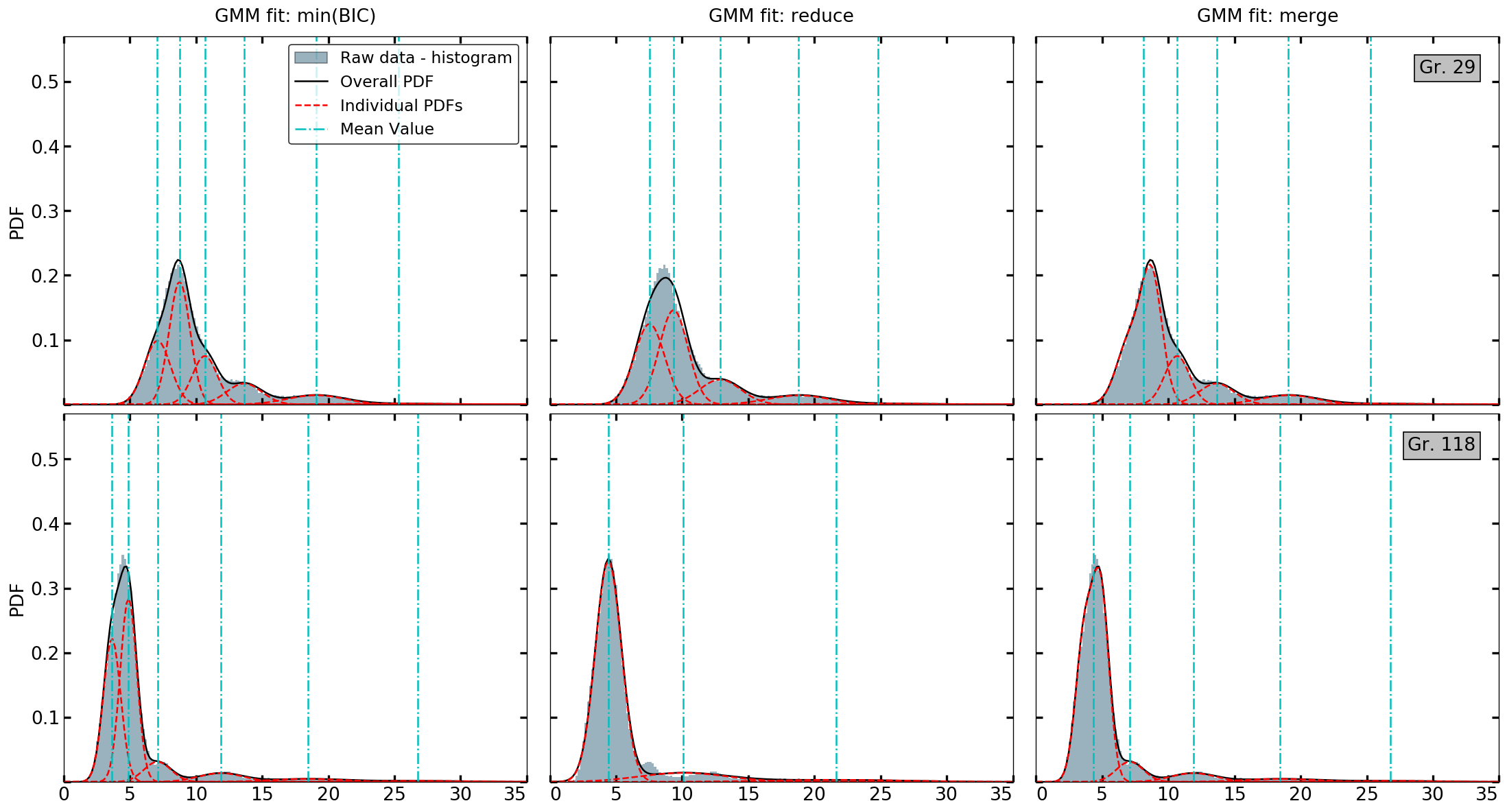}
\end{center}
\caption{Results of unsupervised learning using a GMM on the multi-modal distributions of \Evm{} and \drot{}. The three columns show the results of learning with min(BIC) as the evaluation criterion (\emph{left}), optimum number of clusters obtained by the reduction method (\emph{center}) and merge method (\emph{right}).} \label{fig:GMMfit}
\end{figure}

\begin{figure}[htbp!]
\centering
\includegraphics[width=0.99\textwidth]{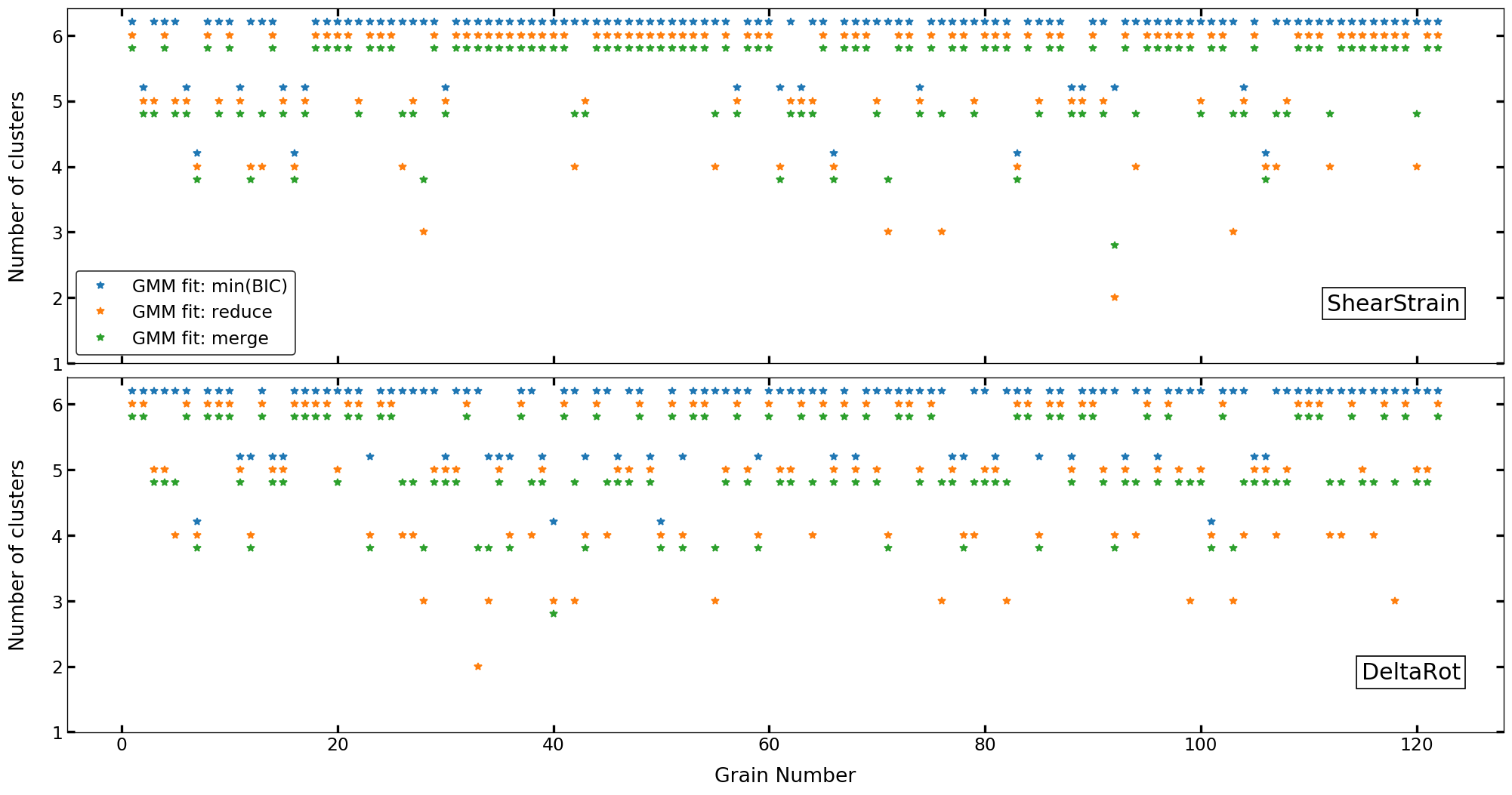}
\caption{Number of GMM clusters predicted in each individual grain by the three different criteria.} \label{fig:GMM_nclusters}
\end{figure}

\Fref{fig:GMMfit} shows the result of such a fit, exemplarily, for two grains 29 and 118. The results for all grains are shown in supplementary figures \ref{sfig:GMM_ShrStr} and \ref{sfig:GMM_DeltaRot}. To assess the best fit, we use the Bayesian information criterion (BIC) and choose the fit that shows the least value of BIC (referred to as \emph{min(BIC)}). For most grains, however, the BIC value decreases continuously with increasing number of clusters, and the best fit is obtained with 6 clusters (see \fref{fig:GMM_nclusters}). This description, in some cases is indicative of overfitting, or a high variance fit, as seen for example in grain 29 for \drot{} (see \fref{fig:GMMfit}b).

To overcome the problem of overfitting and reduce the number of GMM clusters, we present two different approaches. For both these approaches, we first compute the clusters based on the \emph{min(BIC)} criterion. We then evaluate inter-cluster distances in terms of the mean values of individual clusters. If this distance is less than a specific threshold, then:

\begin{enumerate}
\item[a)] we \emph{reduce} the number of clusters by one and re-evaluate inter-cluster distances. This is done iteratively until all inter-cluster distances are above the specified threshold. 
\item[b)] we \emph{merge} all clusters whose inter-cluster distances are below the threshold. The mean values, weights and variances of each cluster are directly recalculated from the original fit.
\end{enumerate}

The chosen threshold in this work is 5\% of the data range of the field variable under consideration. The number of clusters recalculated by the two methods is also shown in \fref{fig:GMM_nclusters}. It is evident that in most cases the number of clusters predicted by all three methods is the same, indicating that the optimal fit based on the least value of BIC suffices for most grains, at least in the current sample. 

For cases where the \emph{min(BIC)} approach results in overfitting , the two methods mentioned above result in an almost identical reduced number of clusters. Although for a few grains, the \emph{reduce} method results in the fewest number of clusters, in such cases the approach seems to sometimes suffer from underfitting as seen for example in the distribution of \drot{} in grain 118 (see \fref{fig:GMMfit}b).

\begin{figure}[htbp!]
\centering
\includegraphics[width=0.75\textwidth]{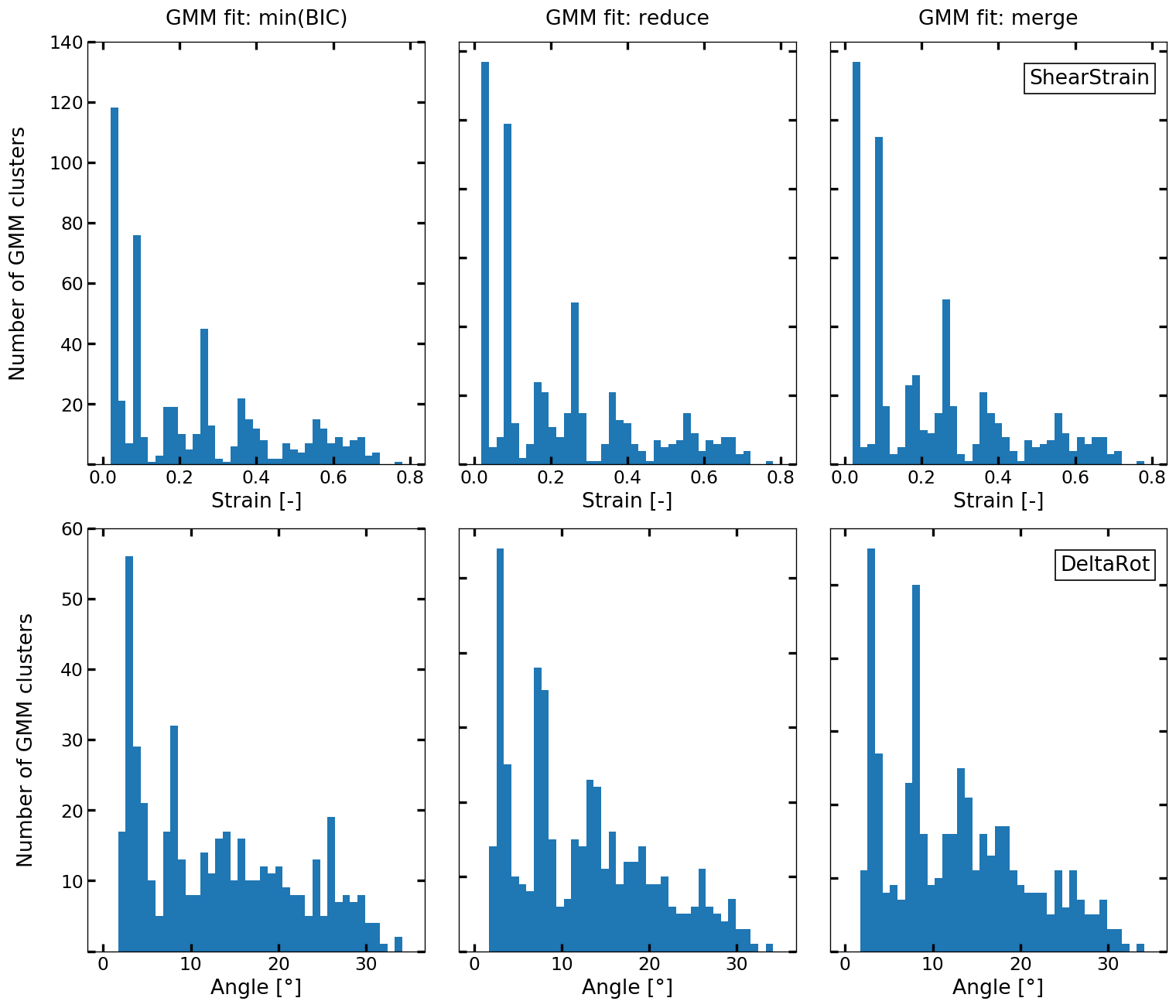}
\caption{Mean values of individual GMM clusters in all grains.} \label{fig:GMM_meanVals}
\end{figure}

The mean values of all clusters as predicted by the three approaches are shown in  \fref{fig:GMM_meanVals}. These distributions provide a general measure of the inhomogeneity of the corresponding field distribution in the polycrystalline sample. For the case of \Evm{}, two primary peaks are observed at values of 0.03 and 0.09. Further peaks are observed at approximate strain values of 0.2, 0.26, 0.36 and 0.56. For the case of \drot{}, two major peaks are observed at approximately 3 and 7 degrees, a further significant peak is observed at 15 degrees and a minor peak at roughly 25 degrees.

\section{Discussion}\label{sec:Discussion}
The aim of the current work is to present a methodology to statistically understand and estimate distributions of continuum fields in atomistic simulations. A fundamental application of the methodology described in the current work is in obtaining an improved mechanistic understanding of the deformation behavior of polycrystalline materials. Specifically, this method allows us to automatically identify various deformation mechanisms like elastic strain, dislocation motion, twinning etc., in individual grains, and quantify their relative contribution to the overall strain. Using the GMM, we have identified the individual peaks in the multi-modal distributions of \Evm{} and \drot{}. These peaks essentially correspond to groups of atoms that evidence values of \Evm{} and \drot{} close to the mean values of the individual Gaussians. This suggests a correlation between the peaks and deformation mechanisms causing such peaks. 

\begin{figure}[htbp!]
\centering
\includegraphics[width=0.5\textwidth]{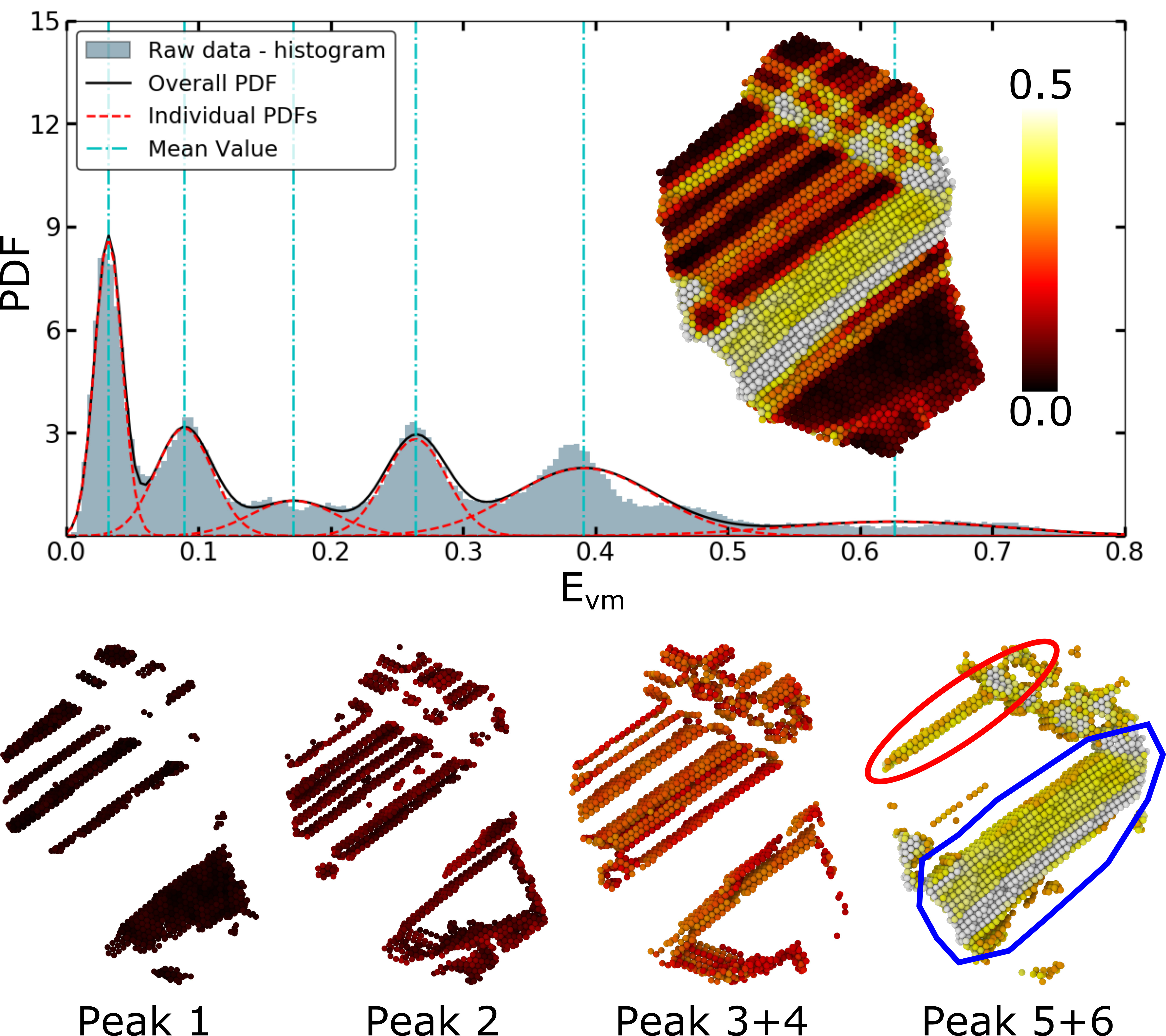}
\caption{Mechanisms associated with the different peaks identified by fitting GMM to the distribution of \Evm{} in grain 65. The inset in the graph shows the distribution of \Evm{} in the atomistic configuration (only a thin slice used here for the purpose of visualization). The pictures below show the atoms of the corresponding peaks.} \label{fig:Discussion}
\end{figure}

We illustrate this exemplarily via an \emph{a-posteriori} analysis of the distribution of \Evm{} in grain 65, see \fref{fig:Discussion}. Note that in the GMM procedure, we are only interested in the values of the individual peaks and not the form of the local distribution itself, which can deviate from a Gaussian distribution. The first peak in the distribution denotes the elastic strain in the grain and is predominantly governed by atoms that have not yet experienced any dislocation activity or GB motion. The second peak is seemingly governed by atoms that form the boundary between the purely elastic region and atoms that have experienced the activity of at least one dislocation. A small contribution to this peak is also from atoms that have experienced GB migration. The third and fourth peaks denotes those atoms that have seen the activity of a single dislocation. The fifth and sixth peaks denote regions with multiple dislocation activity. These include not only planes where two or more dislocations have traversed (encircled in red in \fref{fig:Discussion}), but also twinned regions (encircled in blue in \fref{fig:Discussion}). A small portion of these atoms, particularly those associated with peak~6 include dislocation movement within twinned regions and the intersection nodes of slip planes. 

A second application pertains to bridging length scales with computational frameworks using information transfer. Higher scale models such as, e.g.,  crystal plasticity frameworks, strain gradient models or dislocation dynamics simulations often require information from lower length scales. These models are typically based on simplifications with respect to various aspects of the microstructure, e.g. grain boundaries, slip transfer through GBs, and influence of grain neighborhood. For instance, in classical crystal plasticity simulations, influence of GB and GB processes is usually neglected. Atomistic simulations, on the other hand, contain this information intrinsically. The distribution of field variables from atomistic simulations can be used as input in crystal plasticity frameworks to better reflect the influence of GBs and grain neighborhood on the deformation behavior in any individual grain. Such an approach is used for instance in the quantized crystal plasticity approach \cite{li2009QCP}. The methodology presented in the current work allows one to input distributions as mathematical functions obtained from statistics of multiple grains and configurations. For instance, the parameters of \codeword{johnsonsb}, i.e. the bounded S$_\textrm{b}$ distribution of the Johnson family of distributions, can be used to input the distribution of elastic strain in individual grains, e.g. via FE2AT \cite{moeller2013}, allowing for a better comparison and bridging of atomistic and continuum scale simulations.

Our results show that the \codeword{johnsonsb} function is able to well describe the elastic strain distribution in individual grains. This function has the following mathematical form \cite{scipyWebsite}:
\begin{equation}
f(x,a,b) = \fracdis{b}{x(1-x)} \phi \Bigl( a+b \, \log \left(\fracdis{x}{1-x}\right)  \Bigr),
\end{equation}
where the $a$ and $b$ are parameters with $a,b >0$ and $\phi$ is the normal probability distribution function. It is thus evident that the distribution of elastic strain follows a log-normal nature. Taking a look at the mathematical forms of top five best fit functions tabulated in \Tref{tab:SSEstatsTop5} shows in fact a close-to-lognormal nature of the data, suggesting a multiplicative influence of sources of variation. This lognormal nature of elastic strain in nanocrystalline materials is in line with experimental observations of the total strain \cite{tang2020} but seemingly in contrast to the normal distribution nature of elastic strain in coarse-grained materials \cite{chen2020} observed after 10\% applied strain in crystal plasticity simulations. This suggests a stronger coupling between elastic and plastic strains in nanocrystalline materials, and the need for more accurate constitutive models for continuum crystal plasticity simulations at the nanoscale. The coupling between elastic and plastic strains is ostensibly a result of the discrete nature of dislocation nucleation and propagation due to the absence of dislocation multiplication sources in grains.

We hence conclude that the methodology presented in the current work can lead to improved mechanistic understanding of elasto-plastic deformation behavior, and help bridge length scales in computational frameworks via information passing. The method also helps automatically identify the deformation mechanisms in individual grains, and helps quantify the relative strain contribution of individual mechanisms.

\section{Conclusions}\label{sec:Conclusions}
In this work, we present a methodology for automated analysis of field distributions using statistical machine learning and data mining algorithms. The application of the algorithms is demonstrated on the distributions of elastic strain, total strain and microrotation that have developed in individual grains in a nanocrystalline Al sample after 10\% tensile strain. The distribution of elastic strain is of log-normal nature and is identified as the bounded $\textrm{S}_\textrm{b}$ system of the Johnson family of distributions. The peak values in the multi-modal distributions of total strain and microroration are identified via a Gaussian mixture model. We evaluate the distribution of such peak values over the entire sample and discuss the mechanistic underpinnings in terms of dislocation activity and twinning that result in such local peak values. With this method, the analysis of multiple snapshots across multiple simulations of different materials can be easily performed in an automated manner.

\section*{Acknowledgments}
StS acknowledges funding from the ERC starting grant, ``A Multiscale Dislocation Language for Data-Driven Materials Science'', ERC Grant agreement No. 759419 MuDiLingo. The authors wish to acknowledge the Centre for Information Services and High Performance Computing [Zentrum für Informationsdienste und Hochleistungsrechnen (ZIH)], TU Dresden for providing the computing time for molecular dynamics simulations in the project \emph{NCthinFilms}.

\biboptions{square}

\begin{thebibliography}{10}
\expandafter\ifx\csname url\endcsname\relax
  \def\url#1{\texttt{#1}}\fi
\expandafter\ifx\csname urlprefix\endcsname\relax\def\urlprefix{URL }\fi
\expandafter\ifx\csname href\endcsname\relax
  \def\href#1#2{#2} \def\path#1{#1}\fi

\bibitem{hey2009}
A.~J. Hey, S.~Tansley, K.~M. Tolle, et~al., The fourth paradigm: data-intensive
  scientific discovery, Vol.~1, Microsoft research Redmond, WA, 2009.

\bibitem{sumpter2015}
B.~G. Sumpter, R.~K. Vasudevan, T.~Potok, S.~V. Kalinin, A bridge for
  accelerating materials by design, NPJ Computational Materials 1~(1) (2015)
  1--11.

\bibitem{ramprasad2017}
R.~Ramprasad, R.~Batra, G.~Pilania, A.~Mannodi-Kanakkithodi, C.~Kim, Machine
  learning in materials informatics: recent applications and prospects, npj
  Computational Materials 3~(1) (2017) 1--13.

\bibitem{prakash2018}
A.~Prakash, S.~Sandfeld, Chances and challenges in fusing data science with
  materials science, Practical Metallography 55~(8) (2018) 493--514.

\bibitem{kalidindi2015}
S.~R. Kalidindi, M.~De~Graef, Materials data science: current status and future
  outlook, Annual Review of Materials Research 45 (2015) 171--193.

\bibitem{Farkas2013}
D.~Farkas, Atomistic simulations of metallic microstructures, Current Opinion
  in Solid State and Materials Science 17~(6) (2013) 284--297.

\bibitem{chassagne2011}
M.~Chassagne, M.~Legros, D.~Rodney, Atomic-scale simulation of screw
  dislocation/coherent twin boundary interaction in al, au, cu and ni, Acta
  Materialia 59~(4) (2011) 1456--1463.

\bibitem{pan2014}
Z.~Pan, T.~J. Rupert, Damage nucleation from repeated dislocation absorption at
  a grain boundary, Computational materials science 93 (2014) 206--209.

\bibitem{vaid2019}
A.~Vaid, J.~Gu{\'e}nol{\'e}, A.~Prakash, S.~Korte-Kerzel, E.~Bitzek, Atomistic
  simulations of basal dislocations in mg interacting with mg17al12
  precipitates, Materialia 7 (2019) 100355.

\bibitem{lee2020}
S.~Lee, A.~Vaid, J.~Im, B.~Kim, A.~Prakash, J.~Gu{\'e}nol{\'e}, D.~Kiener,
  E.~Bitzek, S.~H. Oh, In-situ observation of the initiation of plasticity by
  nucleation of prismatic dislocation loops, Nature communications 11~(1)
  (2020) 1--11.

\bibitem{zepeda2017}
L.~A. Zepeda-Ruiz, A.~Stukowski, T.~Oppelstrup, V.~V. Bulatov, Probing the
  limits of metal plasticity with molecular dynamics simulations, Nature
  550~(7677) (2017) 492--495.

\bibitem{Brandstetter2006}
S.~Brandstetter, H.~V. Swygenhoven, S.~V. Petegem, B.~Schmitt, R.~Maass,
  P.~Derlet, From micro to macroplasticity, Advanced Materials 18 (2006)
  1545--1548.

\bibitem{Prakash2017NC}
A.~Prakash, D.~Weygand, E.~Bitzek, Influence of grain boundary structure and
  topology on the plastic deformation of nanocrystalline aluminum as studied by
  atomistic simulations, International Journal of Plasticity 97 (2017)
  107--125.

\bibitem{cao2007}
A.~Cao, Y.~Wei, Atomistic simulations of crack nucleation and intergranular
  fracture in bulk nanocrystalline nickel, Physical Review B 76~(2) (2007)
  024113.

\bibitem{bitzek2015}
E.~Bitzek, J.~R. Kermode, P.~Gumbsch, Atomistic aspects of fracture,
  International Journal of Fracture 191~(1-2) (2015) 13--30.

\bibitem{aidhy2015}
D.~S. Aidhy, C.~Lu, K.~Jin, H.~Bei, Y.~Zhang, L.~Wang, W.~J. Weber, Point
  defect evolution in ni, nife and nicr alloys from atomistic simulations and
  irradiation experiments, Acta Materialia 99 (2015) 69--76.

\bibitem{holmstrom2010}
E.~Holmstr{\"o}m, L.~Toikka, A.~Krasheninnikov, K.~Nordlund, Response of
  mechanically strained nanomaterials to irradiation: Insight from atomistic
  simulations, Physical Review B 82~(4) (2010) 045420.

\bibitem{guenole2016}
J.~Gu{\'e}nol{\'e}, A.~Prakash, E.~Bitzek, Influence of intrinsic strain on
  irradiation induced damage: the role of threshold displacement and surface
  binding energies, Materials \& Design 111 (2016) 405--413.

\bibitem{guenole2017}
J.~Gu{\'e}nol{\'e}, A.~Prakash, E.~Bitzek, Atomistic simulations of focused ion
  beam machining of strained silicon, Applied Surface Science 416 (2017)
  86--95.

\bibitem{kumar2011a}
S.~Kumar, X.~Li, A.~Haque, H.~Gao, Is stress concentration relevant for
  nanocrystalline metals?, Nano Letters 11~(6) (2011) 2510--2516.

\bibitem{li2020}
X.~Li, X.~Zhang, H.~Gao, Atomistic simulations of fracture and fatigue in
  nanotwinned and amorphous materials, Handbook of Materials Modeling:
  Applications: Current and Emerging Materials (2020) 1845--1868.

\bibitem{shreiber2015}
K.~Shreiber, D.~Mordehai, {Dislocation-nucleation-controlled deformation of
  $Ni_3Al$ nanocubes in molecular dynamics simulations}, Modelling and
  Simulation in Materials Science and Engineering 23 (2015) 085004.

\bibitem{sharma2018}
A.~Sharma, J.~Hickman, N.~Gazit, E.~Rabkin, Y.~Mishin, Nickel nanoparticles set
  a new record of strength, Nature communications 9~(1) (2018) 1--9.

\bibitem{houlle2018}
F.~Houll{\'e}, F.~Walsh, A.~Prakash, E.~Bitzek, Atomistic simulations of
  compression tests on $\gamma$-precipitate containing ni 3 al nanocubes,
  Metallurgical and Materials Transactions A 49~(9) (2018) 4158--4166.

\bibitem{weinberger2012}
C.~R. Weinberger, A.~T. Jennings, K.~Kang, J.~R. Greer, Atomistic simulations
  and continuum modeling of dislocation nucleation and strength in gold
  nanowires, Journal of the Mechanics and Physics of Solids 60~(1) (2012)
  84--103.

\bibitem{xie2020}
Z.~Xie, J.~Shin, J.~Renner, A.~Prakash, D.~S. Gianola, E.~Bitzek, Origins of
  strengthening and failure in twinned au nanowires: Insights from in- situ
  experiments and atomistic simulations, Acta Materialia 187 (2020) 166--175.

\bibitem{moody2014}
M.~P. Moody, A.~V. Ceguerra, A.~J. Breen, X.~Y. Cui, B.~Gault, L.~T.
  Stephenson, R.~K. Marceau, R.~C. Powles, S.~P. Ringer, Atomically resolved
  tomography to directly inform simulations for structure--property
  relationships, Nature communications 5~(1) (2014) 1--10.

\bibitem{prakash2015}
A.~Prakash, J.~Gu{\'{e}}nol{\'{e}}, J.~Wang, J.~M{\"{u}}ller, E.~Spiecker,
  M.~J. Mills, I.~Povstugar, P.~Choi, D.~Raabe, E.~Bitzek, {Atom probe informed
  simulations of dislocation--precipitate interactions reveal the importance of
  local interface curvature }, Acta Materialia 92 (2015) 33--45.
\newblock \href {http://dx.doi.org/10.1016/j.actamat.2015.03.050}
  {\path{doi:10.1016/j.actamat.2015.03.050}}.

\bibitem{prakash2017}
A.~Prakash, E.~Bitzek, {Idealized vs. realistic microstructures: An atomistic
  simulation case study on $\gamma/\gamma'$ microstructures}, Materials 10
  (2017) 88.

\bibitem{Hahn2015}
E.~N. Hahn, M.~A. Meyers, Grain-size dependent mechanical behavior of
  nanocrystalline metals, Materials Science and Engineering: A 646 (2015)
  101--134.

\bibitem{Panzarino2016}
J.~F. Panzarino, Z.~Pan, T.~J. Rupert, Plasticity-induced restructuring of
  nanocrystalline grain boundary network, Acta Materialia 120 (2016) 1--13.

\bibitem{schiotz2004a}
J.~Schi{\o}tz, Atomic-scale modeling of plastic deformation of nanocrystalline
  copper, Scripta Materialia 51~(8) (2004) 837--841.

\bibitem{vanswygenhoven.etal.2006}
H.~Van~Swygenhoven, P.~M. Derlet, A.~G. Fr{\o}seth, Nucleation and propagation
  of dislocations in nanocrystalline fcc metals, Acta Materialia 54~(7) (2006)
  1975 -- 1983.

\bibitem{van-swygenhoven2001a}
H.~Van~Swygenhoven, P.~Derlet, Grain-boundary sliding in nanocrystalline fcc
  metals, Physical Review B 64~(22) (2001) 224105.

\bibitem{dewald2006}
M.~Dewald, W.~Curtin, Multiscale modelling of dislocation/grain-boundary
  interactions: I. edge dislocations impinging on $\sigma$11 (1 1 3) tilt
  boundary in al, Modelling and Simulation in Materials Science and Engineering
  15~(1) (2006) S193.

\bibitem{tucker2011}
G.~J. Tucker, J.~A. Zimmerman, D.~L. McDowell, Continuum metrics for
  deformation and microrotation from atomistic simulations: Application to
  grain boundaries, International journal of engineering science 49~(12) (2011)
  1424--1434.

\bibitem{Gunkelmann2017_ComputMaterSci135}
N.~Gunkelmann, I.~A. Alhafez, D.~Steinberger, H.~M. Urbassek, S.~Sandfeld,
  \href{http://www.sciencedirect.com/science/article/pii/S0927025617301908}{Nanoscratching
  of iron: A novel approach to characterize dislocation microstructures},
  Computational Materials Science 135 (2017) 181 -- 188.
\newblock \href
  {http://dx.doi.org/https://doi.org/10.1016/j.commatsci.2017.04.008}
  {\path{doi:https://doi.org/10.1016/j.commatsci.2017.04.008}}.
\newline\urlprefix\url{http://www.sciencedirect.com/science/article/pii/S0927025617301908}

\bibitem{KOSITSKI2018_ComputMaterSci149}
R.~Kositski, D.~Steinberger, S.~Sandfeld, D.~Mordehai,
  \href{http://www.sciencedirect.com/science/article/pii/S0927025618301460}{Shear
  relaxation behind the shock front in <110> molybdenum – from the atomic
  scale to continuous dislocation fields}, Computational Materials Science 149
  (2018) 125 -- 133.
\newblock \href
  {http://dx.doi.org/https://doi.org/10.1016/j.commatsci.2018.02.058}
  {\path{doi:https://doi.org/10.1016/j.commatsci.2018.02.058}}.
\newline\urlprefix\url{http://www.sciencedirect.com/science/article/pii/S0927025618301460}

\bibitem{bitzek.etal.2008}
E.~Bitzek, P.~M. Derlet, P.~M. Anderson, H.~Van~Swygenhoven, {The stress-strain
  response of nanocrystalline metals: A statistical analysis of atomistic
  simulations}, Acta Materialia 56 (2008) 4845--4857.

\bibitem{gupta2020}
A.~Gupta, J.~Gruber, S.~S. Rajaram, G.~B. Thompson, D.~L. McDowell, G.~J.
  Tucker, On the mechanistic origins of maximum strength in nanocrystalline
  metals, npj Computational Materials 6~(1) (2020) 1--12.

\bibitem{Roberts2019}
G.~Roberts, S.~Y. Haile, R.~Sainju, D.~J. Edwards, B.~Hutchinson, Y.~Zhu,
  \href{https://doi.org/10.1038/s41598-019-49105-0}{Deep learning for semantic
  segmentation of defects in advanced {STEM} images of steels}, Scientific
  Reports 9~(1).
\newblock \href {http://dx.doi.org/10.1038/s41598-019-49105-0}
  {\path{doi:10.1038/s41598-019-49105-0}}.
\newline\urlprefix\url{https://doi.org/10.1038/s41598-019-49105-0}

\bibitem{Ma2020}
W.~Ma, E.~J. Kautz, A.~Baskaran, A.~Chowdhury, V.~Joshi, B.~Yener, D.~J. Lewis,
  \href{https://doi.org/10.1063/5.0013720}{Image-driven discriminative and
  generative machine learning algorithms for establishing
  microstructure{\textendash}processing relationships}, Journal of Applied
  Physics 128~(13) (2020) 134901.
\newblock \href {http://dx.doi.org/10.1063/5.0013720}
  {\path{doi:10.1063/5.0013720}}.
\newline\urlprefix\url{https://doi.org/10.1063/5.0013720}

\bibitem{Trampert2021}
P.~Trampert, D.~Rubinstein, F.~Boughorbel, C.~Schlinkmann, M.~Luschkova,
  P.~Slusallek, T.~Dahmen, S.~Sandfeld,
  \href{https://doi.org/10.3390/cryst11030258}{Deep neural networks for
  analysis of microscopy images{\textemdash}synthetic data generation and
  adaptive sampling}, Crystals 11~(3) (2021) 258.
\newblock \href {http://dx.doi.org/10.3390/cryst11030258}
  {\path{doi:10.3390/cryst11030258}}.
\newline\urlprefix\url{https://doi.org/10.3390/cryst11030258}

\bibitem{Kalidindi2020}
S.~R. Kalidindi, \href{https://doi.org/10.1063/5.0011258}{Feature engineering
  of material structure for {AI}-based materials knowledge systems}, Journal of
  Applied Physics 128~(4) (2020) 041103.
\newblock \href {http://dx.doi.org/10.1063/5.0011258}
  {\path{doi:10.1063/5.0011258}}.
\newline\urlprefix\url{https://doi.org/10.1063/5.0011258}

\bibitem{Steinberger2019}
Steinberger, Song, Sandfeld,
  \href{https://www.frontiersin.org/articles/10.3389/fmats.2019.00141/full}{Machine
  learning-based classification of dislocation microstructes}, Frontiers in
  Materials 6, article 141.
\newblock \href {http://dx.doi.org/https://doi.org/10.3389/fmats.2019.00141}
  {\path{doi:https://doi.org/10.3389/fmats.2019.00141}}.
\newline\urlprefix\url{https://www.frontiersin.org/articles/10.3389/fmats.2019.00141/full}

\bibitem{Zheng2020}
B.~Zheng, J.~Yang, B.~Liang, J.~chun Cheng,
  \href{https://doi.org/10.1063/5.0012392}{Inverse design of acoustic
  metamaterials based on machine learning using a gauss{\textendash}bayesian
  model}, Journal of Applied Physics 128~(13) (2020) 134902.
\newblock \href {http://dx.doi.org/10.1063/5.0012392}
  {\path{doi:10.1063/5.0012392}}.
\newline\urlprefix\url{https://doi.org/10.1063/5.0012392}

\bibitem{gomberg2017}
J.~A. Gomberg, A.~J. Medford, S.~R. Kalidindi, Extracting knowledge from
  molecular mechanics simulations of grain boundaries using machine learning,
  Acta Materialia 133 (2017) 100--108.

\bibitem{vimal2022}
M.~Vimal, S.~Sandfeld, A.~Prakash, Grain segmentation in atomistic simulations
  using orientation-based iterative self-organizing data analysis, Materialia
  (2022) 101314.

\bibitem{ceriotti2019}
M.~Ceriotti, Unsupervised machine learning in atomistic simulations, between
  predictions and understanding, The Journal of chemical physics 150~(15)
  (2019) 150901.

\bibitem{bock2019}
F.~E. Bock, R.~C. Aydin, C.~J. Cyron, N.~Huber, S.~R. Kalidindi, B.~Klusemann,
  \href{https://www.frontiersin.org/article/10.3389/fmats.2019.00110}{A review
  of the application of machine learning and data mining approaches in
  continuum materials mechanics}, Frontiers in Materials 6.
\newblock \href {http://dx.doi.org/10.3389/fmats.2019.00110}
  {\path{doi:10.3389/fmats.2019.00110}}.
\newline\urlprefix\url{https://www.frontiersin.org/article/10.3389/fmats.2019.00110}

\bibitem{tan2016}
P.-N. Tan, M.~Steinbach, V.~Kumar, Introduction to data mining, Pearson
  Education India, 2016.

\bibitem{Xu2009}
T.~Xu, M.~Li, Topological and statistical properties of a constrained voronoi
  tessellation, Philosophical Magazine 89~(4) (2009) 349--374.

\bibitem{serrao2021}
P.~Serrao, S.~Sandfeld, A.~Prakash, Optimic: A tool to generate optimized
  polycrystalline microstructures for materials simulations, SoftwareX 15.

\bibitem{prakash2016b}
A.~Prakash, M.~Hummel, S.~Schmauder, E.~Bitzek, {NanoSCULPT: A methodology for
  generating complex realistic configurations for atomistic simulations},
  MethodsX 3 (2016) 1--9.

\bibitem{mishin.etal.1999}
Y.~Mishin, D.~Farkas, M.~J. Mehl, D.~A. Papaconstantopoulos, {Interatomic
  potentials for Al and Ni from experimental data and ab-initio calculations},
  Materials Research Symposium Proceedings 538 (1999) 535--540.

\bibitem{roth2019}
J.~Roth, E.~Eisfeld, D.~Klein, S.~Hocker, H.~Lipp, H.-R. Trebin, Imd--the itap
  molecular dynamics simulation package, The European Physical Journal Special
  Topics 227~(14) (2019) 1831--1836.

\bibitem{ovito}
A.~Stukowski, {Visualization and analysis of atomistic simulation data with
  OVITO-the Open Visualization Tool}, {Modelling and Simulation in Materials
  Science and Engineering} {18}~({1}).
\newblock \href {http://dx.doi.org/{10.1088/0965-0393/18/1/015012}}
  {\path{doi:{10.1088/0965-0393/18/1/015012}}}.

\bibitem{honeycutt.andersen.1987}
J.~D. Honeycutt, H.~C. Andersen, {Molecular Dynamics Study of Melting and
  Freezing of Small Lennard-Jones Clusters}, Journal of Chemical Physics 91
  (1987) 4950--4963.

\bibitem{Shimizu2007}
F.~Shimizu, S.~Ogata, J.~Li, Theory of shear banding in metallic glasses and
  molecular dynamics calculations, Materials Transactions 48~(11) (2007)
  2923--2927.

\bibitem{2020SciPy}
P.~Virtanen, R.~Gommers, T.~E. Oliphant, M.~Haberland, T.~Reddy, D.~Cournapeau,
  E.~Burovski, P.~Peterson, W.~Weckesser, J.~Bright, S.~J. {van der Walt},
  M.~Brett, J.~Wilson, K.~J. Millman, N.~Mayorov, A.~R.~J. Nelson, E.~Jones,
  R.~Kern, E.~Larson, C.~J. Carey, {\.I}.~Polat, Y.~Feng, E.~W. Moore,
  J.~{VanderPlas}, D.~Laxalde, J.~Perktold, R.~Cimrman, I.~Henriksen, E.~A.
  Quintero, C.~R. Harris, A.~M. Archibald, A.~H. Ribeiro, F.~Pedregosa, P.~{van
  Mulbregt}, {SciPy 1.0 Contributors}, {{SciPy} 1.0: Fundamental Algorithms for
  Scientific Computing in Python}, Nature Methods 17 (2020) 261--272.
\newblock \href {http://dx.doi.org/10.1038/s41592-019-0686-2}
  {\path{doi:10.1038/s41592-019-0686-2}}.

\bibitem{bishop2006}
C.~M. Bishop, Pattern recognition and machine learning, springer, 2006.

\bibitem{press2007}
W.~H. Press, S.~A. Teukolsky, W.~T. Vetterling, B.~P. Flannery, Numerical
  recipes 3rd edition: The art of scientific computing, Cambridge university
  press, 2007.

\bibitem{scikit-learn}
F.~Pedregosa, G.~Varoquaux, A.~Gramfort, V.~Michel, B.~Thirion, O.~Grisel,
  M.~Blondel, P.~Prettenhofer, R.~Weiss, V.~Dubourg, J.~Vanderplas, A.~Passos,
  D.~Cournapeau, M.~Brucher, M.~Perrot, E.~Duchesnay, Scikit-learn: Machine
  learning in {P}ython, Journal of Machine Learning Research 12 (2011)
  2825--2830.

\bibitem{johnson1949}
N.~L. Johnson, Systems of frequency curves generated by methods of translation,
  Biometrika 36~(1/2) (1949) 149--176.

\bibitem{li2009QCP}
L.~Li, P.~M. Anderson, M.-G. Lee, E.~Bitzek, P.~Derlet, H.~Van~Swygenhoven, The
  stress--strain response of nanocrystalline metals: a quantized crystal
  plasticity approach, Acta Materialia 57~(3) (2009) 812--822.

\bibitem{moeller2013}
J.~J. M{\"o}ller, A.~Prakash, E.~Bitzek, {FE2AT} -- finite element informed
  atomistic simulations, Modelling and Simulation in Materials Science and
  Engineering 21~(5) (2013) 055011.

\bibitem{scipyWebsite}
{Scipy Stats Package},
  \url{https://docs.scipy.org/doc/scipy/reference/stats.html}, accessed:
  2021-03-07.

\bibitem{tang2020}
A.~Tang, H.~Liu, G.~Liu, Y.~Zhong, L.~Wang, Q.~Lu, J.~Wang, Y.~Shen, Lognormal
  distribution of local strain: A universal law of plastic deformation in
  material, Phys. Rev. Lett. 124 (2020) 155501.
\newblock \href {http://dx.doi.org/10.1103/PhysRevLett.124.155501}
  {\path{doi:10.1103/PhysRevLett.124.155501}}.

\bibitem{chen2020}
J.~Chen, A.~M. Korsunsky, Why is local stress statistics normal, and strain
  lognormal?, Materials \& Design 198 (2020) 109319.

\end{thebibliography}

\newpage
\section{Appendix}
\begin{table}[htbp!]
\caption{Statistics of SSE obtained for the fit of individual functions}\label{tab:SSEallFunc}
\centering
\begin{tabular}{lccccc}
\hline \\[-2ex]
\textbf{Function Name} & \textbf{Mean} ($\bar{M}$) & \textbf{Median} $\tilde{M}$& $\fracdis{\bar{M} \tilde{M}}{\bar{M}+\tilde{M}}$ & \textbf{Std. Dev.} & \textbf{Kurtosis} \\[2ex] \hline \hline \\[-1ex]
alpha                  & 4243.73       & 2588.33         & 1607.74   & 6321.11            & 54.38         \\
beta                   & 2103.13       & 1594.28         & 906.84    & 1806.70            & 7.61          \\
betaprime              & 19799.69      & 3279.92         & 2813.80   & 37975.08           & 9.41          \\
cauchy                 & 20535.29      & 19873.80        & 10099.56  & 4565.77            & 1.18          \\
chi2                   & 82891.06      & 3725.39         & 3565.16   & 342916.10          & 106.58        \\
erlang                 & 2736.39       & 1790.18         & 1082.20   & 2740.05            & 10.51         \\
exponnorm              & 3149.72       & 2523.49         & 1401.02   & 2634.91            & 7.85          \\
exponweib              & 27300.25      & 2325.51         & 2142.97   & 86182.90           & 22.05         \\
fatiguelife            & 2878.82       & 1867.99         & 1132.89   & 2673.36            & 7.29          \\
frechet\_l             & 54099.54      & 3525.85         & 3310.12   & 86510.67           & 0.88          \\
gamma                  & 2732.12       & 1780.97         & 1078.16   & 2755.70            & 10.49         \\
gausshyper             & 6029.60       & 3160.76         & 2073.71   & 10937.98           & 49.56         \\
genextreme             & 6963.81       & 2087.40         & 1606.00   & 19587.21           & 18.97         \\
gennorm                & 5038.20       & 3167.90         & 1944.96   & 4335.82            & 1.20          \\
invgamma               & 3160.37       & 2131.07         & 1272.81   & 3204.16            & 10.43         \\
invgauss               & 5134.56       & 2639.74         & 1743.42   & 10354.83           & 48.58         \\
invweibull             & 33330.54      & 11788.77        & 8708.60   & 39006.47           & 0.74          \\
johnsonsb              & 1869.76       & 1406.28         & 802.62    & 1653.27            & 9.27          \\
johnsonsu              & 2465.08       & 1716.04         & 1011.73   & 2203.57            & 8.34          \\
loggamma               & 6232.27       & 5063.40         & 2793.68   & 4629.79            & 0.97          \\
lognorm                & 3060.14       & 2267.71         & 1302.50   & 2724.10            & 6.57          \\
maxwell                & 7997.27       & 5455.23         & 3243.04   & 7386.89            & 2.98          \\
nakagami               & 110203.01     & 3024.75         & 2943.95   & 600178.01          & 99.75         \\
norm                   & 6172.50       & 4985.81         & 2758.03   & 4556.46            & 0.66          \\
pearson3               & 2471.07       & 1786.44         & 1036.86   & 2089.86            & 5.73          \\
powerlognorm           & 2955.93       & 1743.78         & 1096.77   & 4751.09            & 60.77         \\
powernorm              & inf           & 2029.85         & --       & --                & --           \\
rayleigh               & 24373.71      & 20302.90        & 11076.42  & 13987.62           & 0.57          \\
recipinvgauss          & 3936.44       & 2509.62         & 1532.56   & 3892.25            & 3.78          \\
rice                   & 5002.95       & 3680.83         & 2120.62   & 3948.73            & 1.37          \\
t                      & 5561.50       & 4553.68         & 2503.69   & 4120.68            & 0.80          \\
weibull\_max           & 54099.54      & 3525.85         & 3310.12   & 86510.67           & 0.88          \\ \hline
\end{tabular}
\end{table}

\newpage
\title{Automated analysis of continuum fields from atomistic simulations using statistical machine learning\\Supplementary Material}
\maketitle

\section*{Automated determination of plot limits}
\setcounter{section}{0}
\setcounter{table}{0}
\setcounter{figure}{0}
\setcounter{equation}{0}
\renewcommand{\tablename}{Supplementary Table}
\renewcommand{\figurename}{Supplementary Figure}
\renewcommand{\subfigurename}{Supplementary Figure}
\renewcommand\thefigure{S\arabic{figure}}
\renewcommand\thesubfigure{S\arabic{figure}\alph{subfigure}:}

The first step in our analysis is to visualize the distribution of the continuum fields (here: \Evm{}, \Eel{} and \drot{}) in individual grains.
Usually, visualization of distributions is done by plotting them as histograms or kernel density estimates (KDEs). However, when dealing with hundreds or thousands of snapshots of MD simulations, each of which may contain hundreds of grains, this results in two major issues: a) Significant amount of space is needed to plot the distributions of all grains, which makes comparison of distributions between grains a cumbersome operation. For instance, in the case of the current sample, which corresponds to a particular snapshot at 10\% strain from a single simulation, we obtain 366 plots for the three fields in 122 grains, with each plot containing the distribution as a histogram or KDE, together with the best-fit function or GMM. b) In order to make all plots comparable, the set of units used on the ordinate and abscissa must be identical in all plots. Since the distributions are siginficantly inhomogeneous, manual tuning of the plot limits, which is usually done by inspecting the distributions of only a few grains, can result in truncation of outlying peaks in multimodal distributions. On the other hand, merely using the maximum and minimum value of the data array would result in tails of a distribution with frequency values close to zero dominating the plots of the distributions.

To overcome these obstacles,  we propose a two-step approach as an alternative: a) Automate the calculation of ranges for the distributions, and b) plot the distributions as 1-D heatmaps instead of histograms. The \emph{Step a)} overcomes the drawback of missing out on outlying peaks beyond a fixed range chosen by random inspection of only a subset of data, in addition to the obvious advantage of having greater automation. \emph{Step b)} allows for easy visualization of the data in minimal space, whilst ensuring hotspots in the distributions are not missed.

Here, we propose an automated method for obtaining the data ranges to plot the distributions of field quantities in individual grains. The method follows the procedure below:

\begin{enumerate}
\item[i)] Fix the number of bins of the histogram, \emph{nbins}, and minimum number of atoms in a bin, $N_b^{min}$.
\item[ii)] Discretize data into bins. 
\item[iii)] Obtain maximum and minimum values of the data range for each grain. Herein, discard those bins that do not have at least $N_b^{min}$ atoms. Store calculated values in the arrays \codeword{DataMaxValues} and \codeword{DataMinValues}.
\item[iv)] Choose a $n_p^d$ percentile value of the \codeword{DataMaxValues} and \codeword{DataMinValues} arrays. This ensures that long tails of distributions are cut off. In this work, $n_p^d$ is chosen as 90 percentile for \codeword{DataMaxValues} and 10 percentile for \codeword{DataMinValues}. For minimum values very close to zero, we explicitly set the values to zero. The so-calculated values denote the range of data to be used for further purposes.
\item[v)] Re-discretize the distribution with the chosen number of bins and the data range defined in the previous step.
\item[vi)] For each grain, obtain the maximum values of the frequency density and store them in the array \codeword{FreqDensMaxValues}.
\item[vii)] Choose a $n_p^f$ percentile value of the \codeword{FreqDensMaxValues} array. This value denotes the frequency density to be used for plotting the distributions of all grains. In this work, the value of $n_p^f$ is also taken to be 90 percentile.
\end{enumerate}

\begin{figure}[htbp!]
\centering
\includegraphics[width=0.99\textwidth]{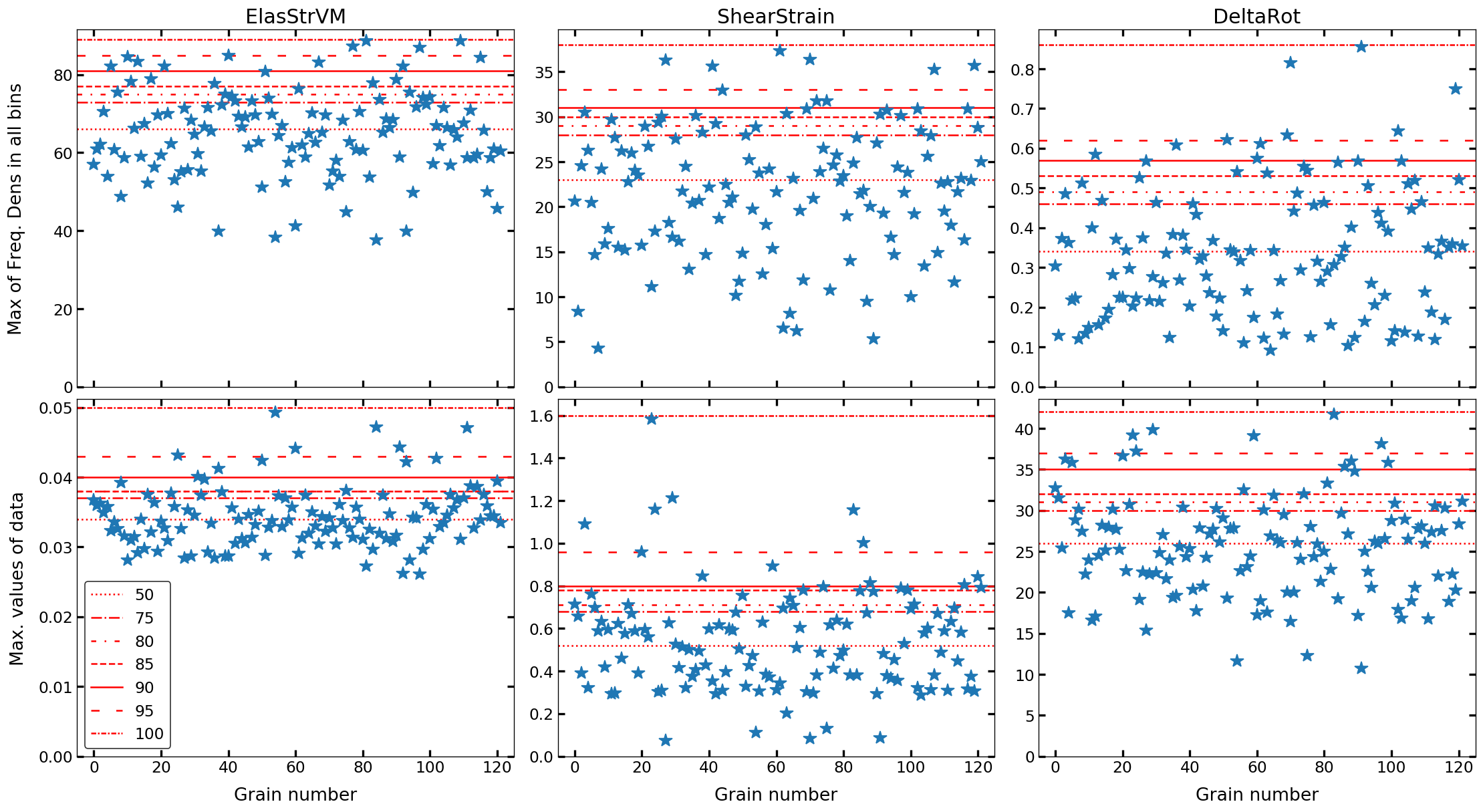}
\caption{Maximum values of frequency density and data ranges of probability distribution function when plotted as histograms in individual grains. The red horizontal lines denote the corresponding percentile rounded values.} \label{fig:dataRangeVals}
\end{figure}

\Fref{fig:dataRangeVals} shows the maximum values of the frequency density	and the data ranges in each grain, for all the three field variables under consideration. The minimum values of the data and frequency density ranges is chosen to be zero, since the data being used contains positive values only.

The obtained ranges for the three field quantities are listed in \tref{tab:dataRanges}. But for \Evm{}, the calculated quantities are identical to the data ranges obtained by visual inspection of only a few grains, indicating the stability of the approach. The higher value for \Evm{} is primarily due to the large kurtosis in the distributions in individual grains.

\begin{table}[htbp!]
\caption{Data ranges of the three field quantities obtained by automation and visual inspection of a few grains. The ranges obtained from the automated extraction are subsequently used for plotting and visualization.}\label{tab:dataRanges}

\centering
\begin{tabular}{lcc}
\hline \\[-2ex]
\textbf{Field quantity} & \textbf{Automated extraction} & \textbf{Visual inspection} \\[2ex] \hline \hline \\[-1ex]
\Eel{} &  (0.0, 0.04) & (0.0, 0.04)\\
\Evm{} & (0.0, 0.8) & (0.0, 0.5)\\
\drot{} & (0.0, 40.0) & (0.0, 40.0)\\ \hline
\end{tabular}
\end{table}

\pagestyle{empty}

\begin{figure}[htbp!]
\centering
    \begin{subfigure}[b]{1.0\textwidth}
        \includegraphics[width=1.0\textwidth]{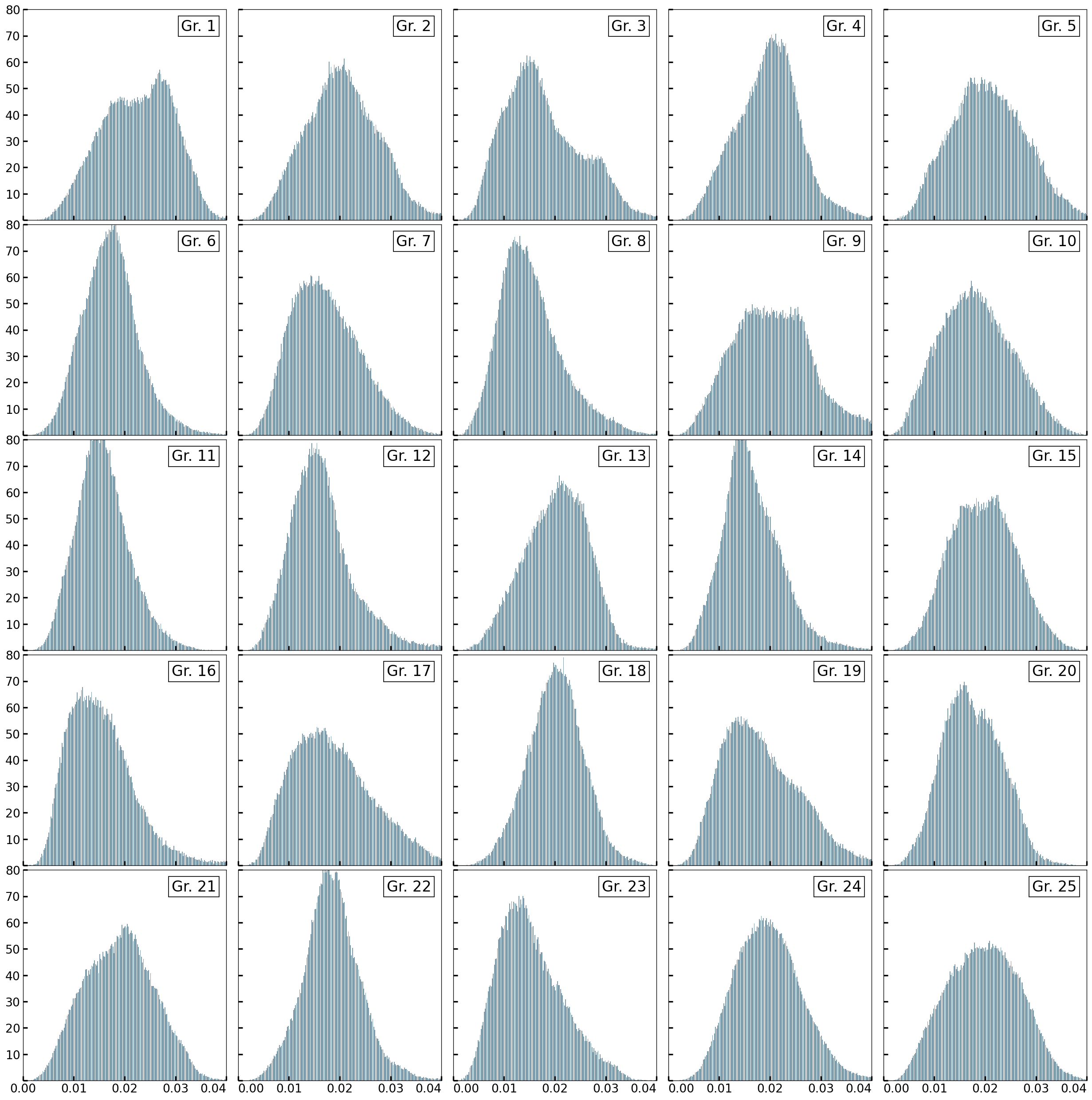}
        \subcaption{Distribution of elastic strain \Eel{} as histograms in grains 1 to 25}
        \label{sfig:EelHist1-25}
    \end{subfigure}
\end{figure}
\begin{figure}[htbp!]\ContinuedFloat
    \begin{subfigure}[b]{1.0\textwidth}
        \includegraphics[width=1.0\textwidth]{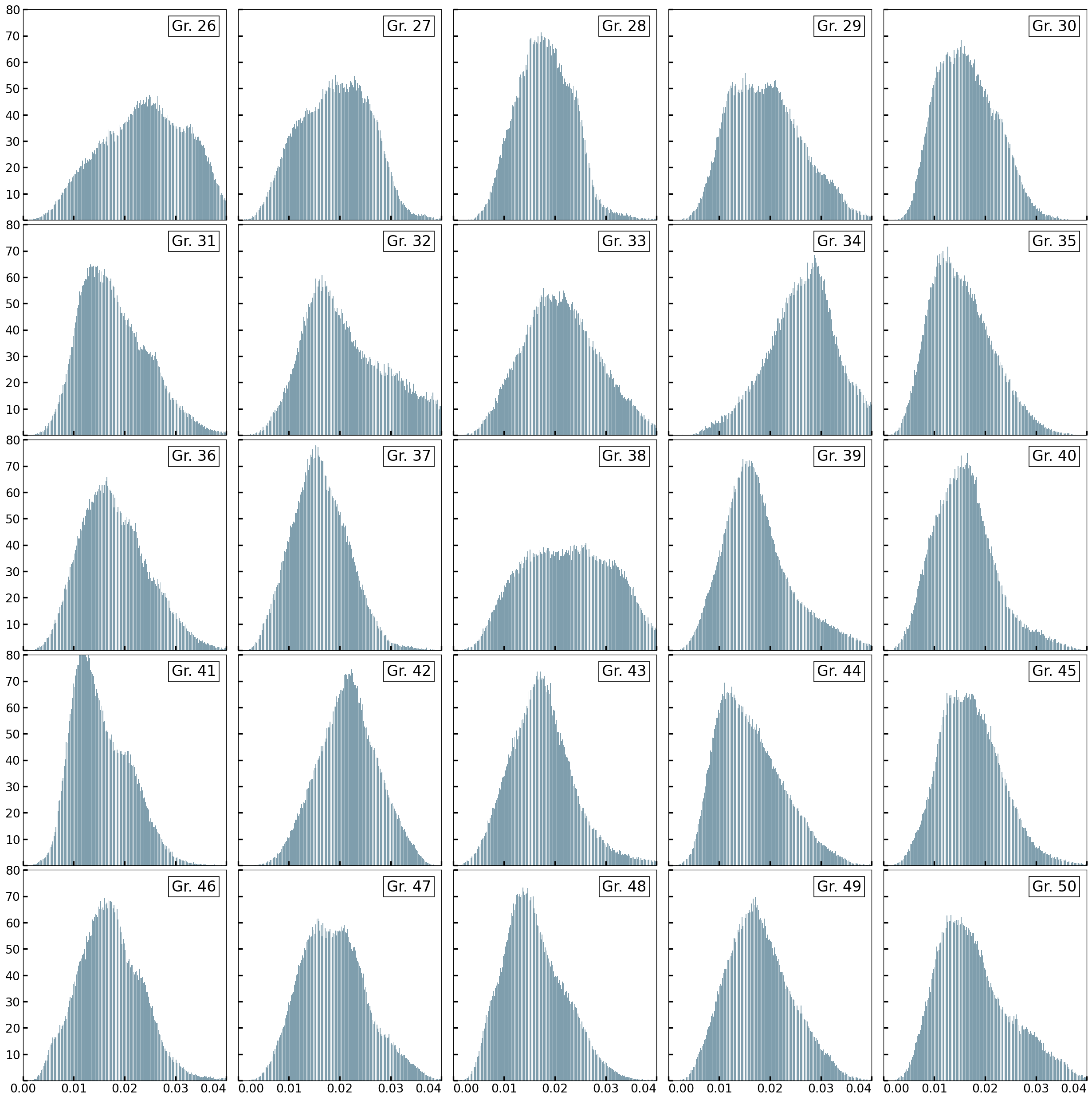}
        \subcaption{Distribution of elastic strain \Eel{} as histograms in grains 26 to 50}
        \label{sfig:EelHist26-50}
    \end{subfigure}
\end{figure}
\begin{figure}[htbp!]\ContinuedFloat
    \begin{subfigure}[b]{1.0\textwidth}
        \includegraphics[width=1.0\textwidth]{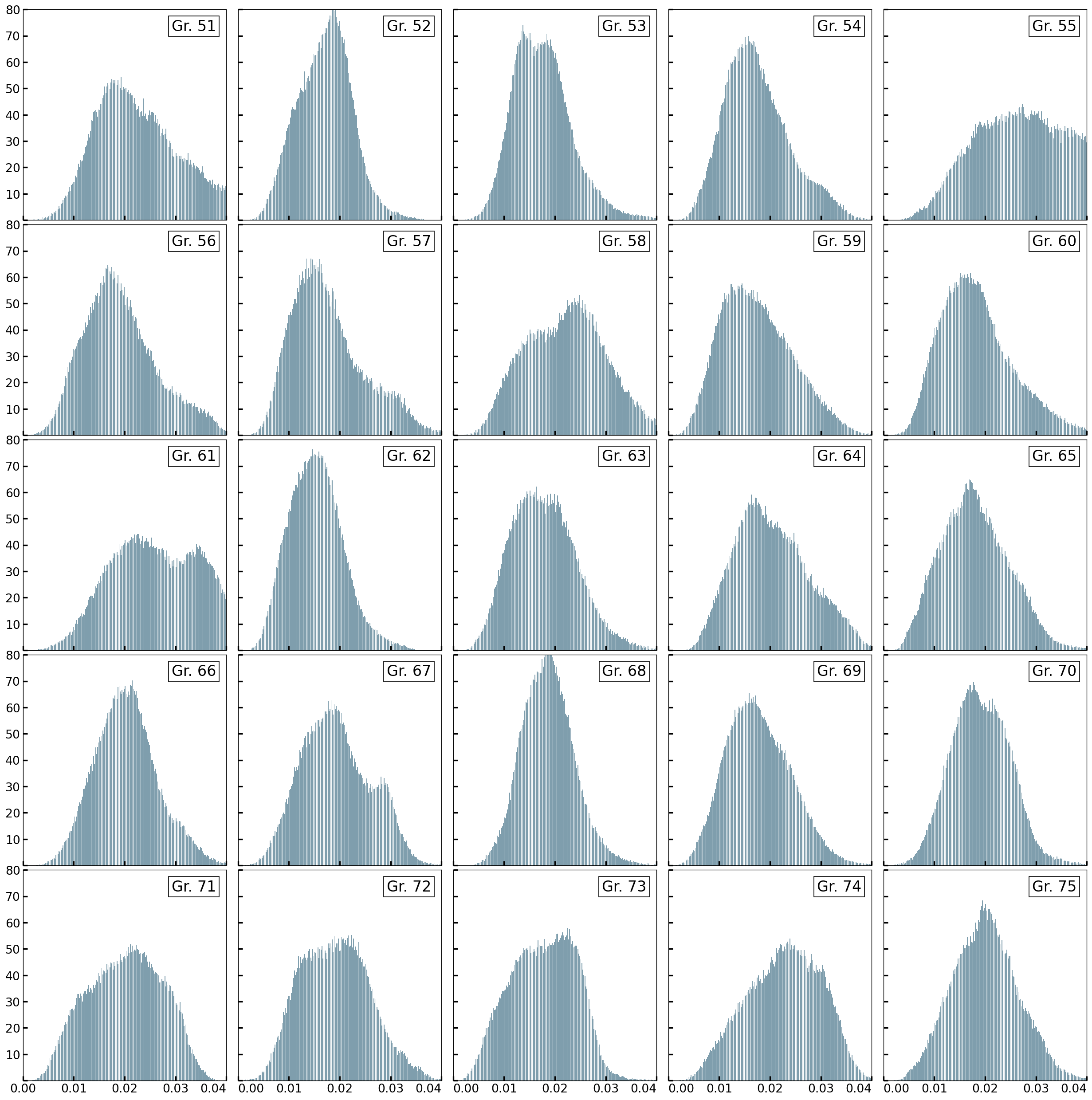}
        \subcaption{Distribution of elastic strain \Eel{} as histograms in grains 51 to 75}
        \label{sfig:EelHist51-75}
    \end{subfigure}
\end{figure}
\begin{figure}[htbp!]\ContinuedFloat
    \begin{subfigure}[b]{1.0\textwidth}
        \includegraphics[width=1.0\textwidth]{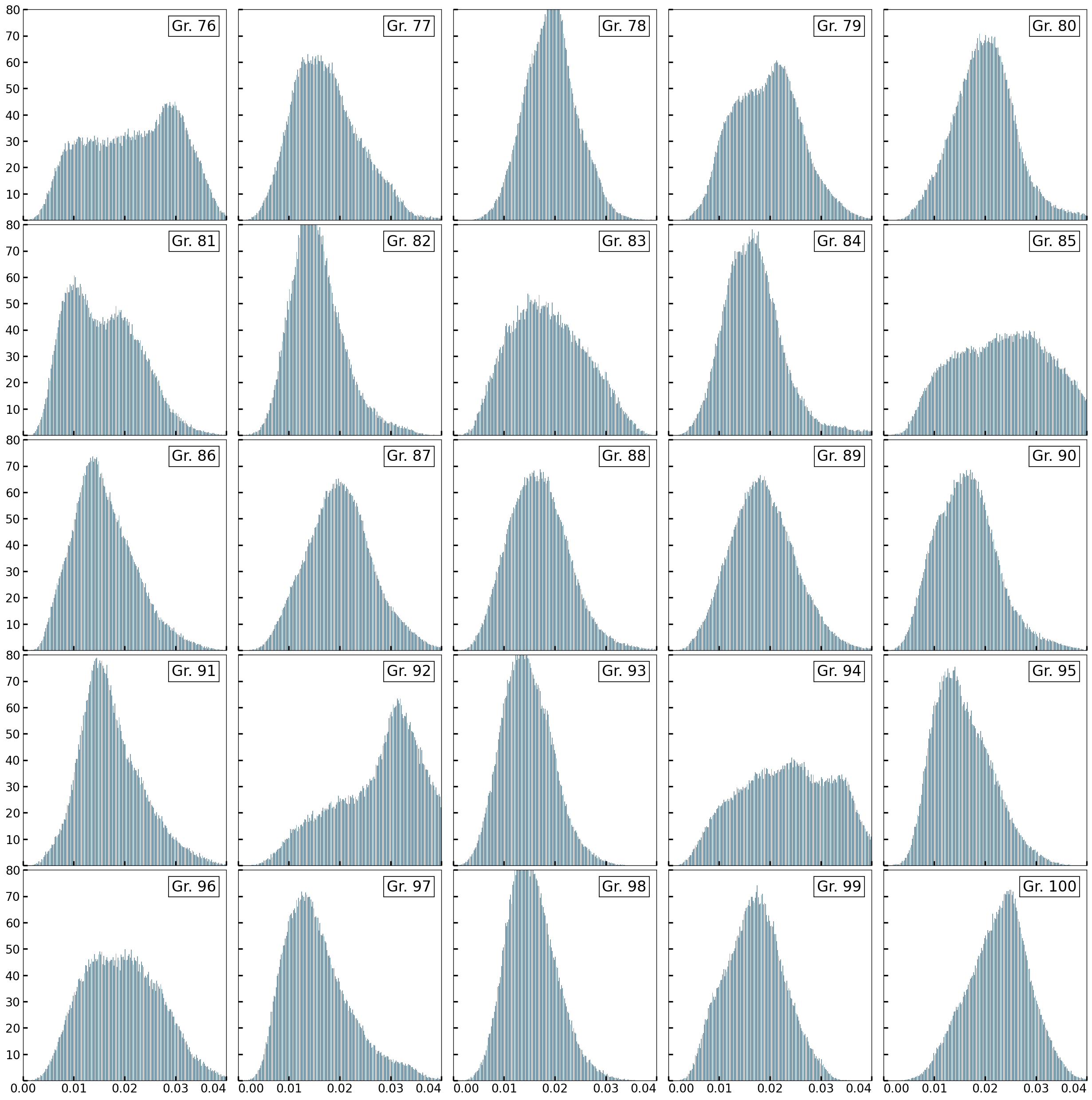}
        \subcaption{Distribution of elastic strain \Eel{} as histograms in grains 76 to 100}
        \label{sfig:EelHist76-100}
    \end{subfigure}
\end{figure}
\begin{figure}[htbp!]\ContinuedFloat
    \begin{subfigure}[b]{1.0\textwidth}
        \includegraphics[width=1.0\textwidth]{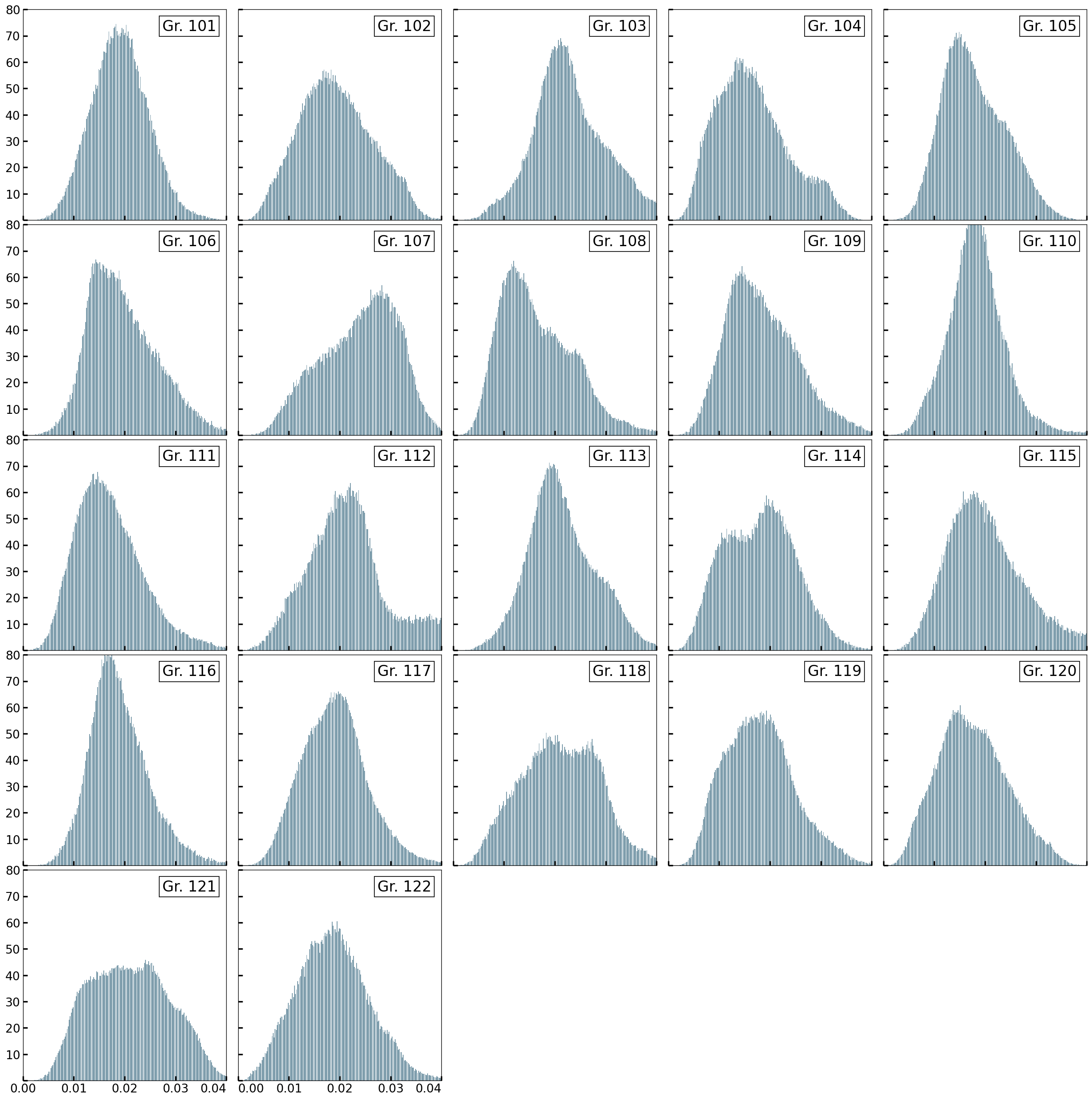}
        \subcaption{Distribution of elastic strain \Eel{} as histograms in grains 101 to 122}
        \label{sfig:EelHist101-122}
    \end{subfigure}
    \caption{Distribution of elastic strain \Eel{} in individual grains as histograms.}\label{sfig:EelHist}
\end{figure}

\begin{figure}[htbp!]
\centering
    \begin{subfigure}[b]{1.0\textwidth}
        \includegraphics[width=1.0\textwidth]{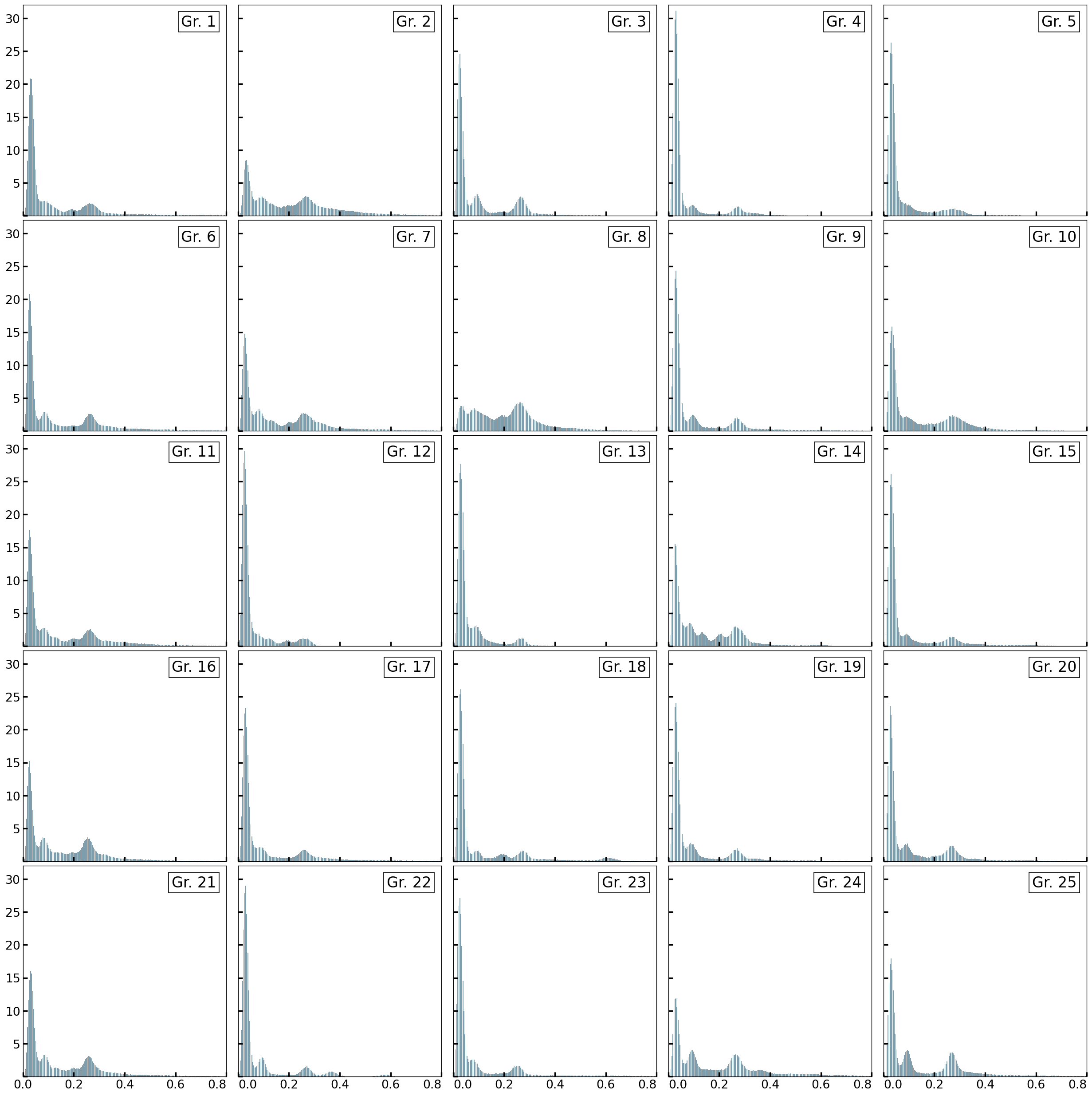}
        \subcaption{Distribution of \Evm{} as histograms in grains 1 to 25}
        \label{sfig:EvmHist1-25}
    \end{subfigure}
\end{figure}
\begin{figure}[htbp!]\ContinuedFloat
    \begin{subfigure}[b]{1.0\textwidth}
        \includegraphics[width=1.0\textwidth]{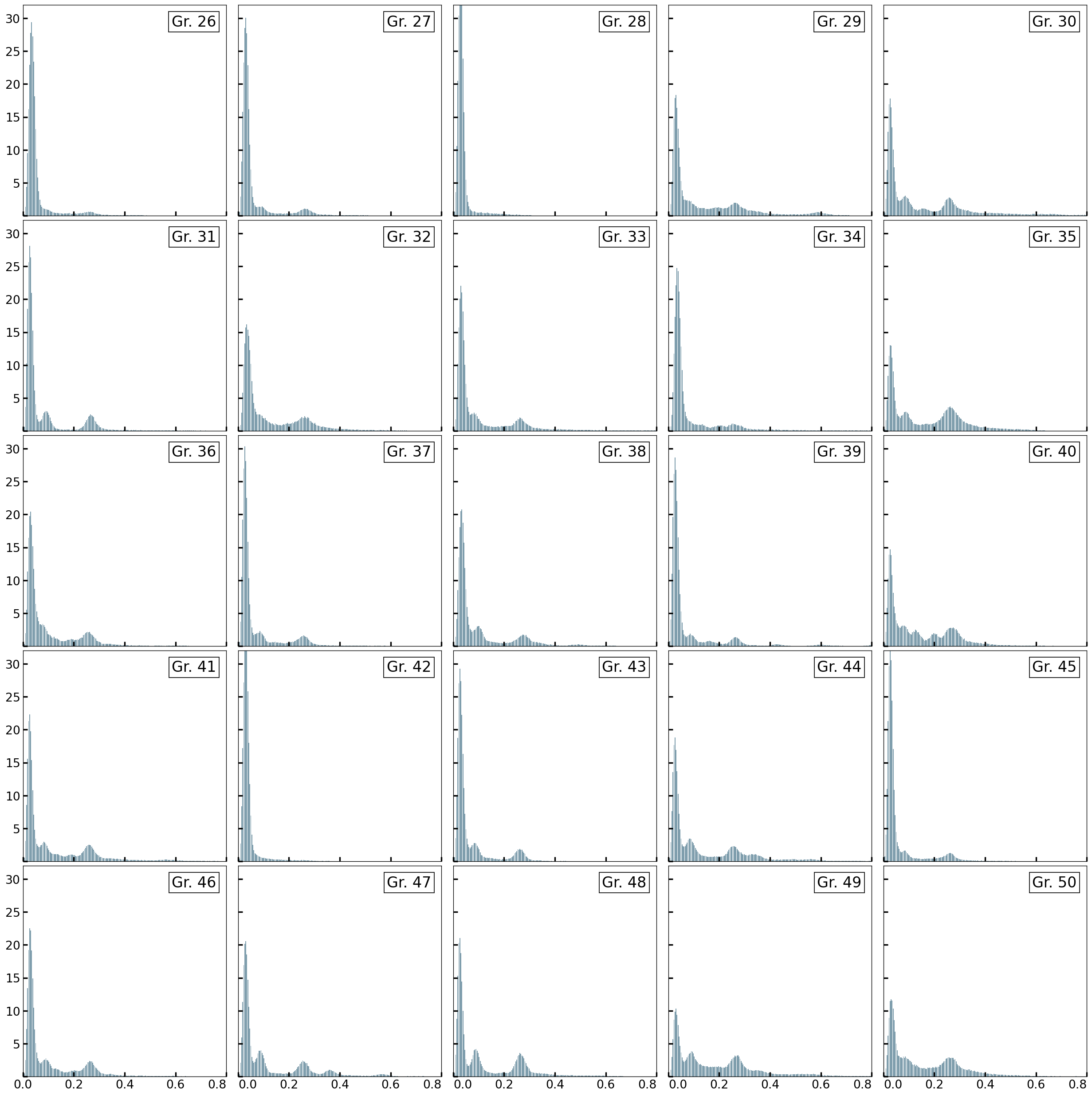}
        \subcaption{Distribution of \Evm{} as histograms in grains 26 to 50}
        \label{sfig:EvmHist26-50}
    \end{subfigure}
\end{figure}
\begin{figure}[htbp!]\ContinuedFloat
    \begin{subfigure}[b]{1.0\textwidth}
        \includegraphics[width=1.0\textwidth]{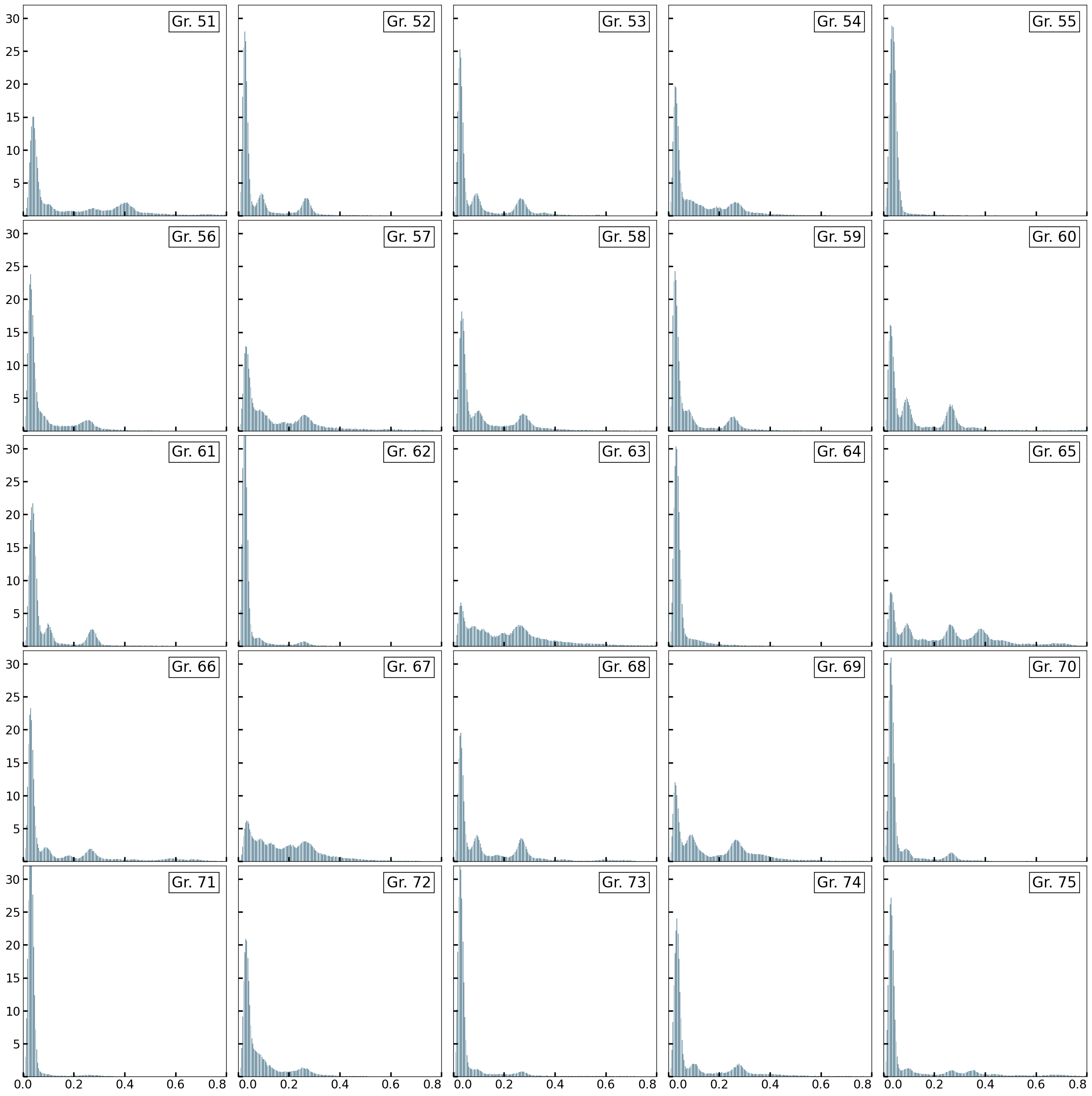}
        \subcaption{Distribution of \Evm{} as histograms in grains 51 to 75}
        \label{sfig:EvmHist51-75}
    \end{subfigure}
\end{figure}
\begin{figure}[htbp!]\ContinuedFloat
    \begin{subfigure}[b]{1.0\textwidth}
        \includegraphics[width=1.0\textwidth]{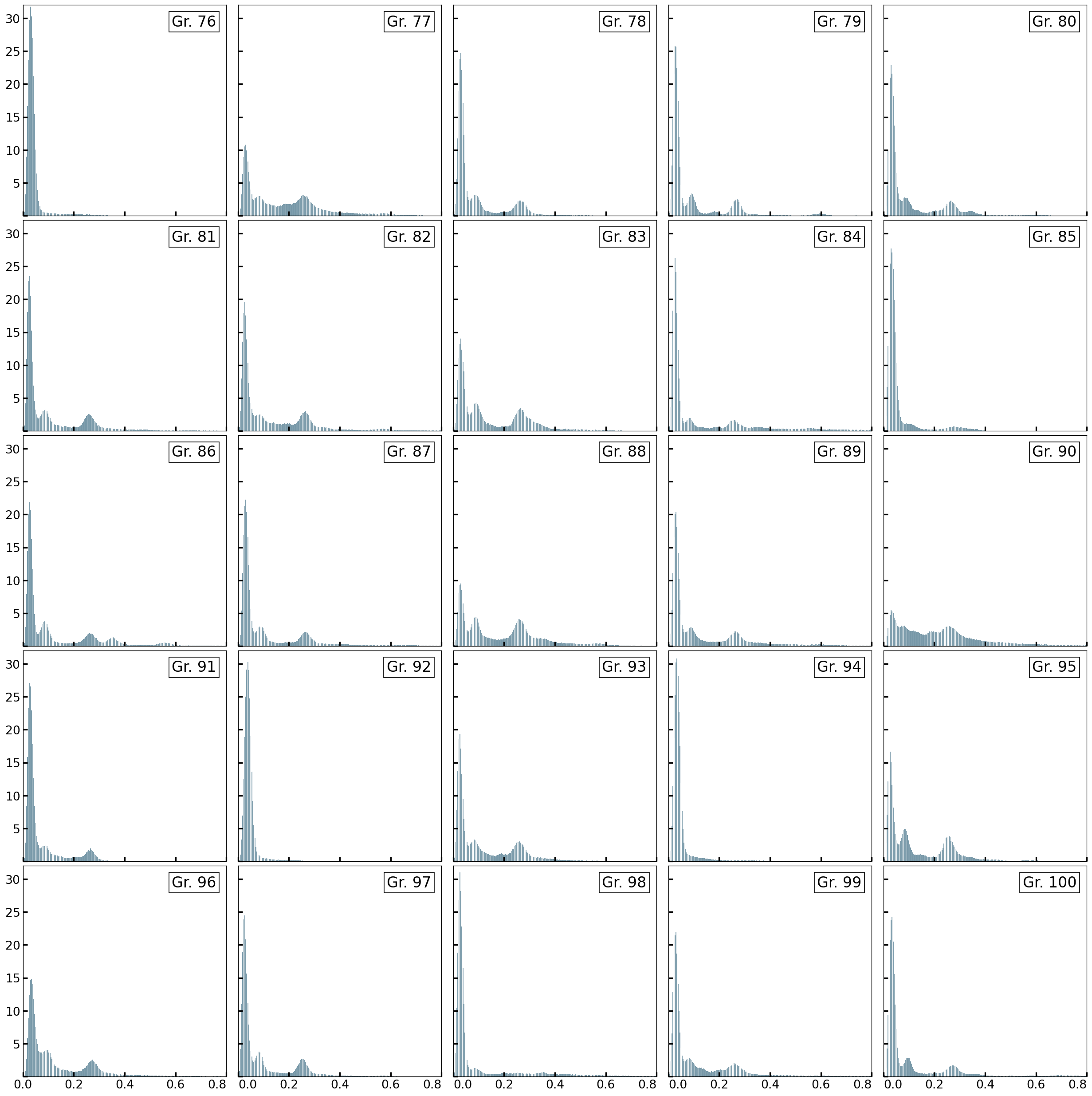}
        \subcaption{Distribution of \Evm{} as histograms in grains 76 to 100}
        \label{sfig:EvmHist76-100}
    \end{subfigure}
\end{figure}
\begin{figure}[htbp!]\ContinuedFloat
    \begin{subfigure}[b]{1.0\textwidth}
        \includegraphics[width=1.0\textwidth]{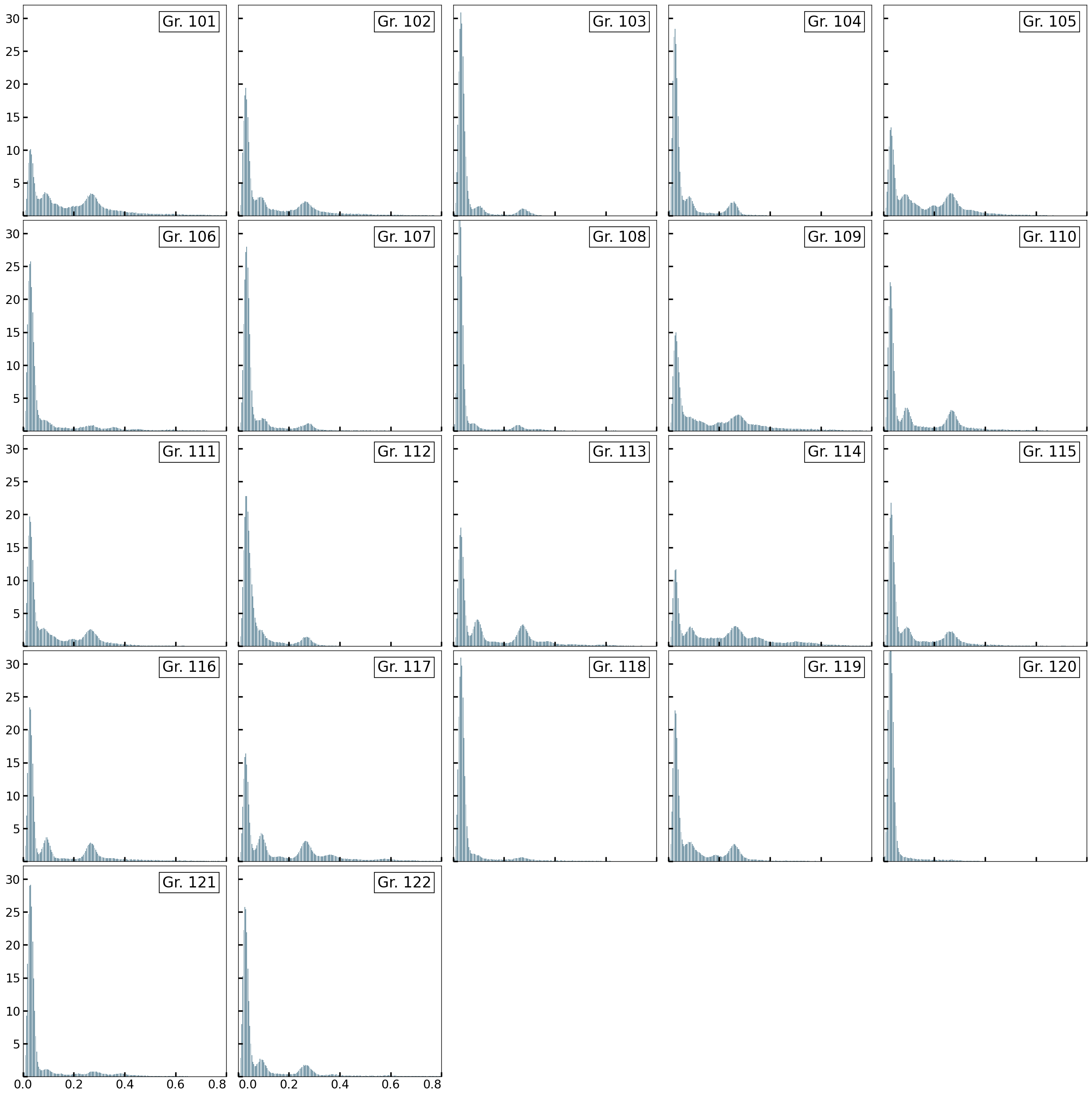}
        \subcaption{Distribution of \Evm{} as histograms in grains 101 to 122}
        \label{sfig:EvmHist101-122}
    \end{subfigure}
    \caption{Distribution of total strain \Evm{} in individual grains as histograms.}\label{sfig:EvmHist}
\end{figure}

\begin{figure}[htbp!]
\centering
    \begin{subfigure}[b]{1.0\textwidth}
        \includegraphics[width=1.0\textwidth]{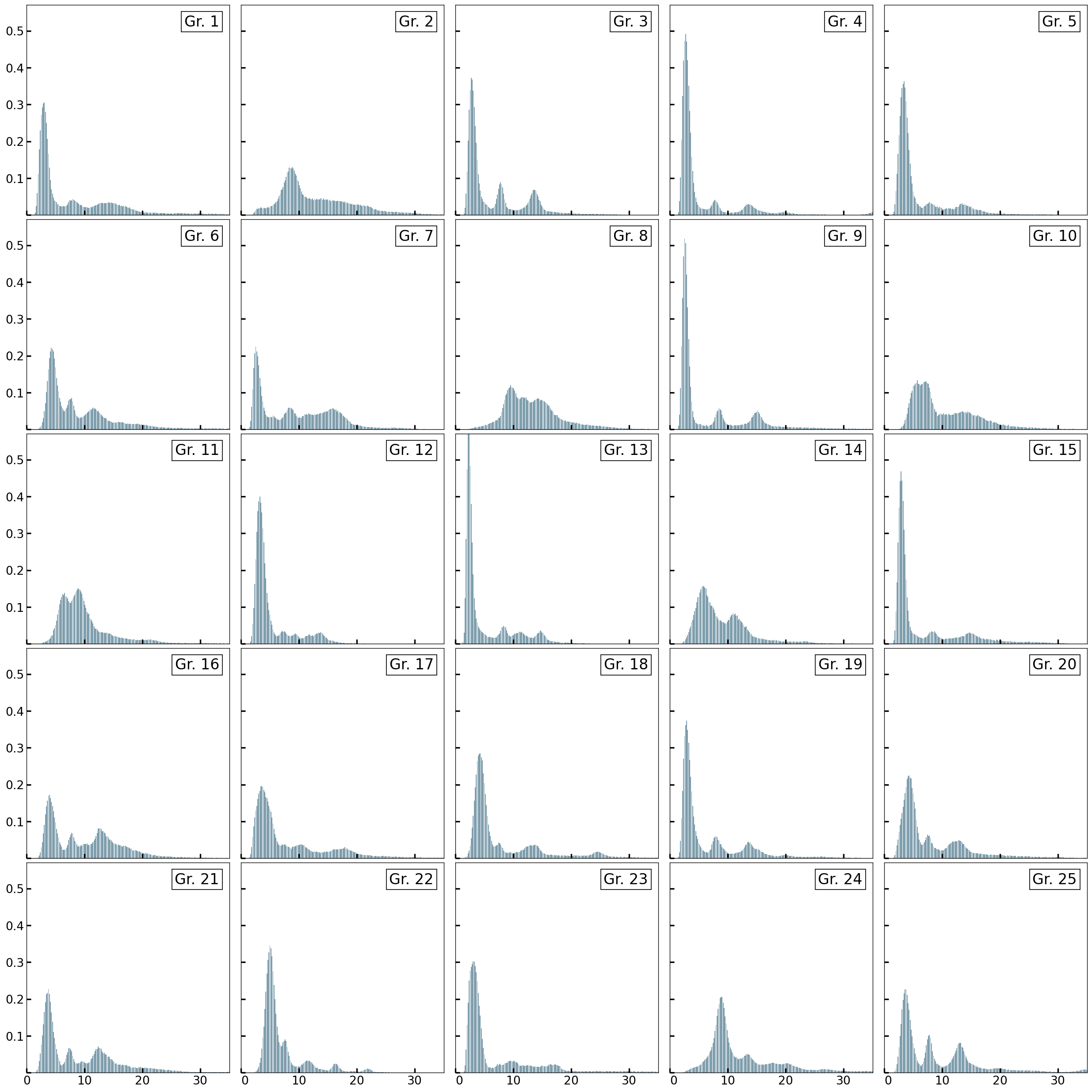}
        \subcaption{Distribution of \drot{} as histograms in grains 1 to 25}
        \label{sfig:DrotHist1-25}
    \end{subfigure}
\end{figure}
\begin{figure}[htbp!]\ContinuedFloat
    \begin{subfigure}[b]{1.0\textwidth}
        \includegraphics[width=1.0\textwidth]{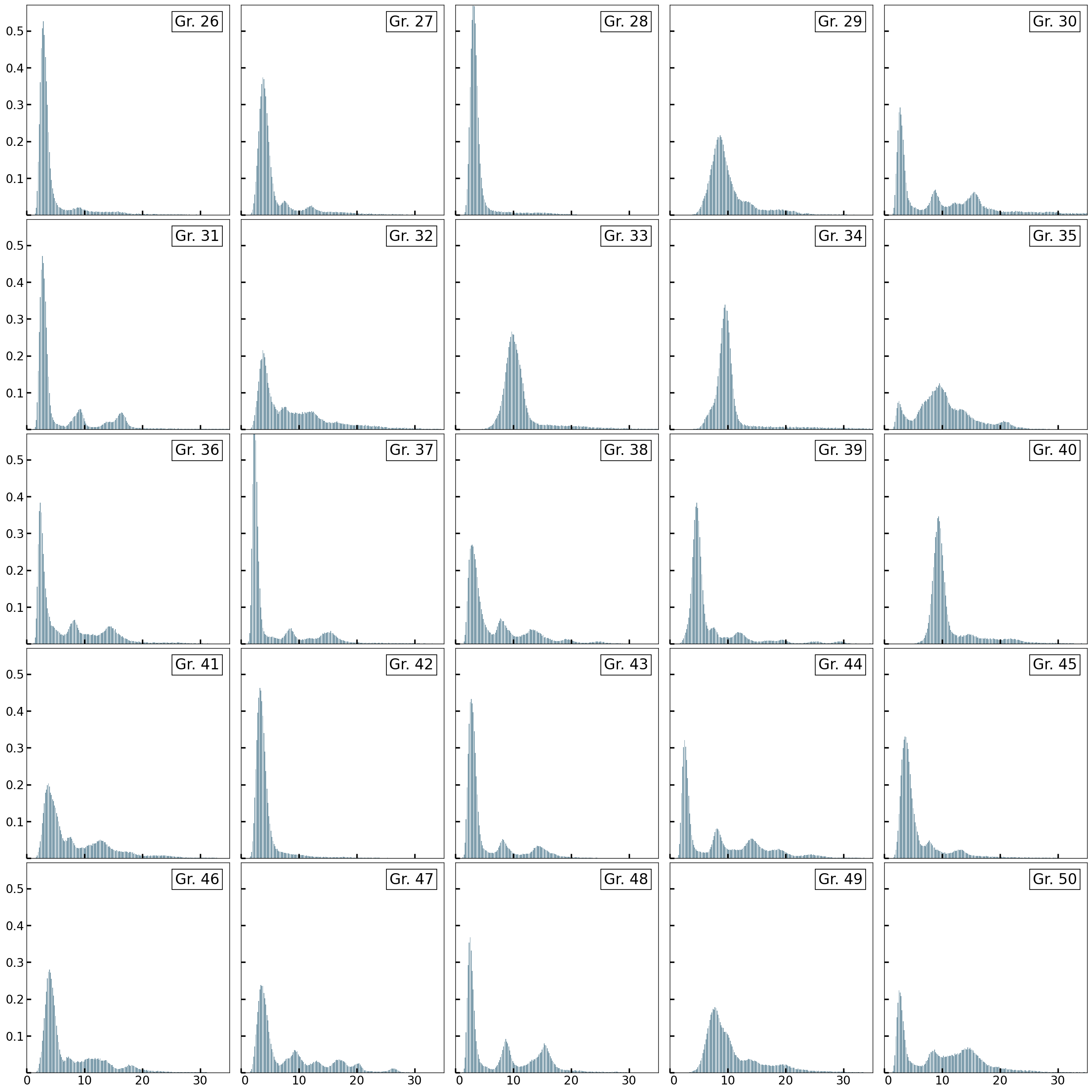}
        \subcaption{Distribution of \drot{} as histograms in grains 26 to 50}
        \label{sfig:DrotHist26-50}
    \end{subfigure}
\end{figure}
\begin{figure}[htbp!]\ContinuedFloat
    \begin{subfigure}[b]{1.0\textwidth}
        \includegraphics[width=1.0\textwidth]{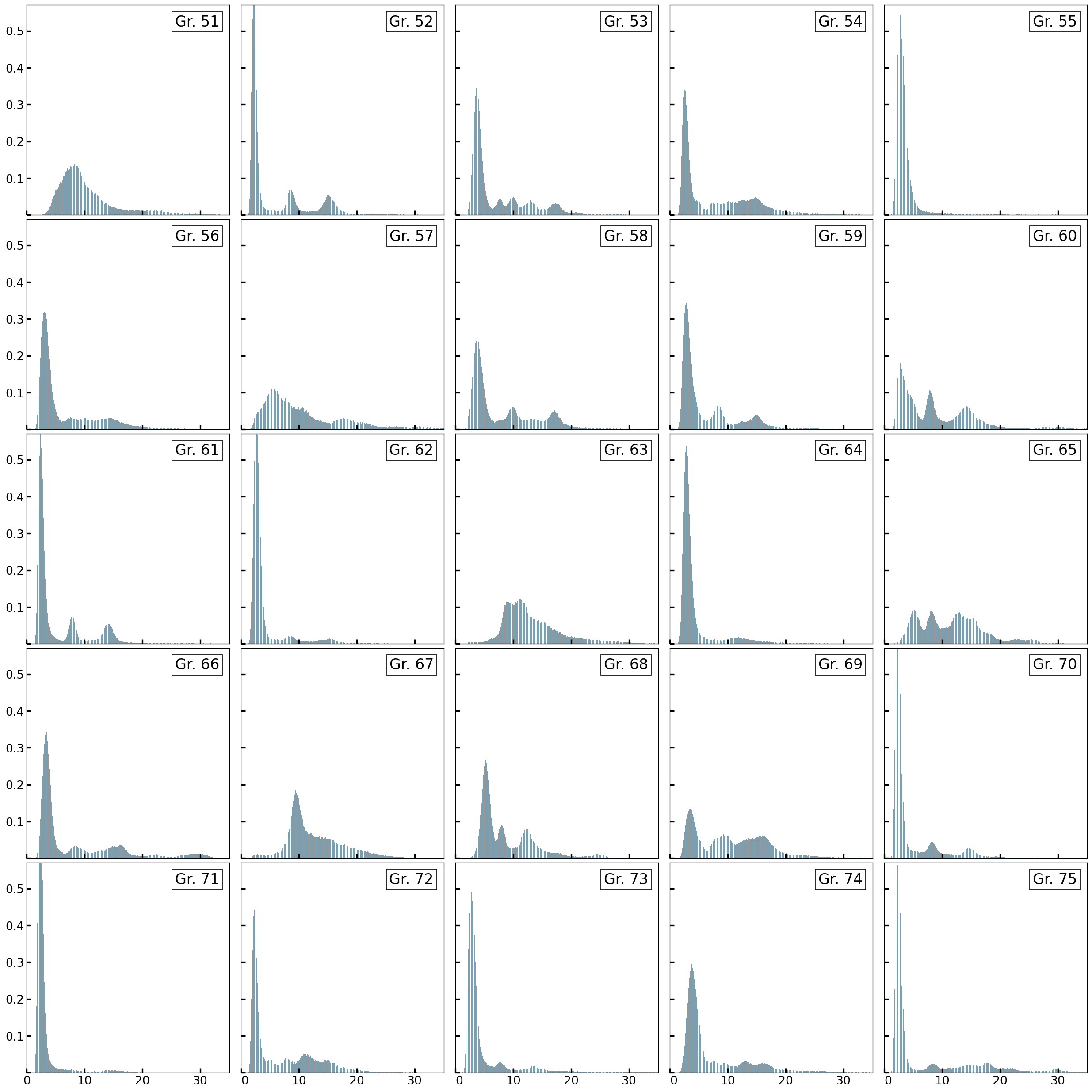}
        \subcaption{Distribution of \drot{} as histograms in grains 51 to 75}
        \label{sfig:DrotHist51-75}
    \end{subfigure}
\end{figure}
\begin{figure}[htbp!]\ContinuedFloat
    \begin{subfigure}[b]{1.0\textwidth}
        \includegraphics[width=1.0\textwidth]{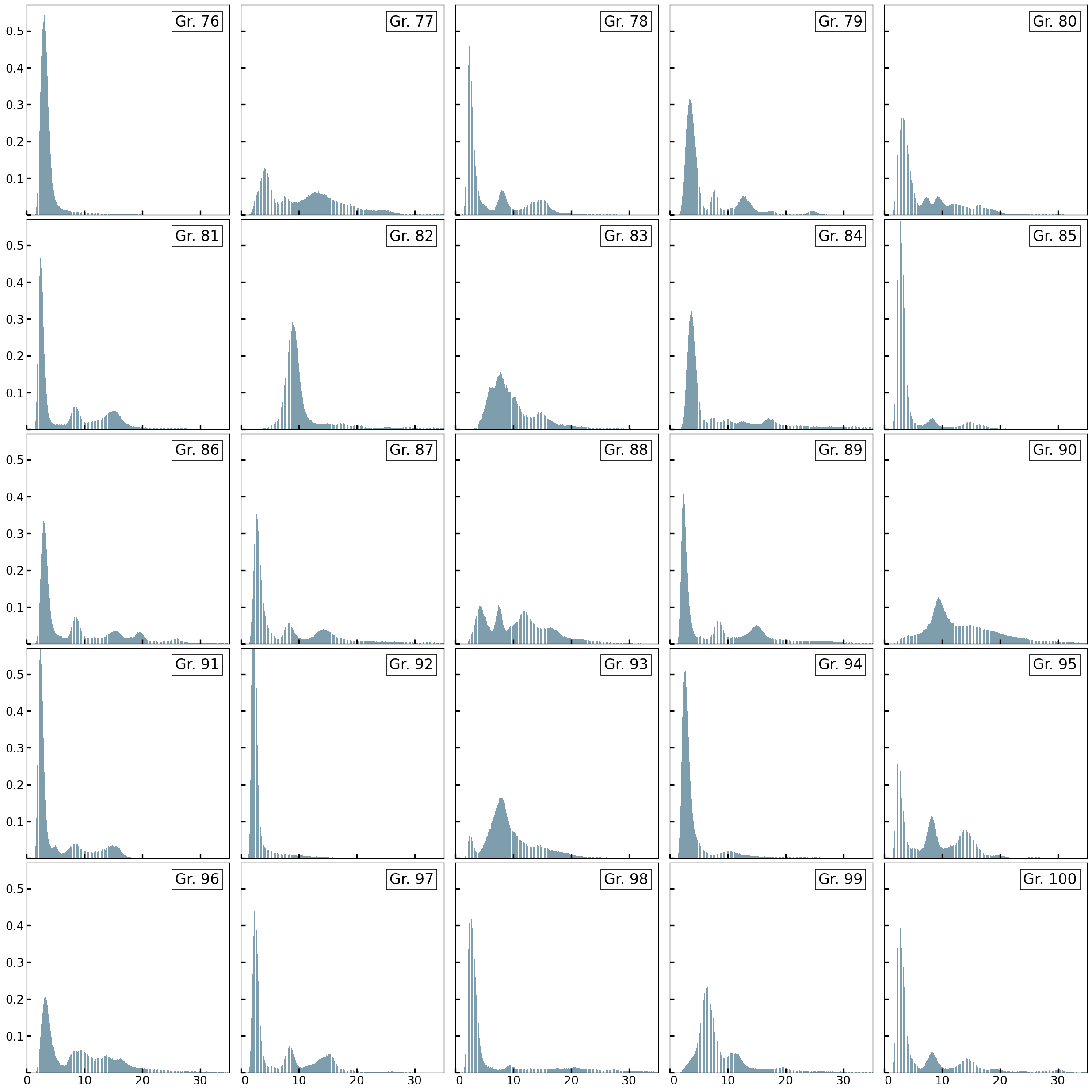}
        \subcaption{Distribution of \drot{} as histograms in grains 76 to 100}
        \label{sfig:DrotHist76-100}
    \end{subfigure}
\end{figure}
\begin{figure}[htbp!]\ContinuedFloat
    \begin{subfigure}[b]{1.0\textwidth}
        \includegraphics[width=1.0\textwidth]{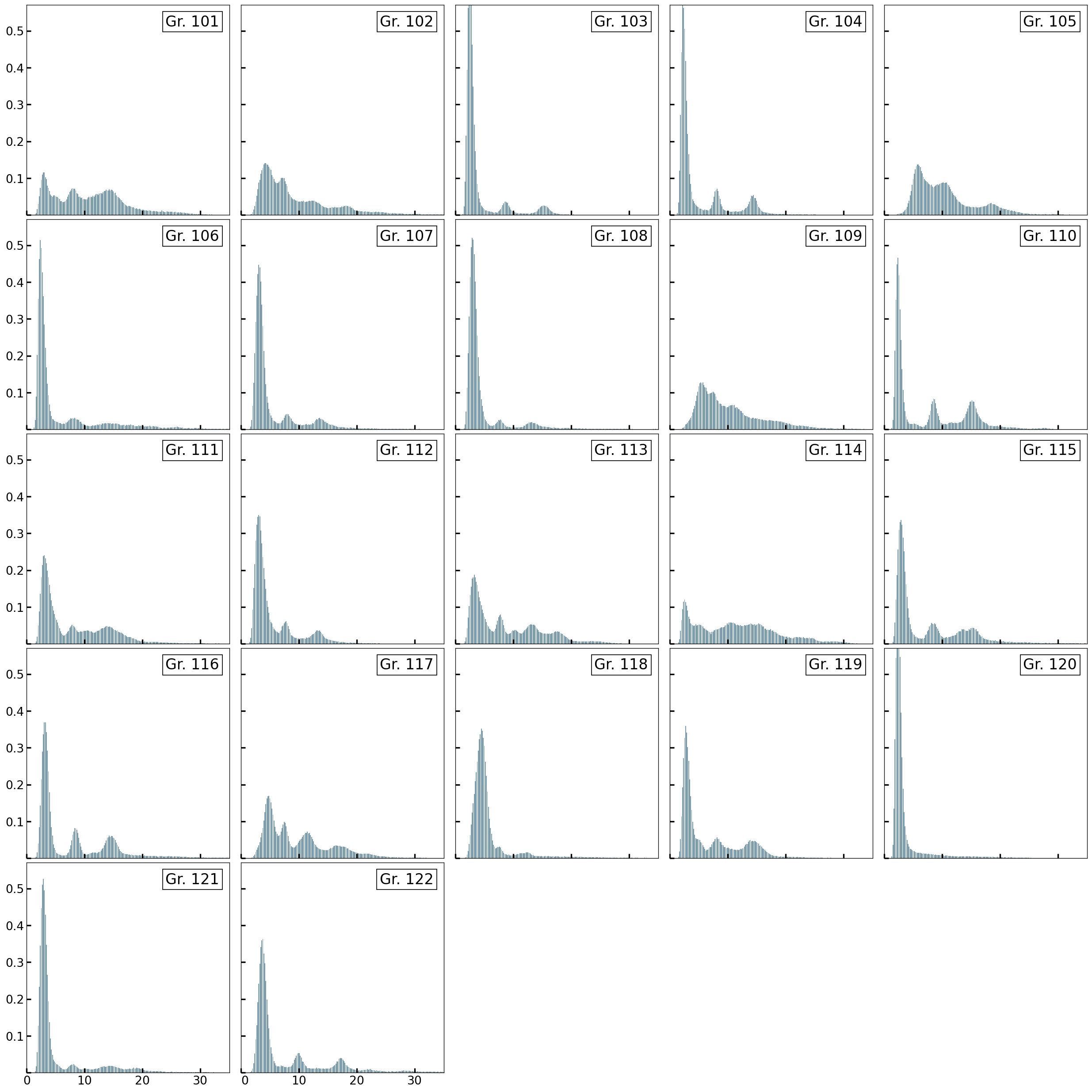}
        \subcaption{Distribution of \drot{} as histograms in grains 101 to 122}
        \label{sfig:DrotHist101-122}
    \end{subfigure}
    \caption{Distribution of the rotation angle \drot{} in individual grains as histograms.}\label{sfig:DrotHist}
\end{figure}

\clearpage

\begin{figure}[htp!]
    \begin{subfigure}[t!]{1.0\textwidth}
        \includegraphics[width=1.0\textheight, angle=90]{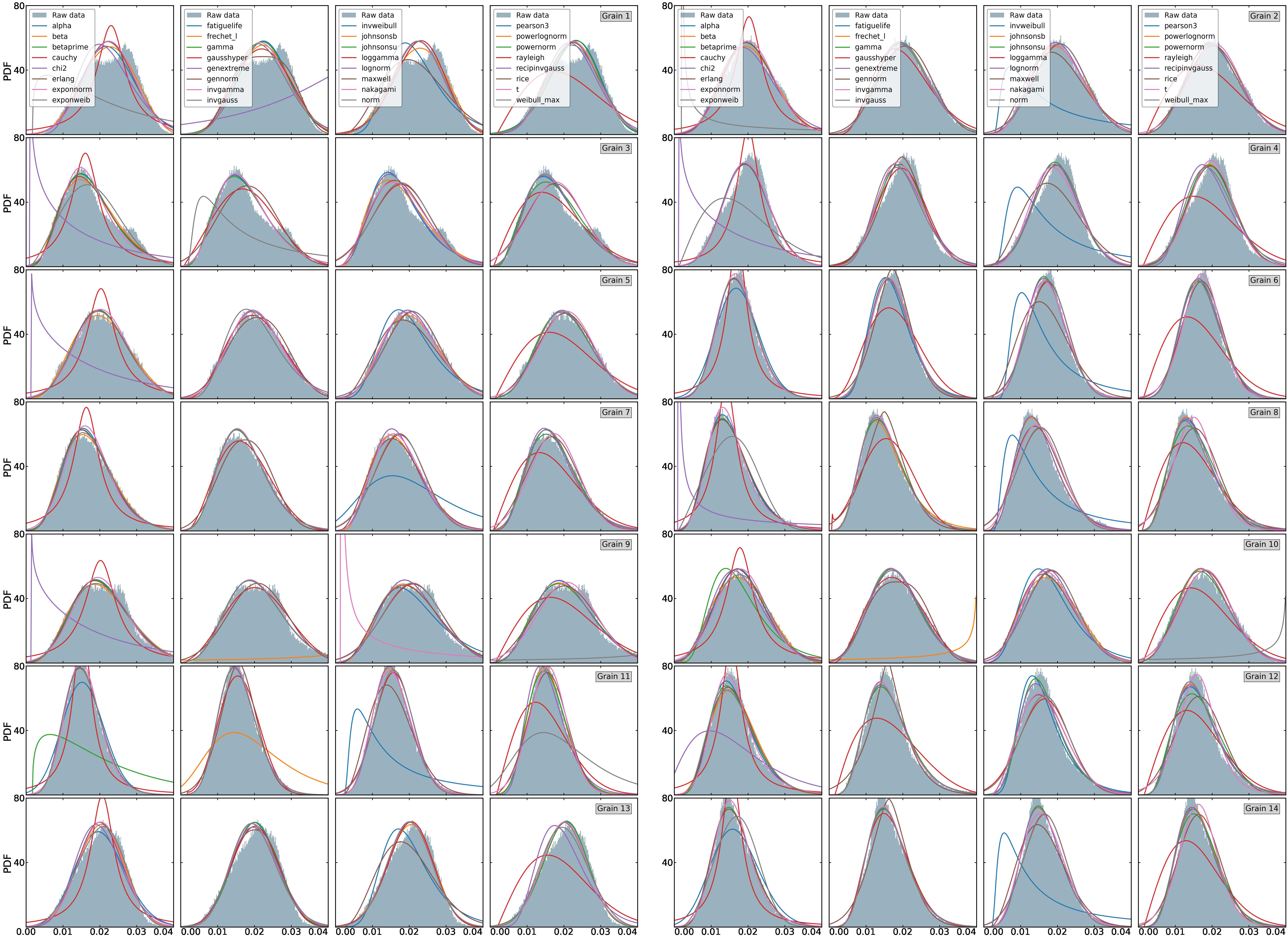}
        \subcaption{Results of identifying the mathematical form of distribution of \Eel{} in grains 1 to 14.}
        \label{sfig:EelWFit1-14}
    \end{subfigure}
\end{figure}
\begin{figure}[htbp!]\ContinuedFloat
    \begin{subfigure}[t!]{1.0\textwidth}
        \includegraphics[width=1.0\textheight, angle=90]{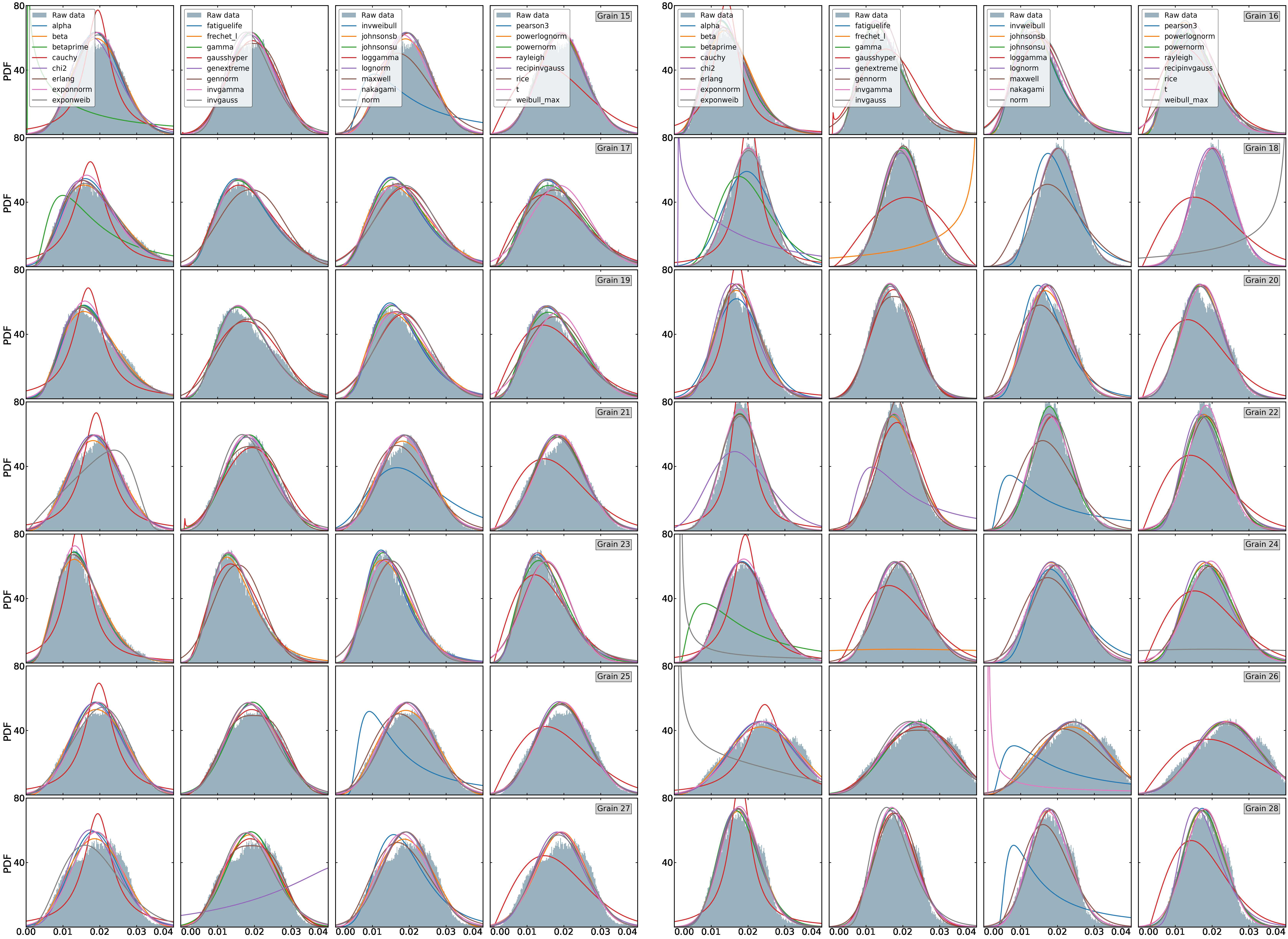}
        \subcaption{Results of identifying the mathematical form of distribution of \Eel{} in grains 15 to 28.}
        \label{sfig:EelWFit15-28}
    \end{subfigure}
\end{figure}

\begin{figure}[htbp!]\ContinuedFloat
    \begin{subfigure}[t!]{1.0\textwidth}
        \includegraphics[width=1.0\textheight, angle=90]{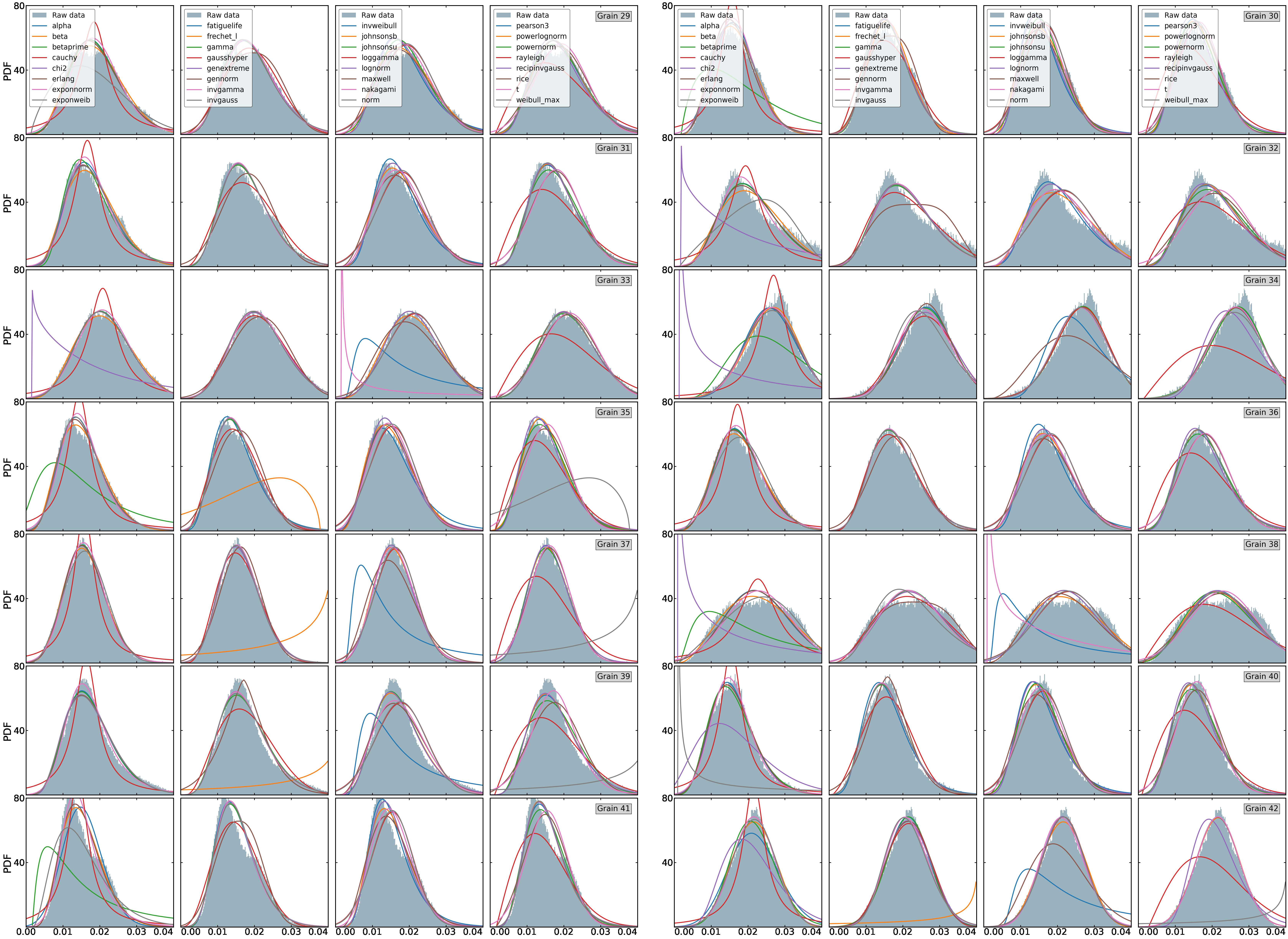}
        \subcaption{Results of identifying the mathematical form of distribution of \Eel{} in grains 29 to 42.}
        \label{sfig:EelWFit29-42}
    \end{subfigure}
\end{figure}

\begin{figure}[htbp!]\ContinuedFloat
    \begin{subfigure}[t!]{1.0\textwidth}
        \includegraphics[width=1.0\textheight, angle=90]{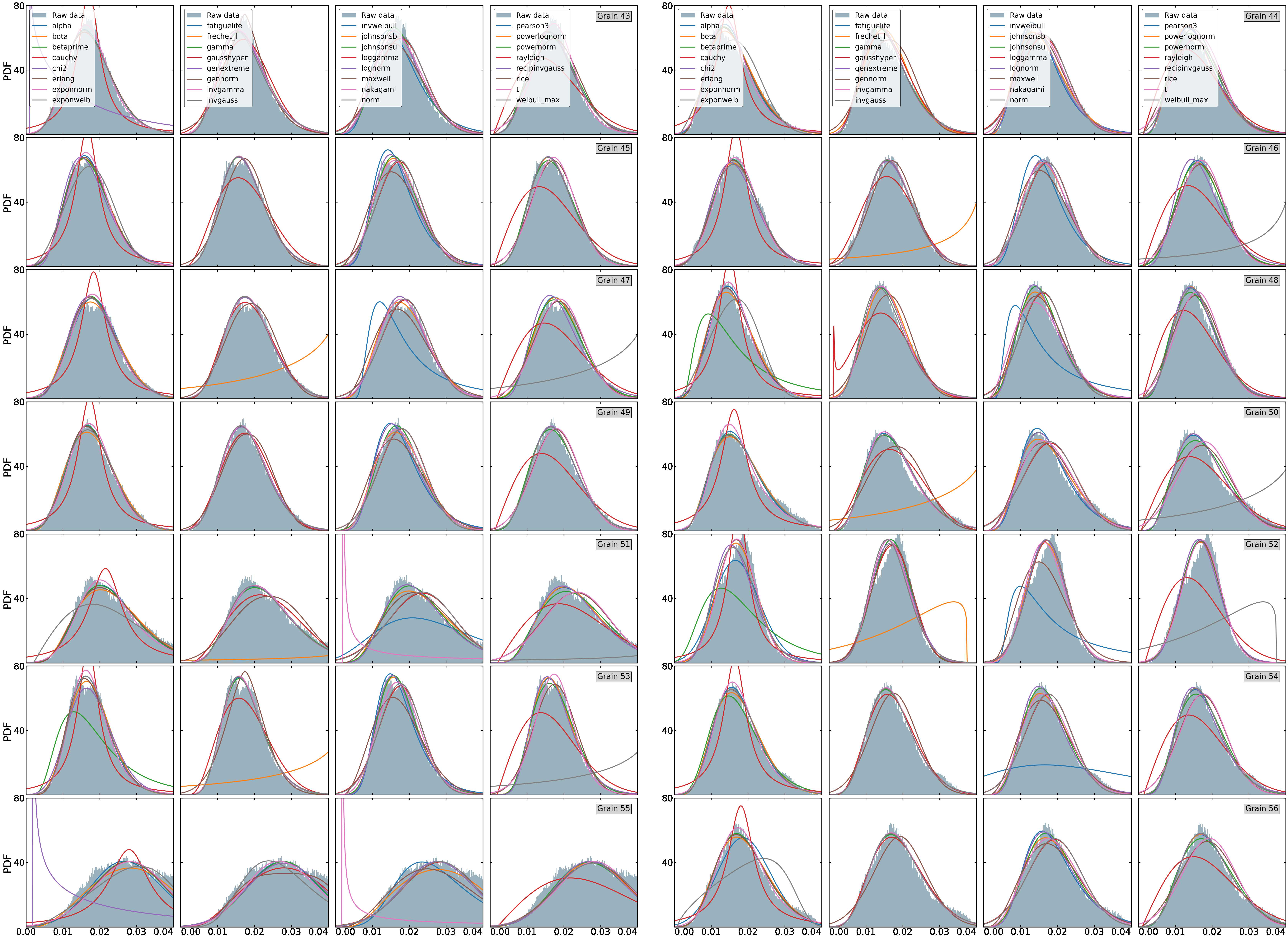}
        \subcaption{Results of identifying the mathematical form of distribution of \Eel{} in grains 43 to 56.}
        \label{sfig:EelWFit43-56}
    \end{subfigure}
\end{figure}
\begin{figure}[htbp!]\ContinuedFloat
    \begin{subfigure}[t!]{1.0\textwidth}
        \includegraphics[width=1.0\textheight, angle=90]{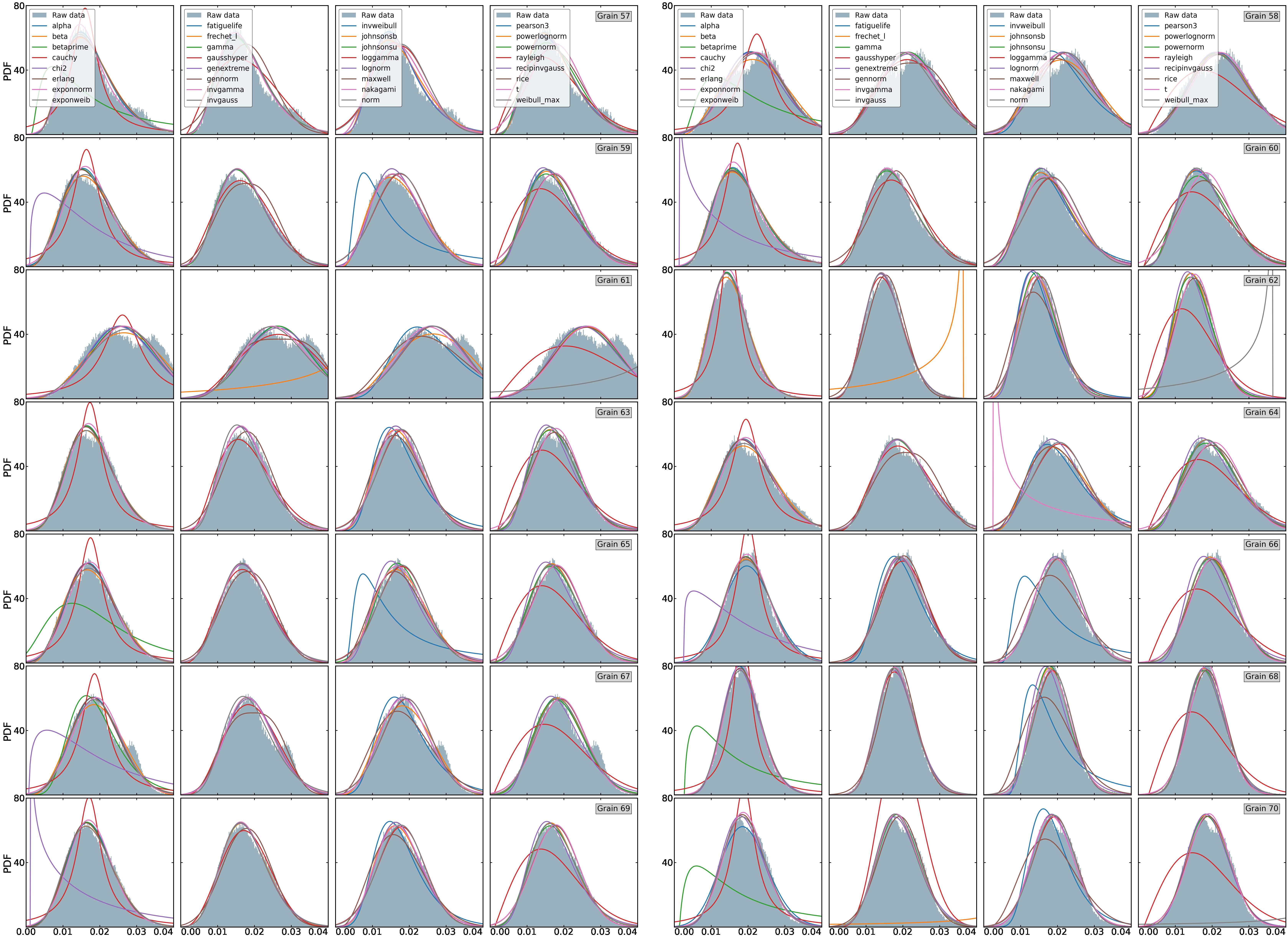}
        \subcaption{Results of identifying the mathematical form of distribution of \Eel{} in grains 57 to 70.}
        \label{sfig:EelWFit57-70}
    \end{subfigure}
\end{figure}
\begin{figure}[htbp!]\ContinuedFloat
    \begin{subfigure}[t!]{1.0\textwidth}
        \includegraphics[width=1.0\textheight, angle=90]{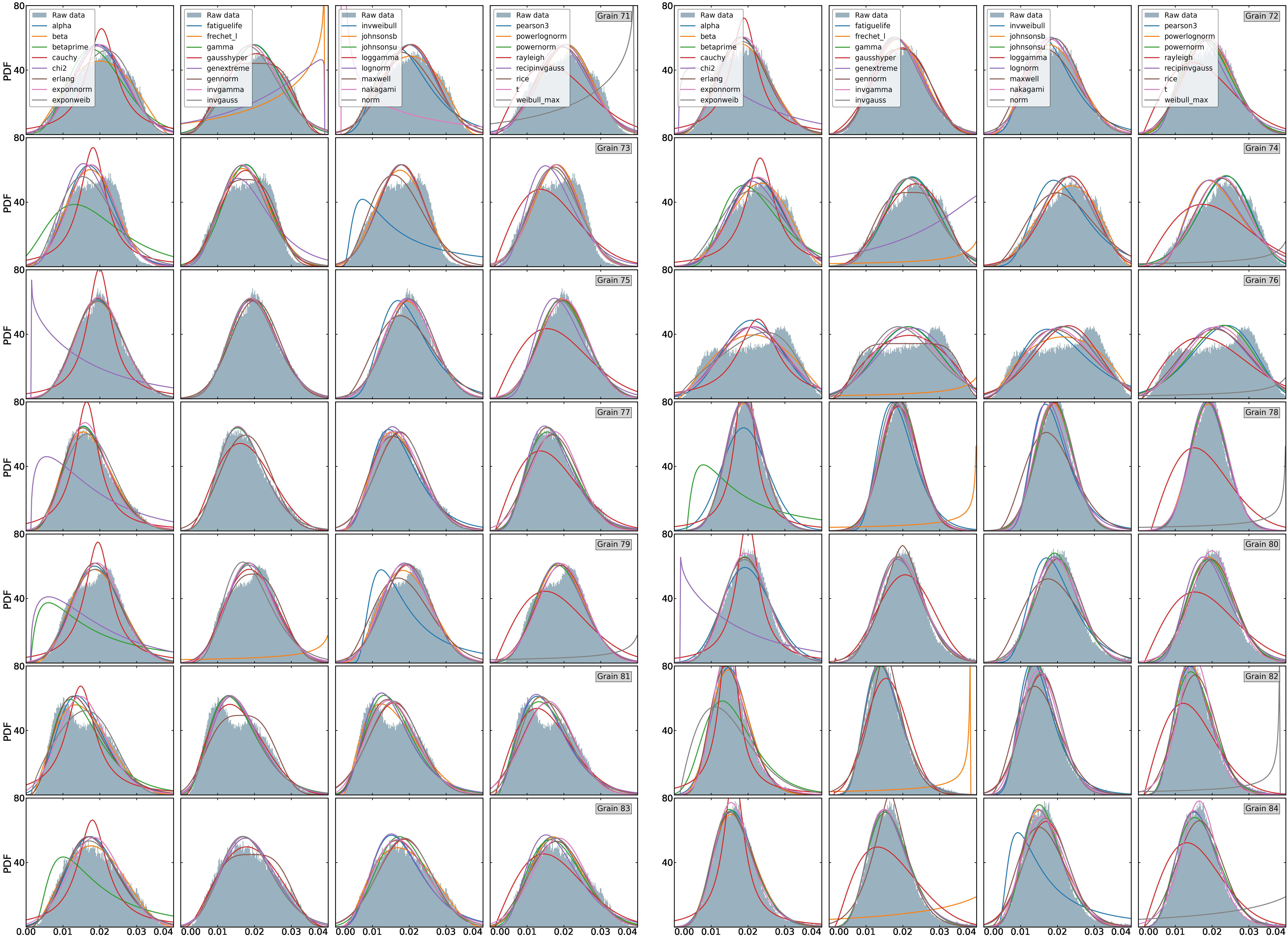}
        \subcaption{Results of identifying the mathematical form of distribution of \Eel{} in grains 71 to 84.}
        \label{sfig:EelWFit71-84}
    \end{subfigure}
\end{figure}
\begin{figure}[htbp!]\ContinuedFloat
    \begin{subfigure}[t!]{1.0\textwidth}
        \includegraphics[width=1.0\textheight, angle=90]{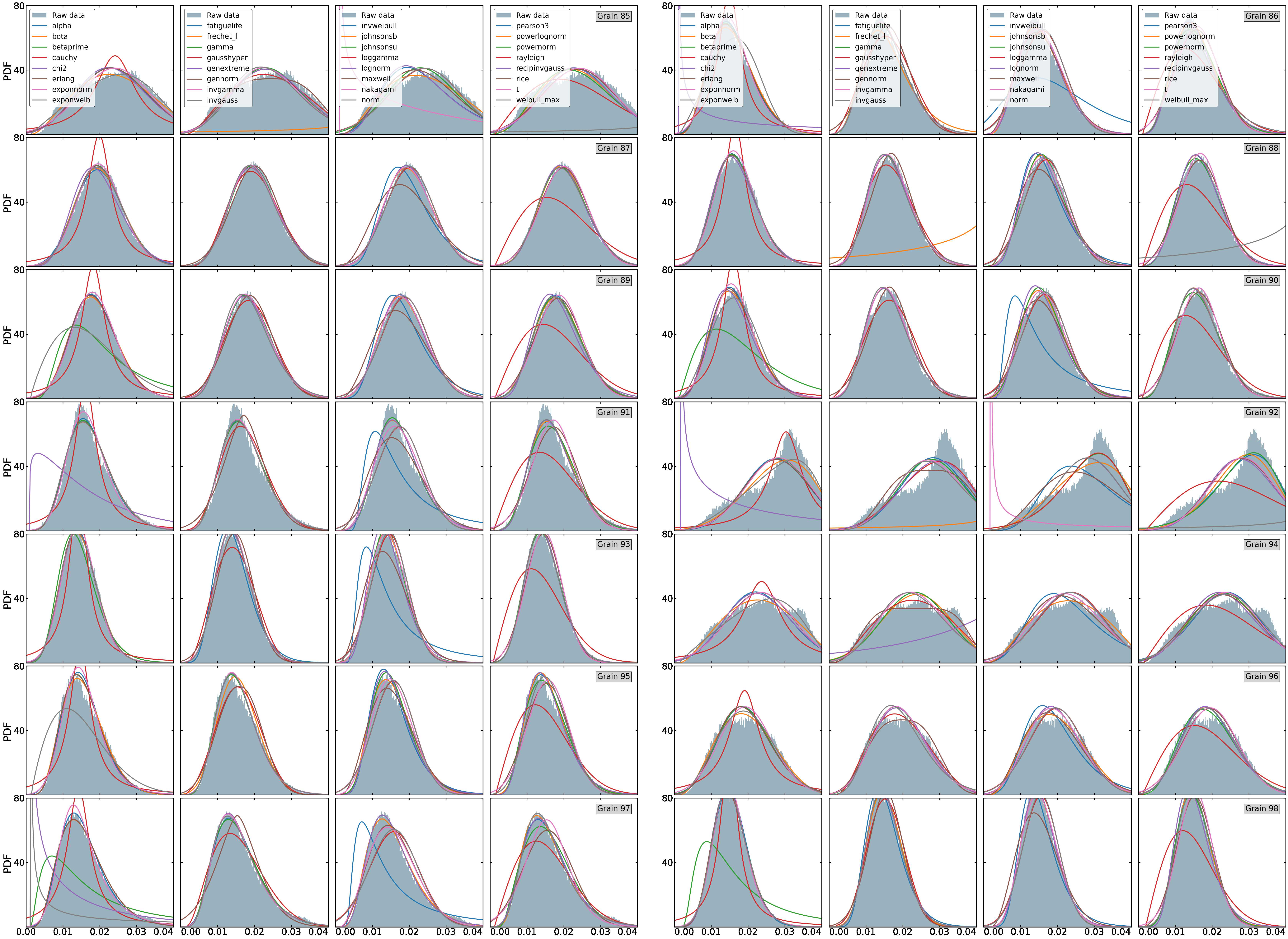}
        \subcaption{Results of identifying the mathematical form of distribution of \Eel{} in grains 85 to 98.}
        \label{sfig:EelWFit85-98}
    \end{subfigure}
\end{figure}
\begin{figure}[htbp!]\ContinuedFloat
    \begin{subfigure}[t!]{1.0\textwidth}
        \includegraphics[width=1.0\textheight, angle=90]{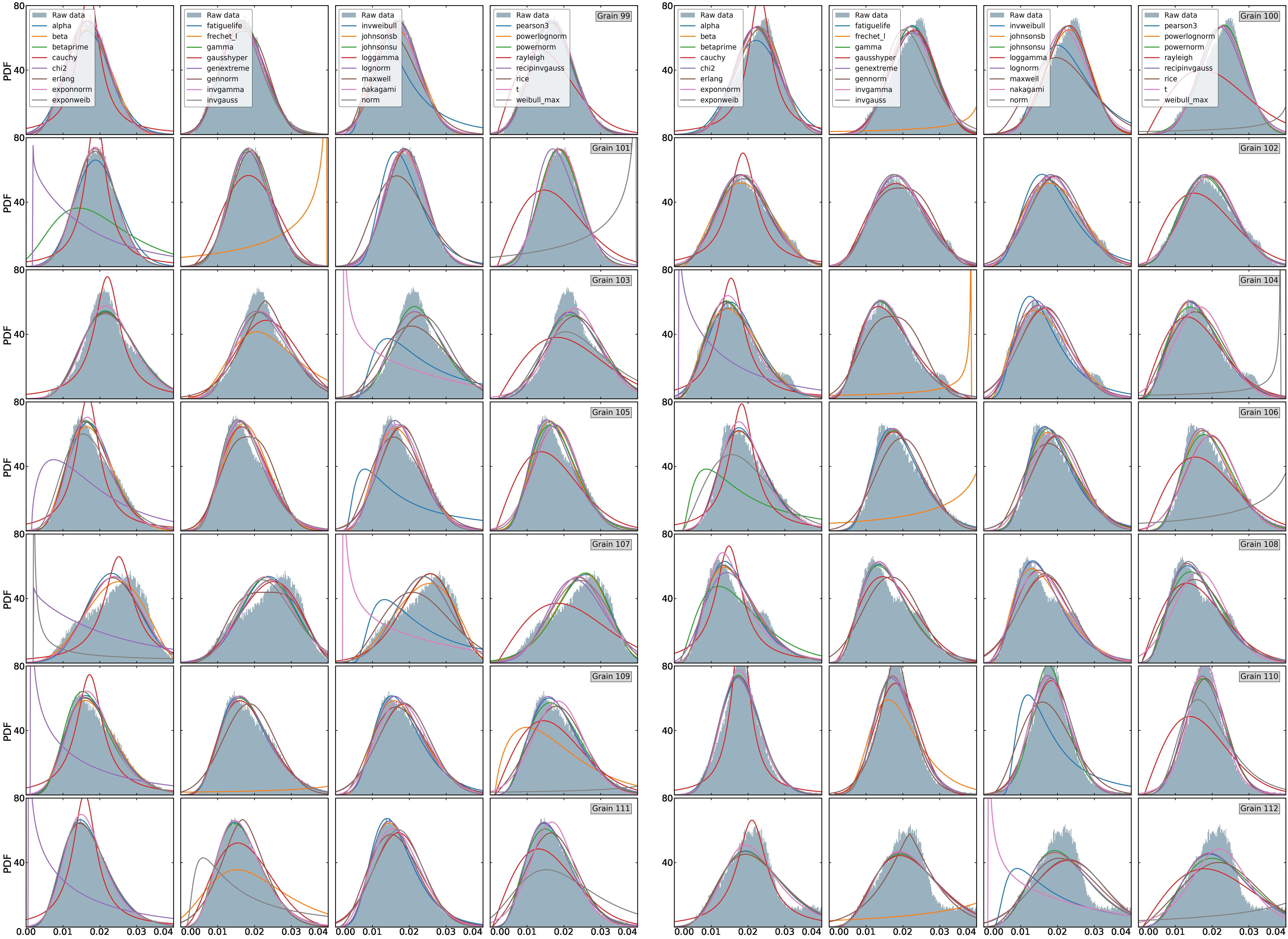}
        \subcaption{Results of identifying the mathematical form of distribution of \Eel{} in grains 99 to 112.}
        \label{sfig:EelWFit99-112}
    \end{subfigure}
\end{figure}
\begin{figure}[htbp!]\ContinuedFloat
    \begin{subfigure}[t!]{1.0\textwidth}
        \includegraphics[width=1.0\textheight, angle=90]{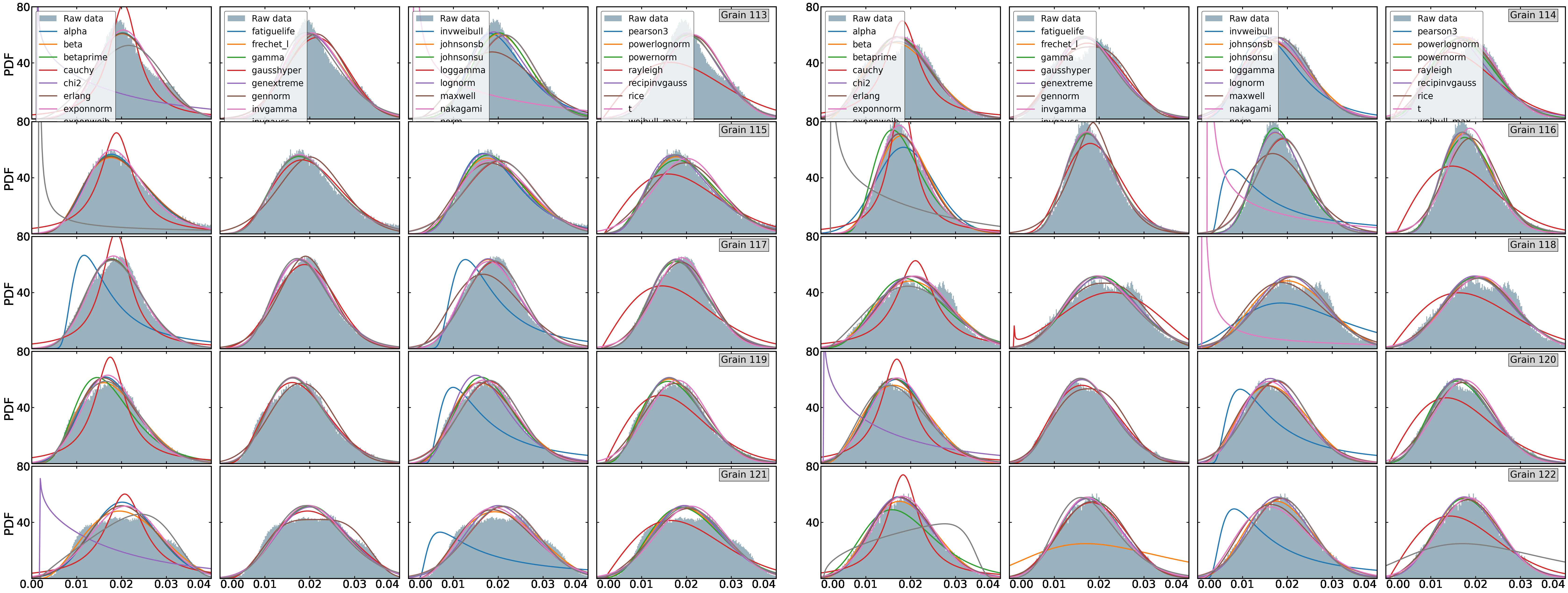}
        \subcaption{Results of identifying the mathematical form of distribution of \Eel{} in grains 113 to 122.}
        \label{sfig:EelWFit113-122}
    \end{subfigure}
    \caption{Results of identifying the mathematical form of distribution of \Eel{} in individual grains.}\label{sfig:EelWFit}
\end{figure}

\clearpage
\begin{figure}[htp!]
    \begin{subfigure}[t!]{1.0\textwidth}
        \includegraphics[width=1.0\textheight, angle=90]{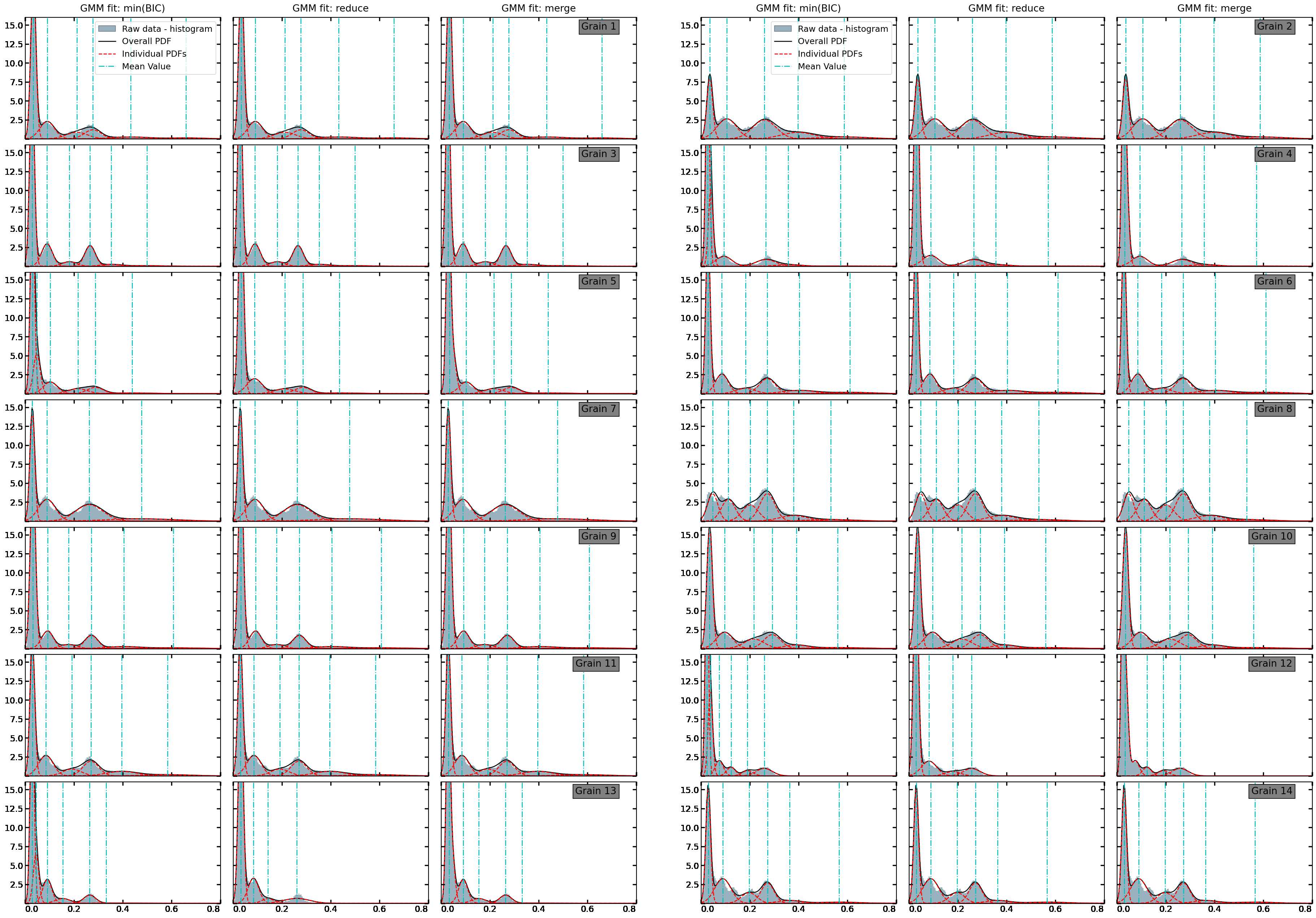}
        \subcaption{Results of Gaussian mixture model on the distribution of \Evm{} in grains 1 to 14.}
        \label{sfig:GMM_ShrStr1-14}
    \end{subfigure}
\end{figure}
\begin{figure}[htp!]\ContinuedFloat
    \begin{subfigure}[t!]{1.0\textwidth}
        \includegraphics[width=1.0\textheight, angle=90]{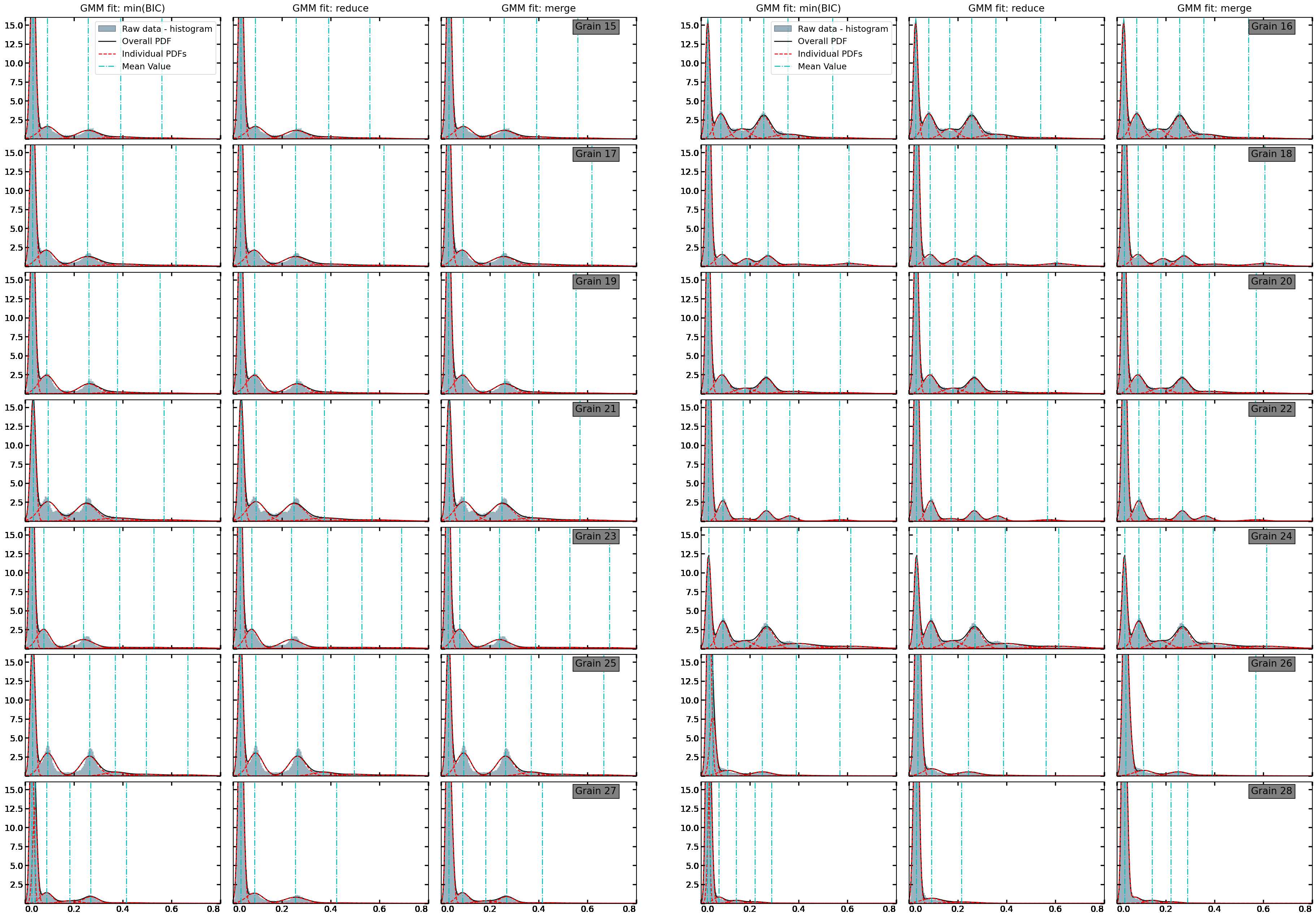}
        \subcaption{Results of Gaussian mixture model on the distribution of \Evm{} in grains 15 to 28.}
        \label{sfig:GMM_ShrStr14-28}
    \end{subfigure}
\end{figure}
\begin{figure}[htp!]\ContinuedFloat
    \begin{subfigure}[t!]{1.0\textwidth}
        \includegraphics[width=1.0\textheight, angle=90]{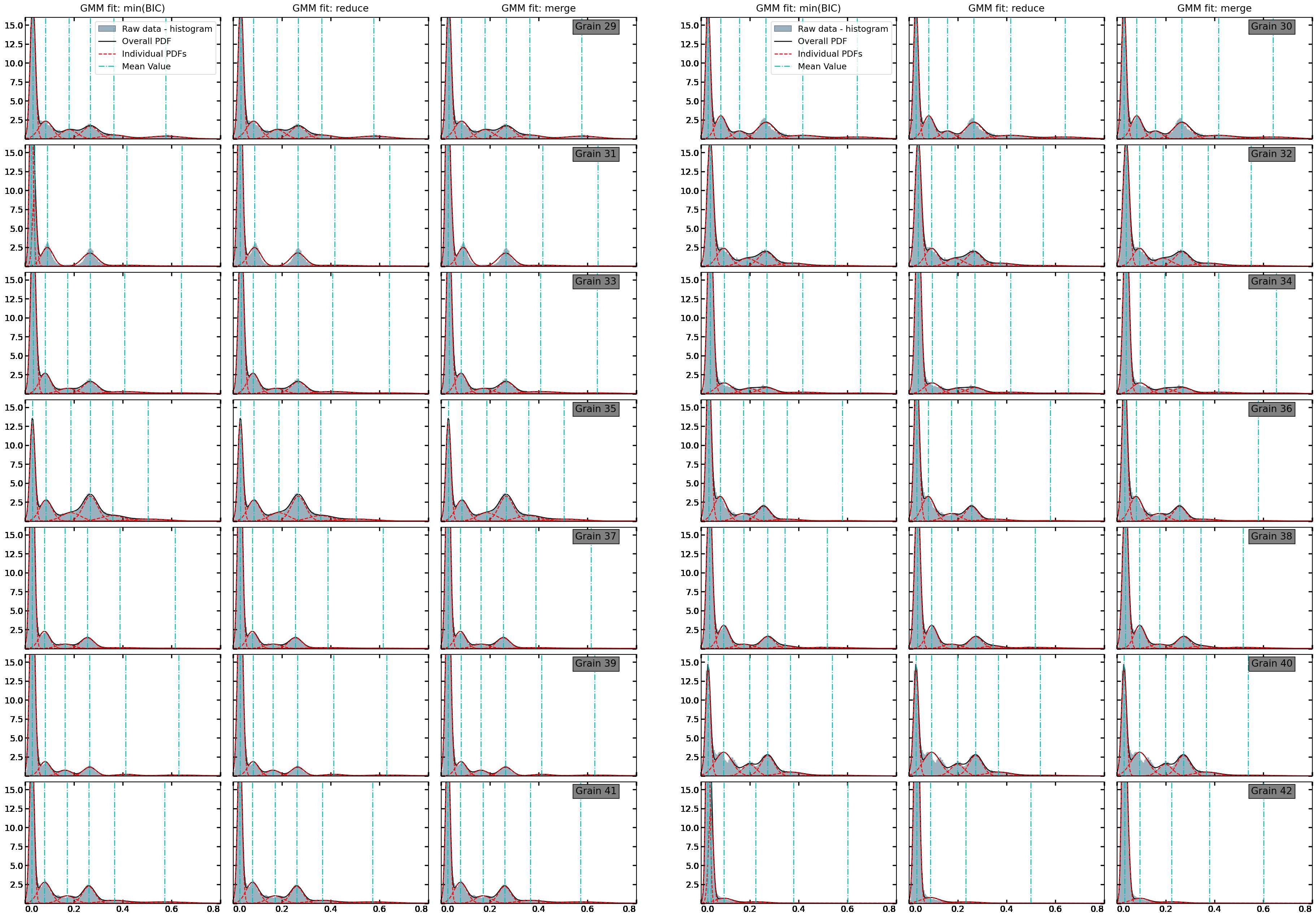}
        \subcaption{Results of Gaussian mixture model on the distribution of \Evm{} in grains 29 to 42.}
        \label{sfig:GMM_ShrStr29-42}
    \end{subfigure}
\end{figure}
\begin{figure}[htp!]\ContinuedFloat
    \begin{subfigure}[t!]{1.0\textwidth}
        \includegraphics[width=1.0\textheight, angle=90]{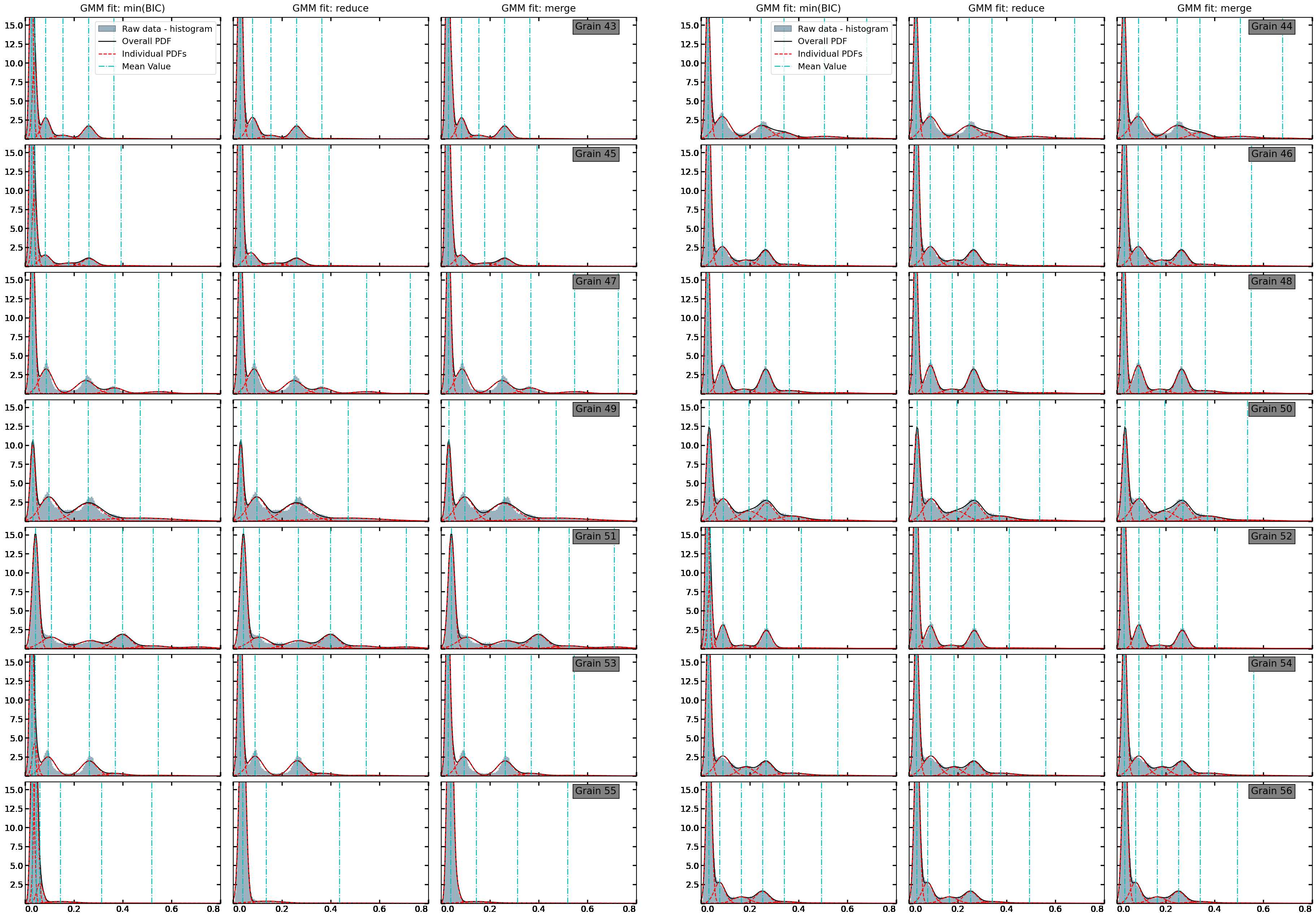}
        \subcaption{Results of Gaussian mixture model on the distribution of \Evm{} in grains 43 to 56.}
        \label{sfig:GMM_ShrStr43-56}
    \end{subfigure}
\end{figure}
\begin{figure}[htp!]\ContinuedFloat
    \begin{subfigure}[t!]{1.0\textwidth}
        \includegraphics[width=1.0\textheight, angle=90]{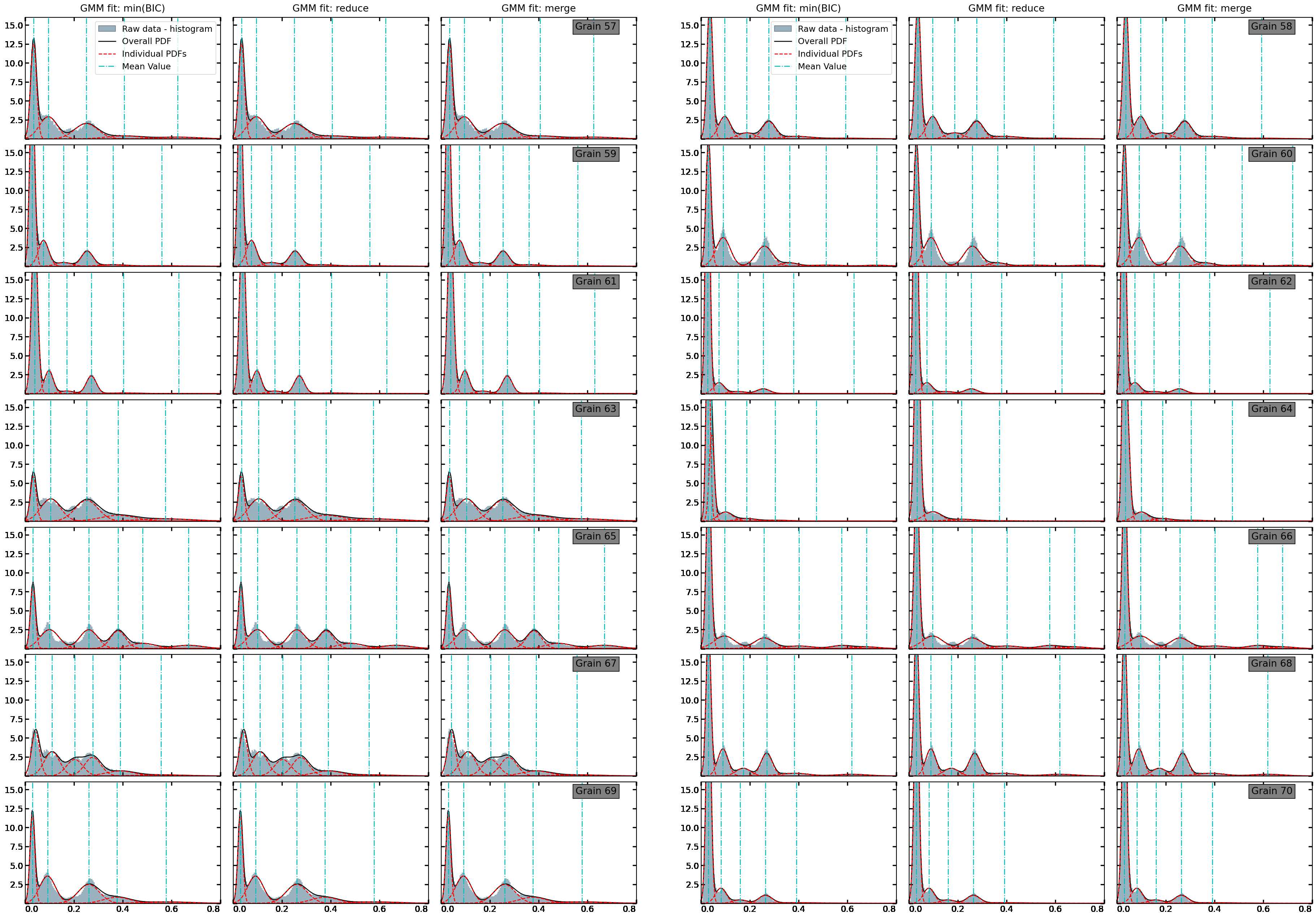}
        \subcaption{Results of Gaussian mixture model on the distribution of \Evm{} in grains 57 to 70.}
        \label{sfig:GMM_ShrStr57-70}
    \end{subfigure}
\end{figure}
\begin{figure}[htp!]\ContinuedFloat
    \begin{subfigure}[t!]{1.0\textwidth}
        \includegraphics[width=1.0\textheight, angle=90]{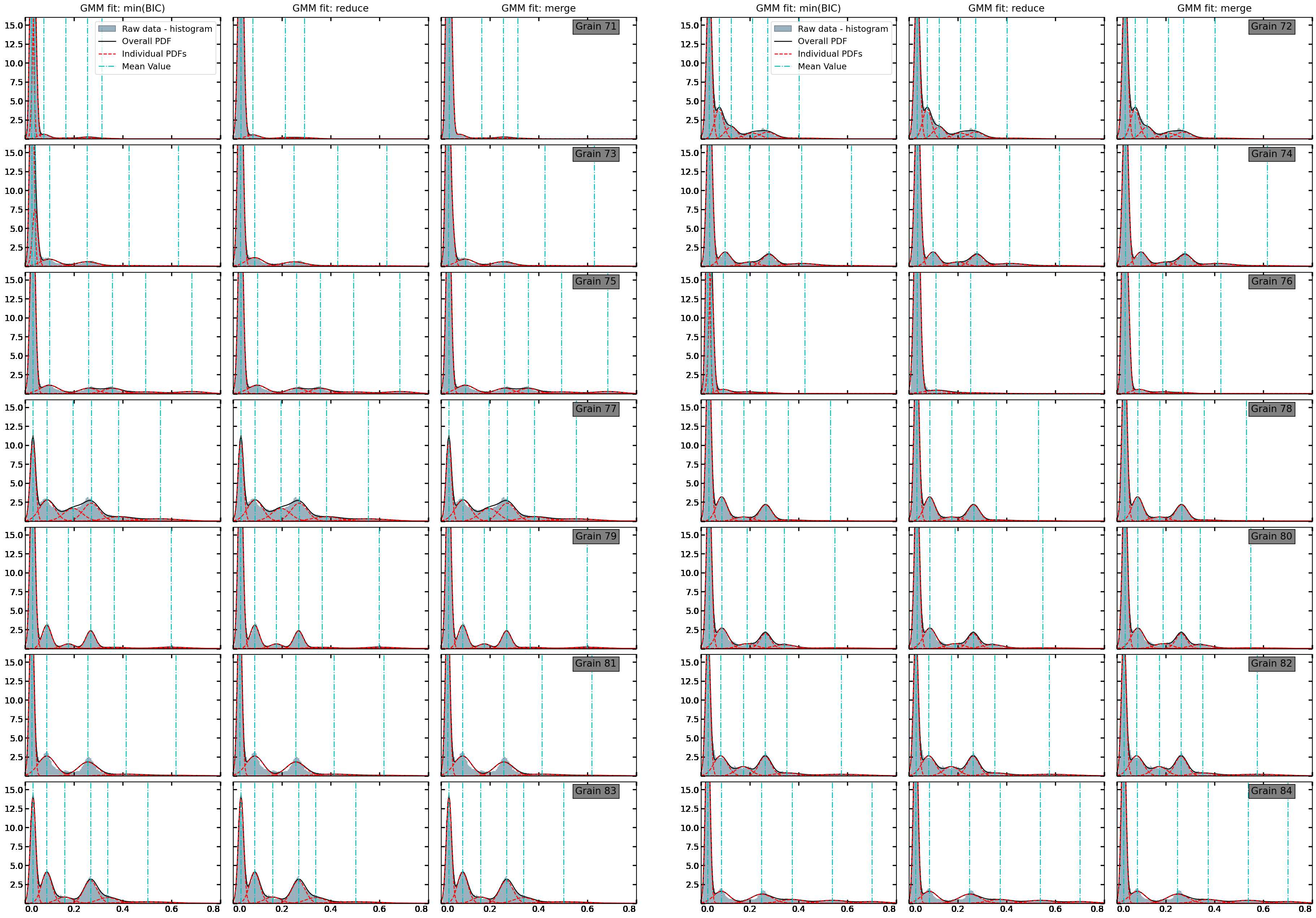}
        \subcaption{Results of Gaussian mixture model on the distribution of \Evm{} in grains 71 to 84.}
        \label{sfig:GMM_ShrStr71-84}
    \end{subfigure}
\end{figure}
\begin{figure}[htp!]\ContinuedFloat
    \begin{subfigure}[t!]{1.0\textwidth}
        \includegraphics[width=1.0\textheight, angle=90]{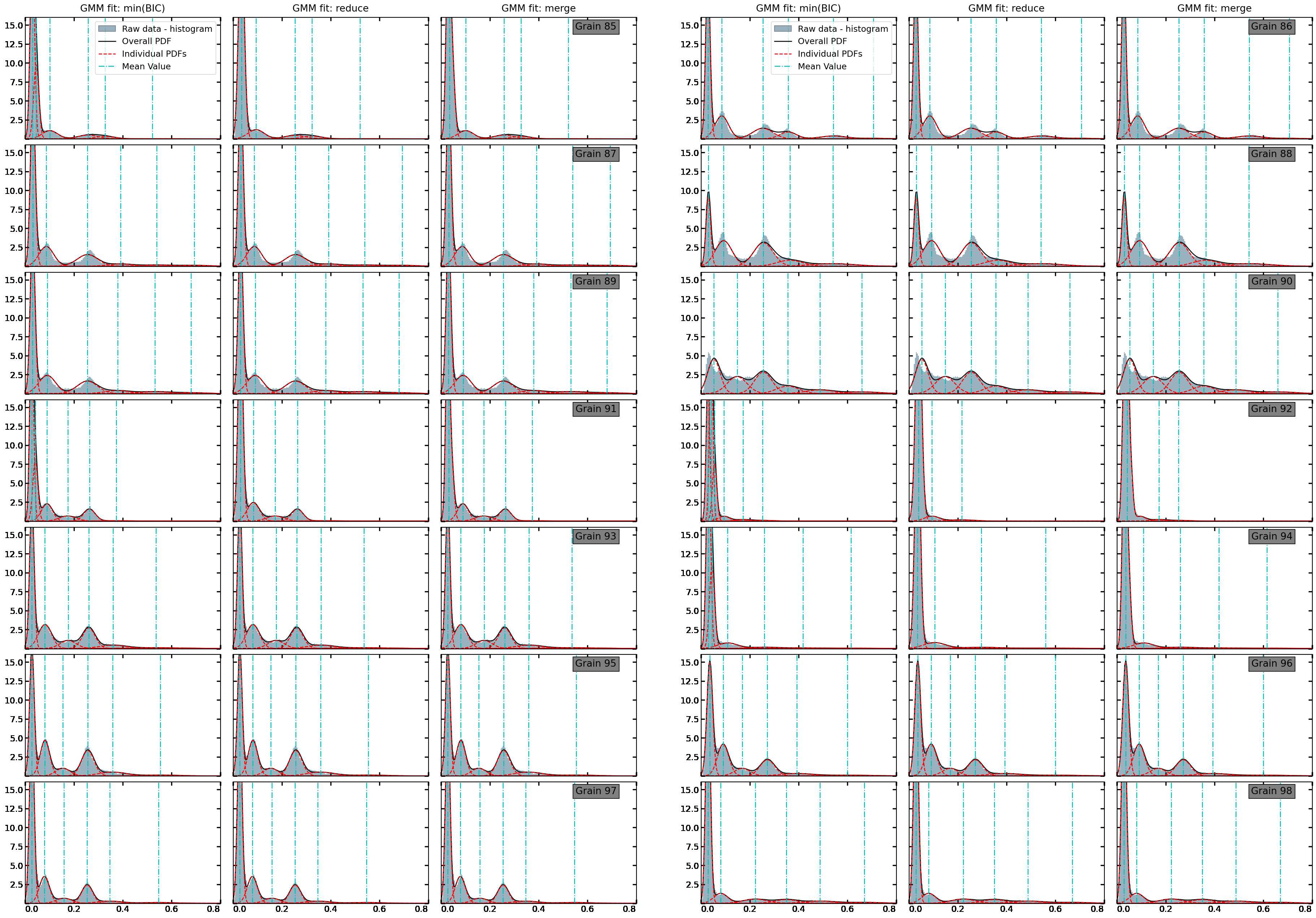}
        \subcaption{Results of Gaussian mixture model on the distribution of \Evm{} in grains 85 to 98.}
        \label{sfig:GMM_ShrStr85-98}
    \end{subfigure}
\end{figure}
\begin{figure}[htp!]\ContinuedFloat
    \begin{subfigure}[t!]{1.0\textwidth}
        \includegraphics[width=1.0\textheight, angle=90]{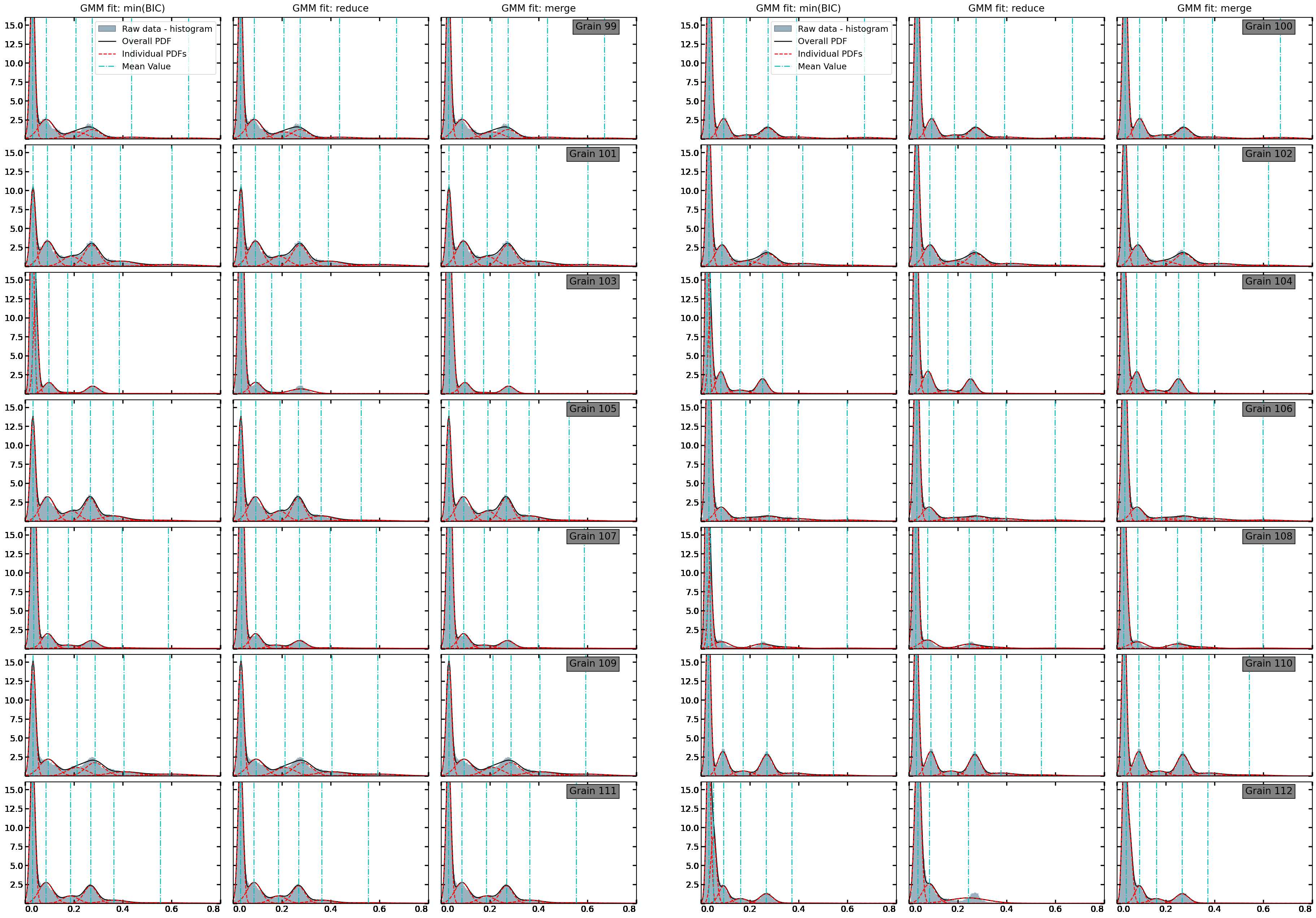}
        \subcaption{Results of Gaussian mixture model on the distribution of \Evm{} in grains 99 to 112.}
        \label{sfig:GMM_ShrStr99-112}
    \end{subfigure}
\end{figure}
\begin{figure}[htp!]\ContinuedFloat
    \begin{subfigure}[t!]{1.0\textwidth}
        \includegraphics[width=1.0\textheight, angle=90]{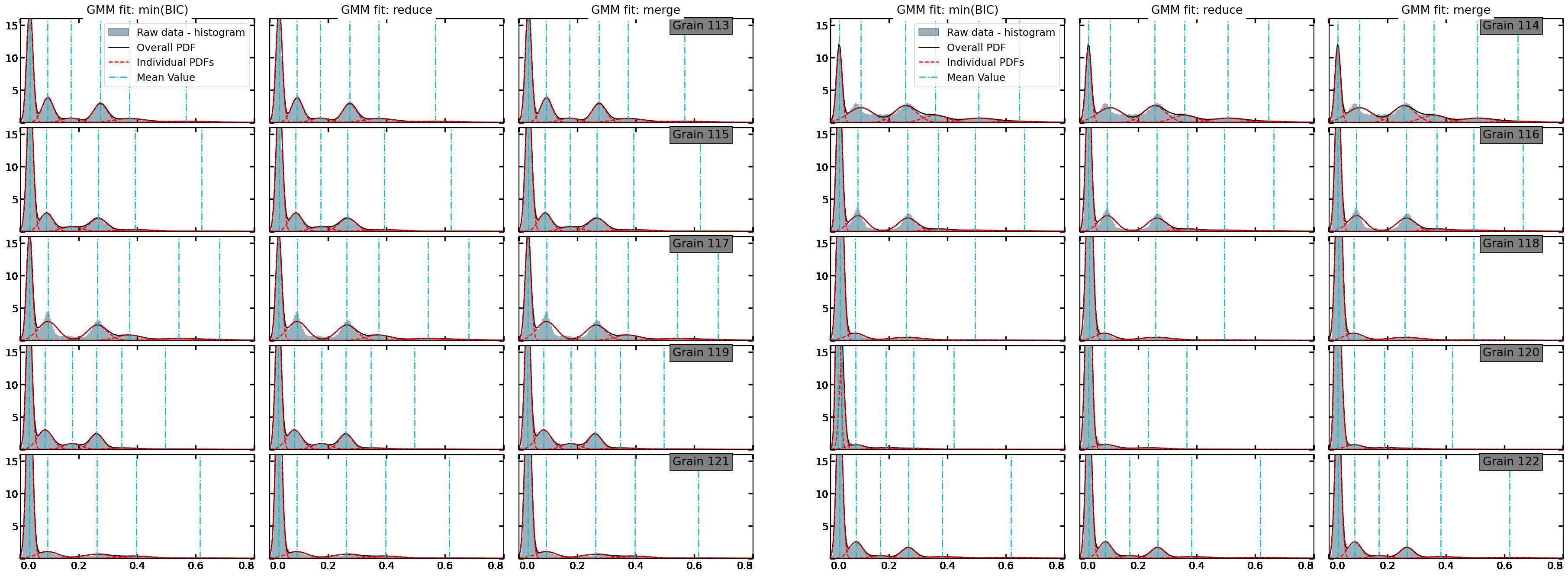}
        \subcaption{Results of Gaussian mixture model on the distribution of \Evm{} in grains 113 to 122.}
        \label{sfig:GMM_ShrStr113-122}
    \end{subfigure}
    \caption{Results of Gaussian mixture model on the distribution of \Evm{} in individual grains.}\label{sfig:GMM_ShrStr}
\end{figure}

\clearpage
\begin{figure}[htp!]
    \begin{subfigure}[t!]{1.0\textwidth}
        \includegraphics[width=1.0\textheight, angle=90]{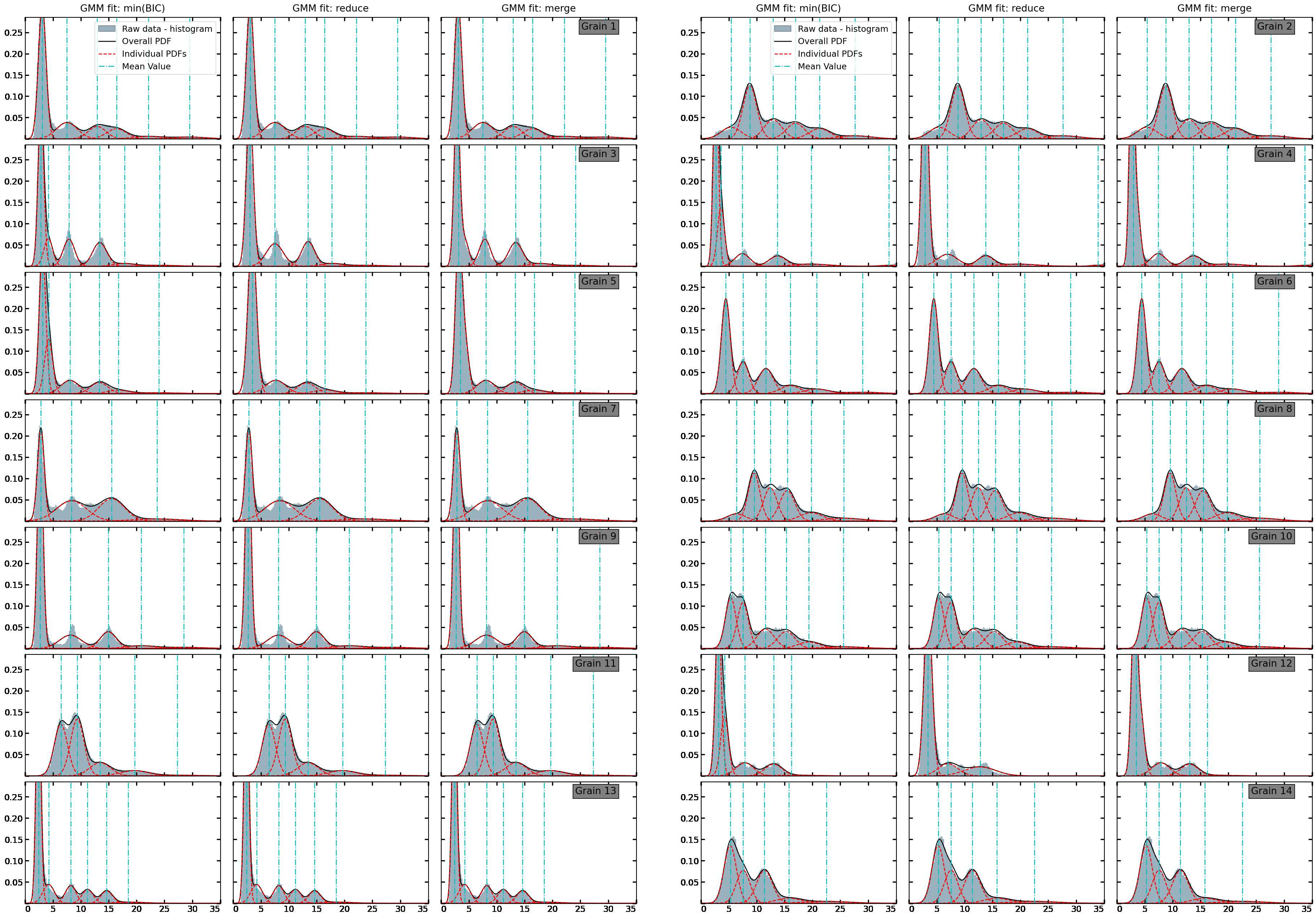}
        \subcaption{Results of Gaussian mixture model on the distribution of \drot{} in grains 1 to 14.}
        \label{sfig:GMM_DeltaRot1-14}
    \end{subfigure}
\end{figure}
\begin{figure}[htp!]\ContinuedFloat
    \begin{subfigure}[t!]{1.0\textwidth}
        \includegraphics[width=1.0\textheight, angle=90]{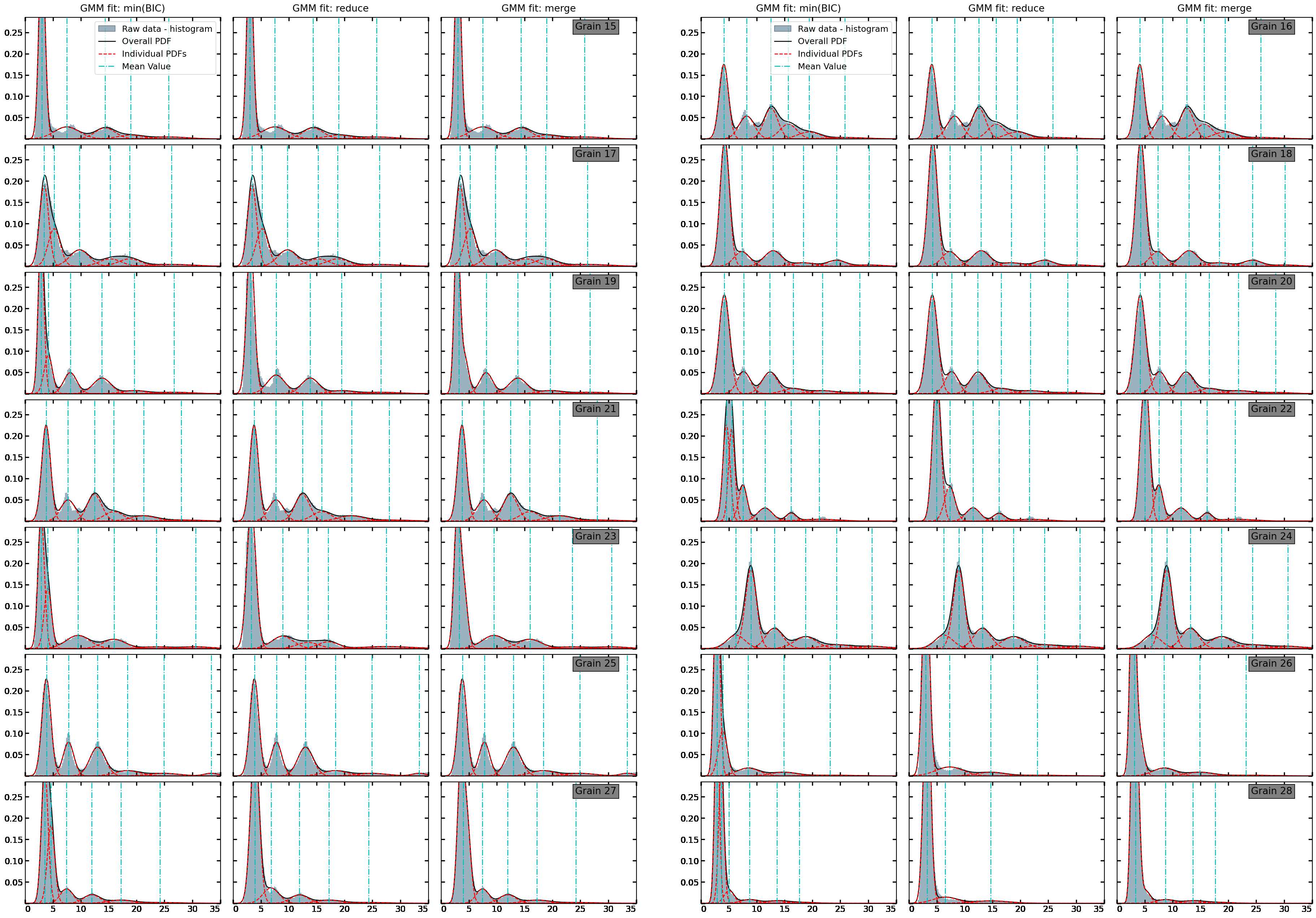}
        \subcaption{Results of Gaussian mixture model on the distribution of \drot{} in grains 15 to 28}
        \label{sfig:GMM_DeltaRot15-28}
    \end{subfigure}
\end{figure}
\begin{figure}[htp!]\ContinuedFloat
    \begin{subfigure}[t!]{1.0\textwidth}
        \includegraphics[width=1.0\textheight, angle=90]{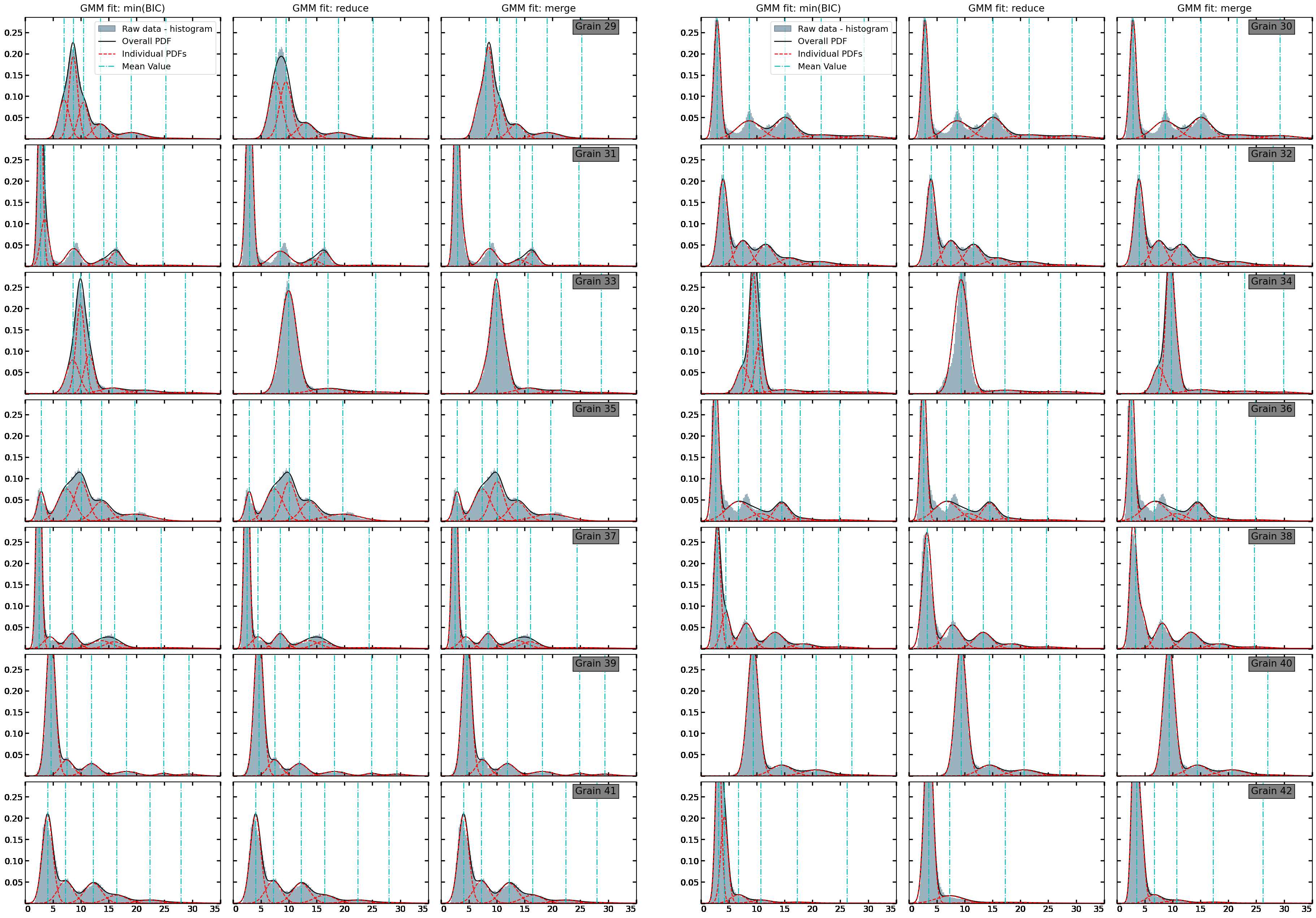}
        \subcaption{Results of Gaussian mixture model on the distribution of \drot{} in grains 29 to 42.}
        \label{sfig:GMM_DeltaRot29-42}
    \end{subfigure}
\end{figure}
\begin{figure}[htp!]\ContinuedFloat
    \begin{subfigure}[t!]{1.0\textwidth}
        \includegraphics[width=1.0\textheight, angle=90]{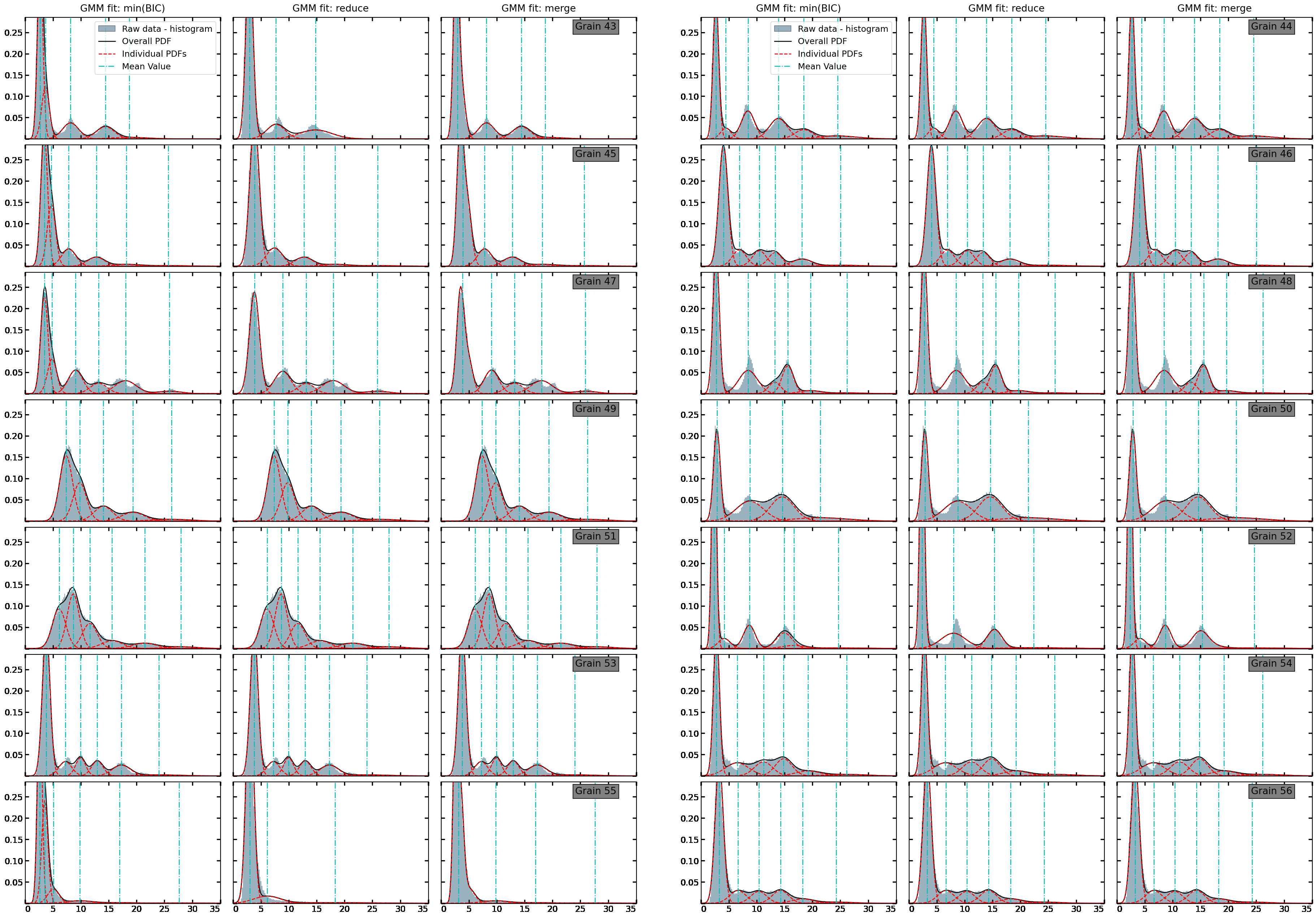}
        \subcaption{Results of Gaussian mixture model on the distribution of \drot{} in grains 43 to 56.}
        \label{sfig:GMM_DeltaRot43-56}
    \end{subfigure}
\end{figure}
\begin{figure}[htp!]\ContinuedFloat
    \begin{subfigure}[t!]{1.0\textwidth}
        \includegraphics[width=1.0\textheight, angle=90]{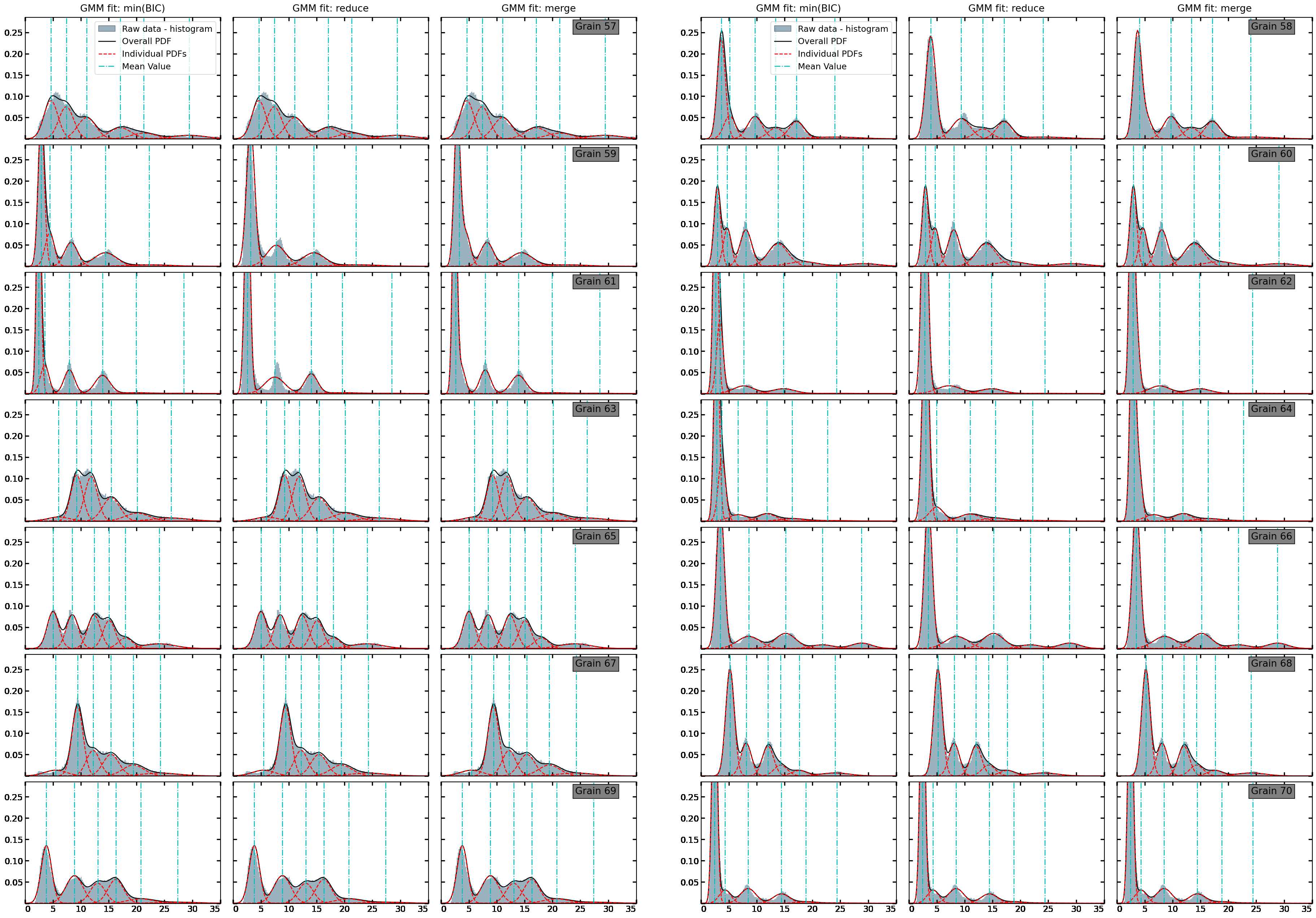}
        \subcaption{Results of Gaussian mixture model on the distribution of \drot{} in grains 57 to 70.}
        \label{sfig:GMM_DeltaRot57-70}
    \end{subfigure}
\end{figure}
\begin{figure}[htp!]\ContinuedFloat
    \begin{subfigure}[t!]{1.0\textwidth}
        \includegraphics[width=1.0\textheight, angle=90]{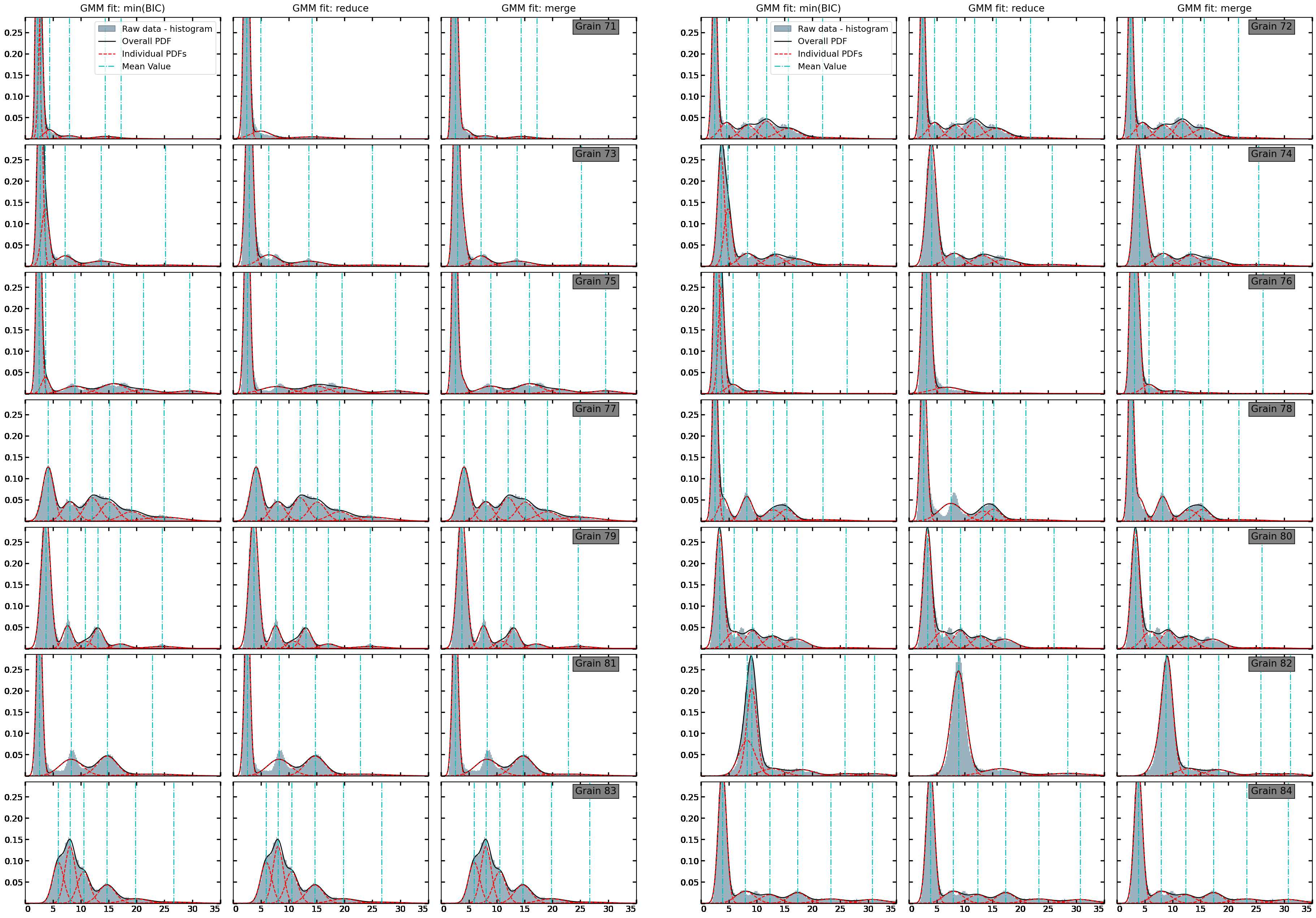}
        \subcaption{Results of Gaussian mixture model on the distribution of \drot{} in grains 71 to 84.}
        \label{sfig:GMM_DeltaRot71-84}
    \end{subfigure}
\end{figure}
\begin{figure}[htp!]\ContinuedFloat
    \begin{subfigure}[t!]{1.0\textwidth}
        \includegraphics[width=1.0\textheight, angle=90]{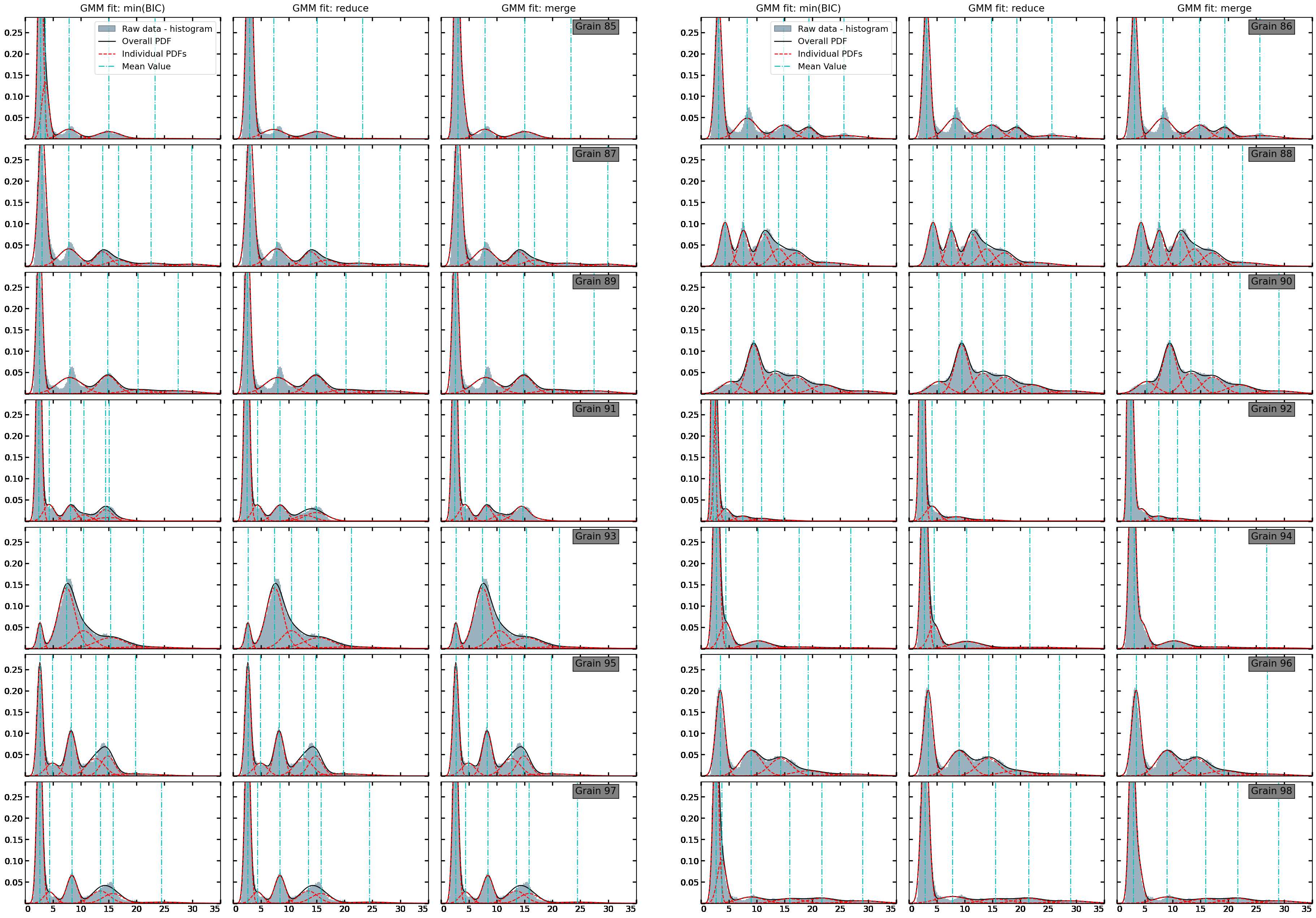}
        \subcaption{Results of Gaussian mixture model on the distribution of \drot{} in grains 85 to 98.}
        \label{sfig:GMM_DeltaRot85-98}
    \end{subfigure}
\end{figure}
\begin{figure}[htp!]\ContinuedFloat
    \begin{subfigure}[t!]{1.0\textwidth}
        \includegraphics[width=1.0\textheight, angle=90]{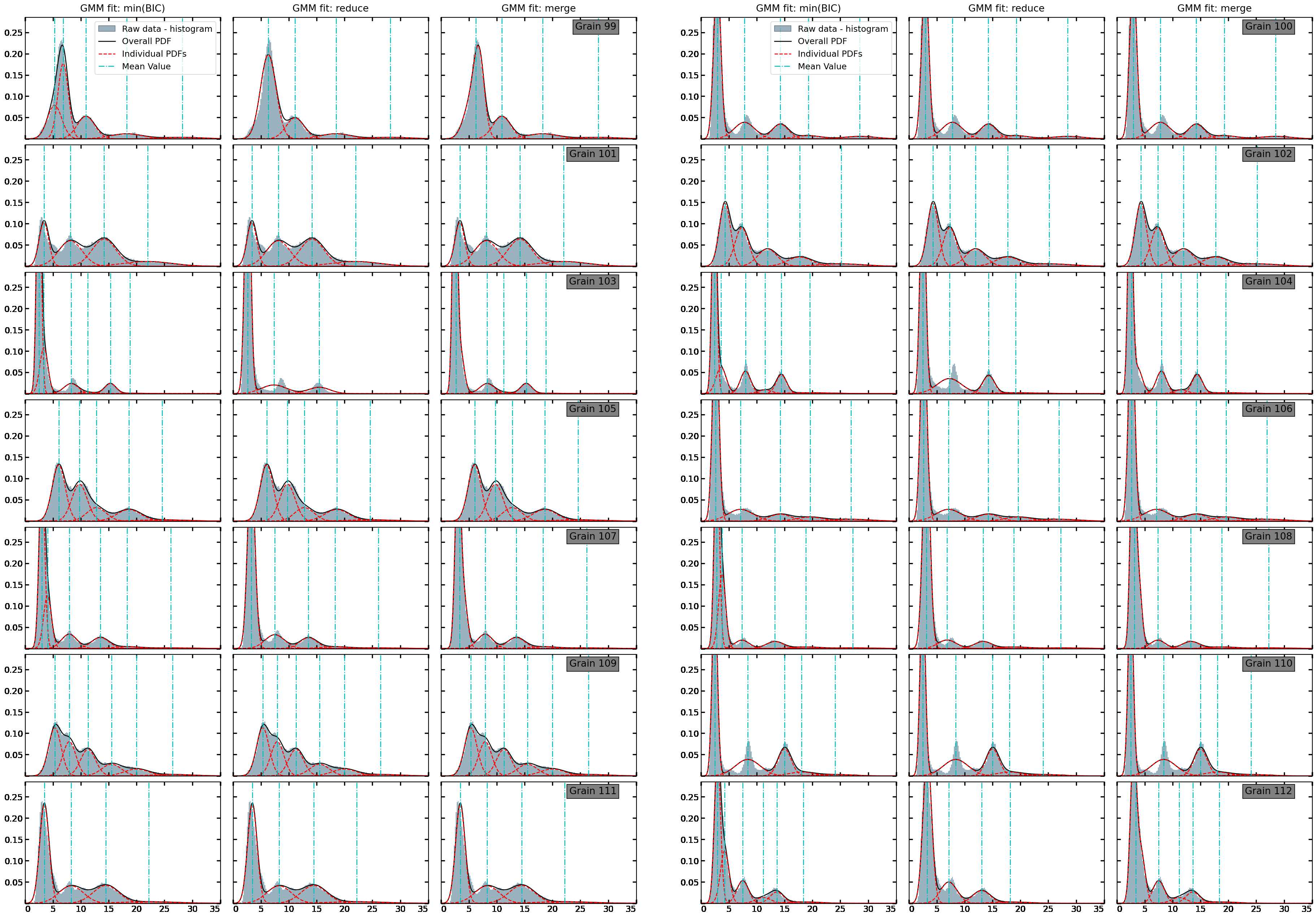}
        \subcaption{Results of Gaussian mixture model on the distribution of \drot{} in grains 99 to 112.}
        \label{sfig:GMM_DeltaRot99-112}
    \end{subfigure}
\end{figure}
\begin{figure}[htp!]\ContinuedFloat
    \begin{subfigure}[t!]{1.0\textwidth}
        \includegraphics[width=1.0\textheight, angle=90]{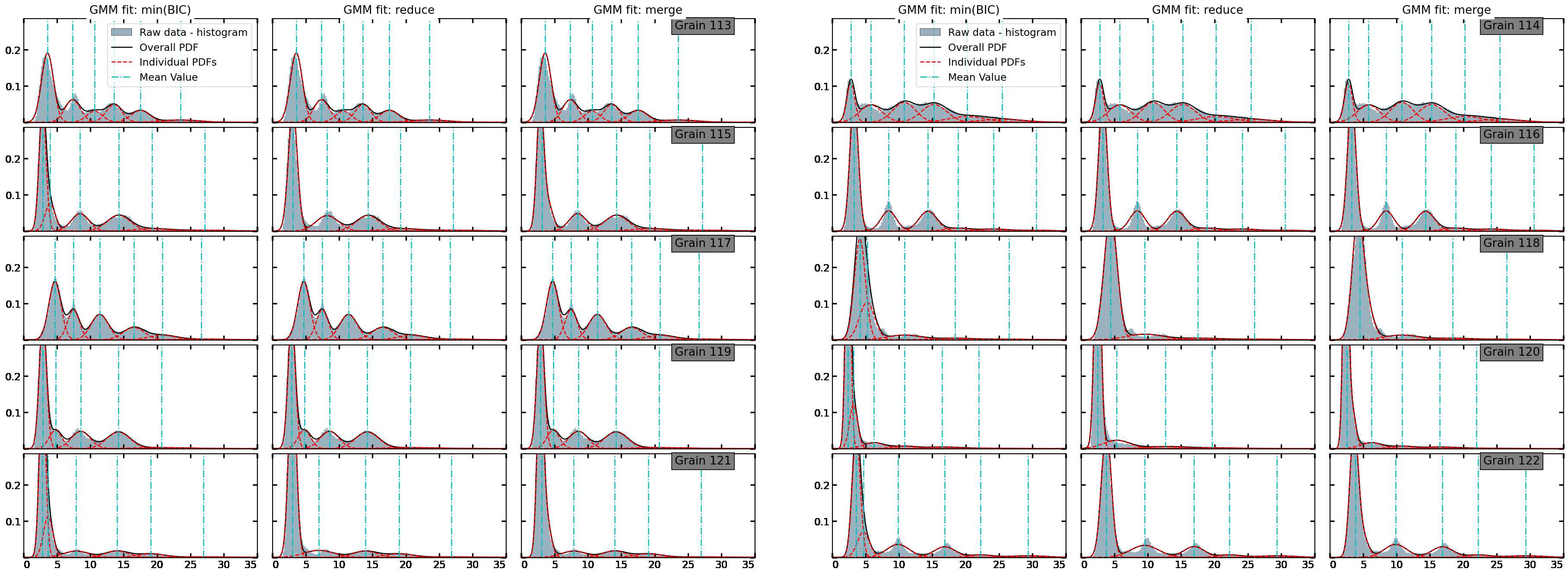}
        \subcaption{Results of Gaussian mixture model on the distribution of \drot{} in grains 113 to 122.}
        \label{sfig:GMM_DeltaRot113-122}
    \end{subfigure}
    \caption{Results of Gaussian mixture model on the distribution of \drot{} in individual grains.}\label{sfig:GMM_DeltaRot}
\end{figure}

\end{document}